\catcode`\@=11
\newif\if@fewtab\@fewtabtrue
{\count255=\time\divide\count255 by 60
\xdef\hourmin{\number\count255}
\multiply\count255 by-60\advance\count255 by\time
\xdef\hourmin{\hourmin:\ifnum\count255<10 0\fi\the\count255}}
\def\ps@draft{\let\@mkboth\@gobbletwo
    \def\@oddfoot{\hbox to 7 cm{\tiny \versionno
       \hfil}\hskip -7cm\hfil\rm\thepage \hfil {\tiny\draftdate}}
    \def\@oddhead{}
    \def\@evenhead{}\let\@evenfoot\@oddfoot}
\def\draftdate{\number\month/\number\day/\number\year\ \ \ \hourmin }

\def\citen#1{\if@filesw \immediate\write \@auxout {\string\citation{#1}}\fi%
\@tempcntb\m@ne \let\@h@ld\relax \def\@citea{}%
\@for \@citeb:=#1\do {\@ifundefined {b@\@citeb}%
    {\@h@ld\@citea\@tempcntb\m@ne{\bf ?}%
    \@warning {Citation `\@citeb ' on page \thepage \space undefined}}%
    {\@tempcnta\@tempcntb \advance\@tempcnta\@ne
    \setbox\z@\hbox\bgroup\ifcat0\csname b@\@citeb \endcsname \relax
    \egroup \@tempcntb\number\csname b@\@citeb \endcsname \relax
    \else \egroup \@tempcntb\m@ne \fi \ifnum\@tempcnta=\@tempcntb
    \ifx\@h@ld\relax \edef \@h@ld{\@citea\csname b@\@citeb\endcsname}%
    \else \edef\@h@ld{\hbox{--}\penalty\@highpenalty
    \csname b@\@citeb\endcsname}\fi
    \else \@h@ld\@citea\csname b@\@citeb \endcsname \let\@h@ld\relax \fi}%
\def\@citea{,\penalty\@highpenalty\hskip.13em plus.13em minus.13em}}\@h@ld}
\def\@citex[#1]#2{\@cite{\citen{#2}}{#1}}%
\def\@cite#1#2{\leavevmode\unskip\ifnum\lastpenalty=\z@\penalty\@highpenalty\fi%
  \ [{\multiply\@highpenalty 3 #1%
  \if@tempswa,\penalty\@highpenalty\ #2\fi}]}   %
\makeatother 
\catcode`\@=12

\def\apo           {{\rm S}}
\newcommand\apppicture[2] { \put(#2,0) {\bPo
                   \scalebox{.29}{\includegraphics{imgs/app#1.eps}}\eP}}

\def\be            {\begin{equation}}
\def\bearl         {\begin{array}{l}}
\def\bearll        {\begin{array}{ll}}
\newcommand\bee[5] {\begin{eqnarray} #5 \nonumber\\[-#1.#2em]~\\[#3.#4em]~\nonumber\end{eqnarray}}

\def\Bl            {\ensuremath{\mathcal B}}
\def\bP            {\begin{picture}}
\def\bPo           {\begin{picture}(0,0)}
\def\C             {{\ensuremath{\mathcal C}}}
\def\CA            {\ensuremath{\mathcal C_{\!\sss A}}}
\def\CAA           {\ensuremath{\mathcal C_{\!\sss A|A}}}
\def\CAB           {\ensuremath{\mathcal C_{\!\sss A|B}}}

\def\Cep           {{\ensuremath{\mathcal C_{(1,p)}}}}
\def\cft           {conformal field theory}

\def\Chi           {\mathcal X}
\def\chii          {\raisebox{.15em}{$\chi$}}
\def\cir           {\,{\circ}\,}

\def\Cob           {\ensuremath{\mathcal C\hspace{-1.1pt}\mbox{\sl ob}}}

\newcommand\coen[1] {\int^{#1}\hspace*{-.3em}}
\newcommand\coend[1]{\int^{#1}\hspace*{-.3em}#1^\vee{\otimes} #1}
\newcommand\Coend[2]{\int^{#2}\hspace*{-.3em}#1(#2,#2)}

\def\complex       {{\ensuremath{\mathbbm C}}}

\def\CopC          {{\ensuremath{\mathcal C\op\Times\mathcal C}}}
\def\COPC          {{\ensuremath{\mathcal C\op{\times}\hspace{1.1pt}\mathcal C}}}
\def\Cor           {\ensuremath{\mbox{\em Cor}}}
\def\CX            {\ensuremath{\mbox{\em Cor}(\X)}}
\def\CXCA          {\ensuremath{\mbox{\em Cor}_{(\C,A)}(\X)}}
\def\D             {\ensuremath{\mathcal D}}
\def\Deltaf        {\Delta_{\rm F}}
\def\dim           {\ensuremath{\mathrm{dim}}}
\def\dimc          {\ensuremath{\mathrm{dim}_\complex^{}}}
\def\drin          {\Phi} 
\def\Drinh         {{\widehat\Phi}} 
\def\dsty          {\displaystyle }

\def\ee            {\end{equation}}
\def\eE            {{\rm e}}
\def\eear          {\end{array}}

\def\End           {{\rm End}}
\def\eP            {\end{picture}}
\def\eps           {\varepsilon}
\def\epsf          {\varepsilon_{\rm F}}
\def\eq            {\,{=}\,}
\newcommand\eqpic[4]{\begin{picture}(#2,#3) #4 \end{picture}}
\newcommand\Eqpic[4]{\begin{eqnarray} \begin{picture}(#2,#3){}\end{picture}\nonumber\\
                   \raisebox{-#3pt}{ \begin{picture}(#2,#3) #4 \end{picture} }
                   \label{#1} \\~\nonumber \end{eqnarray} }
\newcommand\erf[1] {(\ref{#1})}
\def\etab          {\end{tabular}}
\def\F             {\ensuremath{\mathscr F}}
\def\fhmap         {{{\ensuremath{\widehat\Psi}}}}
\def\findim        {fini\-te-di\-men\-si\-o\-nal}
\def\fmap          {{{\ensuremath{\Psi}}}}
\def\fmapi         {{{\ensuremath{\Psi^{-}}}}}
\newcommand\Frac[2]{\mbox{\large$\frac{#1}{#2}$}}
\def\frho          {\rho_{\sss(2p)}}
\newcommand\Fs[6]  {{\sf F}_{\,{#5}\,{#6}}^{\,({#1}\,{#2}\,{#3})\,{#4}}}
\def\FUN           {{\ensuremath{\mathcal F}\!U\!N}}
\def\fus           {\,{*}\,}

\newcommand\Gs[6]  {{\sf G}_{\,{#5}\,{#6}}^{\,({#1}\,{#2}\,{#3})\,{#4}}}
\def\GX            {\ensuremath{\Gamma_{\!\mathrm Y}}}
\def\H             {\ensuremath{\mathscr H}}  
\def\Ha            {{\ensuremath{H^*_{}}}} 

\def\Hig           {\mathrm{Hig}}
\def\Hom           {{\rm Hom}}
\def\HomA          {{\rm Hom}_{\!A}}
\newcommand\Homaa[2]{\ensuremath{\HomAA(#1,#2)}}
\def\HomAA         {{\rm Hom}_{\!A|A}}
\def\HomAB         {{\rm Hom}_{\!A|B}}

\def\Homk          {{\rm Hom}_{\ko}}

\def\Hs            {{\ensuremath{H^\star}}}
\newcommand\hsp[1] {\hspace*{#1em}}
\def\I             {{\mathcal I}}
\def\ia            {{\ensuremath{\imath}}}
\def\ib            {{\ensuremath{{\bar\imath}}}}
\def\id            {{\rm id}}
\def\idA           {{\rm id}_{\!A}}
\def\idH           {\ensuremath{\id_H}}
\def\idHv          {\ensuremath{\id_{H^{\!\vee}}}}
\def\idsm          {\mbox{\tiny\sl id}}

\def\ii            {{\rm i}}

\def\iN            {\,{\in}\,}
\newcommand\Includepichopf[1] {{\begin{picture}(0,0)(0,0)
                   \scalebox{.38}{\includegraphics{imgs/pic_hopf_#1.eps}}\end{picture}}}
\newcommand\Includepichopfsm[1] {{\begin{picture}(0,0)(0,0)
                   \scalebox{.266}{\includegraphics{imgs/pic_hopf_#1.eps}}\end{picture}}}
\newcommand\Includepichtft[1] {{\begin{picture}(0,0)(0,0)
                   \scalebox{.38}{\includegraphics{imgs/pic_htft_#1.eps}}\end{picture}}}
\def\infdim        {in\-fi\-ni\-te-di\-men\-si\-o\-nal}

\def\Itemize       {\def\leftmargini{1.01em}~\\[-2.66em]\begin{itemize}\addtolength\itemsep{-6pt}}
\def\itx           {\item[\raisebox{.08em}{\rule{.44em}{.44em}}]}
\def\J             {\ensuremath{J}}
\def\ja            {{\ensuremath{\jmath}}}
\def\jb            {{\ensuremath{{\bar\jmath}}}}

\def\kap           {\kappa}

\def\kbottle       {\ensuremath{\mathrm K}}
\def\kc            {\ensuremath{\kO(\mathcal C)}}
\def\ko            {\ensuremath{\Bbbk}}
\def\kO            {\ensuremath{G_0}}
\newcommand\labl[1]{\label{#1}\ee}

\def\Mod           {\mbox{-mod}}
\def\mosp          {\mathfrak M} 

\def\MX            {\ensuremath{{\mathscr M}_{\mathrm Y}}}
\newcommand\N[3]   {{N_{#1#2}}^{\!#3}}
\newcommand\Nh[3]  {\widehat N_{#1#2}^{~\,#3}}
 
\newcommand\NNh[1] {\widehat N^{}_{#1}}
\def\nul           {\circ}
\def\mtc           {modular tensor category}
\def\mtcs          {modular tensor categories}
\def\nE            {\,{\ne}\,}
\newcommand\nxl[1] {\\[#1pt]}
\newcommand\nxm[1] {\\~\\[-#1pt]}
\def\nxt           {{\raisebox{.08em}{\rule{.44em}{.44em}}~~}}

\def\one           {{\bf1}}
\def\onedim        {one-dimen\-sio\-nal}

\def\op            {^{\mathrm{op}}}

\def\oti           {\,{\otimes}\,}
\def\Oti           {{\otimes}}
\def\otic          {\,{\otimes^{}_{\sss\complex}}\,}
\def\otim          {\,{\otimes^-}\,}
\def\Otim          {{\otimes^{\!-}}}
\def\otip          {\,{\otimes^+}\,}
\def\Otip          {{\otimes^{\!+}}}

\def\pchi          {\psi}
\def\PicC          {\ensuremath{\mathrm{Pic}(\C)}}
\def\PicCAA        {\ensuremath{\mathrm{Pic}(\CAA)}}

\def\qquand        {\qquad{\rm and}\qquad}

\def\rep           {re\-pre\-sen\-ta\-ti\-on}
\def\Rep           {\ensuremath{\mathcal{R}ep}}

\def\repV          {{\ensuremath{\mathcal{R}ep(\V)}}}
\def\Rey           {\mathrm{Rey}}
\newcommand\Rs[4]  {{\sf R}^{#1\,(#2\,#3)\,#4}}
\def\Sb            {{\ensuremath{\check s}}}
\def\scs           {\scriptstyle}
\def\Schi          {{\ensuremath{S^{\sss\chi}_{\phantom|}}}}
\newcommand\setulen[2]{\setlength\unitlength{.#1#2pt}}
\newcommand\setUlen[2]{\setlength\unitlength{#1.#2pt}}
\def\SF            {\varSigma}

\def\Sl            {{\ensuremath{\hat s}}}
\def\slz           {\ensuremath{\mathrm{SL}(2,\zet)}}
\def\So            {{\ensuremath{S^{\sss\otimes}_{\phantom;}}}}
\def\SO            {S^{\sss\otimes}}
\def\soc           {\mathrm{Soc}}
\def\Soinv         {{\ensuremath{S^{\sss\otimes\,-1}_{\phantom;}}}}
\def\soo           {{\ensuremath{s^{\dsty\circ\!\!\circ}_{\phantom;}}}}
\def\sOO           {s^{\dsty\circ\!\!\circ}}
\def\Soo           {{\ensuremath{S^{\dsty\circ\!\!\circ}_{\phantom;}}}}

\def\SOstar        {S^{\sss\otimes^{\scs *}}}

\def\sse           {\scriptstyle}
\def\ssfa          {symmetric special Fro\-be\-ni\-us algebra} 
\def\ssg           {\scriptstyle}

\def\sss           {\scriptscriptstyle}
\def\sxchi         {{\ensuremath{\widehat S}^{\sss\chi_{\phantom;}}}}
\def\Sxchi         {{\ensuremath{\widehat S}^{\sss\chi_{\phantom;}}_{}}}
\def\tft           {topological field theory}
\def\tftc          {\ensuremath{\mbox{\tt tft}_\C^{}}}
\def\threedim      {three-di\-men\-sio\-nal}
\def\Times         {\,{\times}\,}
\def\To            {\,{\to}\,}
\def\torus         {\ensuremath{\mathrm T}}
\newcommand\turlabg[1]{{\begin{turn}{90}$\scs #1$\end{turn}}}
\newcommand\Turlabg[1]{{\begin{turn}{270}$\scs #1$\end{turn}}}
\def\twodim        {two-di\-men\-sio\-nal}
\def\TX            {\ensuremath{\Gamma_{\!\mathrm Y}^\circ}}
\def\U             {S}  
\def\uiv           {{b^{-1}}}
\def\uqz           {\ensuremath{\overline{\mathrm U_q}(\mathfrak{sl}_2)}}

\def\V             {\ensuremath{\mathscr V}}
\def\Vect          {\ensuremath{{\mathcal V}\!\mbox{\sl ect}^{}_\complex}}
\def\Vectk         {\ensuremath{{\mathcal V}\!\mbox{\sl ect}^{}_\ko}}
\def\Vee           {{}{}^{\vee\!}}
\def\Vep           {{\ensuremath{\mathscr V_{(1,p)}}}}

\newcommand\void[1]{}
\def\vPhi          {\varPhi}

\def\vhi           {\vPhi}

\def\wrho          {\widehat\rho_{\sss(3p-1)}}
\def\wrhp          {\rho_{\sss(3p-1)}}
\def\X             {\ensuremath{\mathrm Y}}
\def\Xh            {{\ensuremath{\widehat{\mathrm Y}}}}

\def\Z             {{\mathrm Z}}
\def\zet           {{\ensuremath{\mathbb Z}}}
\def\zzmatrixS     {\mbox{$\Big(\!\begin{array}{cc}\!\scs 0\!\!\!&\!\scs-1\!%
                    \\[-3pt]\!\scs 1\!\!\!&\!\scs 0\!\eear\!\Big)$}}

\documentclass[12pt]{article}
\usepackage{latexsym, amsmath, amsthm, amsfonts, enumerate, amssymb, bbm, xspace, xypic }
\usepackage[all]{xy}
\usepackage[mathscr]{eucal}
\usepackage[pagewise, modulo]{lineno}
\usepackage{graphicx} \usepackage{rotating}
\usepackage{epstopdf,hyperref}
\usepackage{hyperref}

\begin{document}
\def\cir{\,{\circ}\,} 
\numberwithin{equation}{section}

\thispagestyle{empty} 
\begin{flushright}
   {\sf ZMP-HH/10-11}\\
   {\sf Hamburger$\;$Beitr\"age$\;$zur$\;$Mathematik$\;$Nr.$\;$372}\\[2mm]
   April 2010
\end{flushright}
\vskip 2.0em
\begin{center}\Large
HOPF ALGEBRAS AND FINITE TENSOR CATEGORIES\\[.4em] IN CONFORMAL FIELD THEORY
\end{center}\vskip 1.4em
\begin{center}
  J\"urgen Fuchs $^{a}$
  ~~and~~ Christoph Schweigert $^b$
\end{center}

\vskip 6mm

\begin{center}
  $^a$ Teoretisk fysik, \ Karlstads Universitet
  \\Universitetsgatan 21, \ S\,--\,\,651\,88 Karlstad
 \\[7pt]
  $^b$ Organisationseinheit Mathematik, \ Universit\"at Hamburg\\
  Bereich Algebra und Zahlentheorie\\
  Bundesstra\ss e 55, \ D\,--\,20\,146\, Hamburg
\end{center}

\vskip 3em

\noindent{\sc Abstract}
\\[3pt]
In conformal field theory the understanding of correlation functions can be divided 
into two distinct conceptual levels:
The analytic properties of the correlators endow the representation categories of 
the underlying chiral symmetry algebras with additional structure, which in 
suitable cases is the one of a finite tensor category. The problem of specifying the 
correlators can then be encoded in algebraic structure internal to those categories.
\\
After reviewing results for conformal field theories for which these representation 
categories are 
semisimple, we explain what is known about representation categories of chiral 
symmetry algebras that are not semisimple. We focus on generalizations of the 
Verlinde formula, for which certain finite-dimensional complex Hopf algebras are used as a 
tool, and on the structural importance of the presence of a Hopf algebra internal 
to finite tensor categories.

\newpage

\section{Introduction}

For at least twenty-five years, \twodim\ conformal quantum field theory -- or CFT, for
short -- has engaged physicists (working for instance on critical systems in statistical 
mechanics, on quasi \onedim\ condensed matter systems, or on string theory) and 
mathematicians (concerned with e.g.\ \infdim\ algebra, operator algebras, topology, 
the theory of mo\-du\-lar forms, or algebraic geometry) alike \cite{fuRs13}. 

A crucial technical ingredient in the algebraic study of CFT is a certain finiteness property
which distinguishes among the various models those which are easiest to understand.
More specifically, conserved quantities lead to an algebraic structure \V, called the 
\emph{chiral symmetry algebra}, and much information about a CFT model is encoded in 
the \rep\ category $\C\,{\simeq}\,\repV$ of \V, called the category of chiral 
data. The relevant finiteness property is then that \C\ has the structure of a 
\emph{finite tensor category} in the sense of \cite{etos}. This class of categories
contains in particular the \emph{modular tensor categories}, which are semisimple. 
Another class of examples of (braided) finite tensor categories is provided 
by the categories of modules over \findim\ Hopf algebras, while Hopf algebras 
internal to finite tensor categories allow for a generalization of the notion of
\mtc\ that encompasses also non-semisimple categories. In this report we present an 
overview of some pertinent aspects of CFT related to finite tensor categories; 
semisimple categories, categories of modules over Hopf algebras, and general 
(not necessarily semisimple) modular categories are considered in sections 
\ref{sec.corr.cb}, \ref{sec.verl} and \ref{sec.coend}, respectively.

\medskip

Of central interest in CFT (or, for that matter, in any quantum field theory) are the 
correlation functions. A correlation function is, roughly, a linear functional on an
appropriate tensor product of state spaces that depends on geometric data, compatible 
with various structures on those data. Correlation functions are often of direct
relevance in applications. The relevant state spaces are related, by some state-field 
correspondence, to field operators which, in turn, can describe observable quantities 
like for instance quasi-particle excitations in condensed matter physics. 

In CFT, the geometric data for a correlation function include a \twodim\ manifold \X\ 
(with additional structure) called the \emph{world sheet}. They include in particular 
the genus and the moduli of a conformal structure on \X\ as well as the location of 
field insertion points on \X. To each insertion point there is associated an 
appropriate state space depending on the type of field insertion. These state spaces, 
as well as further aspects of the world sheet like e.g.\ boundary conditions, 
are specified with the help of certain \emph{decoration data}, in addition to the
chiral data that are given by the category \C.

Essential information about CFT correlation functions can already be obtained when 
one knows the category \C\ just as an abstract category (with additional properties,
inherited from the chiral symmetry algebra \V) rather than concretely realized as a 
\rep\ category. In particular, the problem of distinguishing between different CFT
models that possess the same chiral symmetries, and thus the same chiral data, can be 
addressed on this purely categorical level. Indeed, a complete solution of this 
problem has been given in the case that \C\ is a (semisimple) \mtc. The corresponding 
CFT models are called (semisimple) \emph{rational} CFTs, or RCFTs.

\medskip

For a rational CFT, a construction of all correlation functions can be achieved
by expressing them in terms of invariants of suitable three-manifolds (with boundary). 
Apart from the \mtc\ \C\ this construction requires as one additional input datum a 
certain Frobenius algebra internal to \C, and it makes heavy use of a \threedim\ 
topological field theory that is associated to \C. This construction is outlined 
in sections \ref{subsec.TFT}\,--\,\ref{subsec.digest}. Prior to that we provide, 
in sections \ref{subsec.corr}\,--\,\ref{subsec.solu}, some relevant details about 
CFT correlation functions, and in particular about the consistency conditions they 
are required to satisfy.

The analysis of non-rational CFTs turns out to be much harder, and so far no
construction of correlation functions similar to the RCFT case is available. We
discuss two issues which are expected to be relevant for gaining a better 
understanding of non-rational CFTs. First, section \ref{sec.verl} is devoted to
relations between fusion rules and modular transformations of characters. In
section \ref{verl.L1p} we present some Verlinde-like relations that have been 
observed for a specific class of models, while in section \ref{hopf.L1p} 
certain \findim\ Hopf algebras related to these models are described; a
Verlinde-like formula for the Higman ideal of any factorizable ribbon Hopf algebra
is quoted in section \ref{ssec.cowe}.

The second issue, a generalization of the notion of \mtc\ that includes 
non-semisimple finite tensor categories, is the subject of section \ref{sec.coend}. 
This generalization, given in section \ref{subsec.mtcH}, makes use of a categorical
Hopf algebra that is defined as the coend of a functor related to rigidity.
Remarkably, this Hopf algebra gives rise both to three-manifold invariants (section 
\ref{subsec.invH}) and to representations of mapping class groups (section 
\ref{subsec.mtcH}), albeit they do not quite fit together. One may thus suspect 
that in the study of conformal field theory, methods from \threedim\ topological 
field theory can still be relevant also for non-rational CFTs.

\vskip 3.5em

\noindent$~\,${\sc Acknowledgments:}\nxl8
\begin{tabular}{ll}\multicolumn2l{%
\begin{minipage}{37.6em}
JF is partially supported by VR under project no.\ 621-2009-3343. \\
CS is partially supported by the DFG Priority Program 1388 `Representation theory'.
\end{minipage}}
\\[-4pt] \begin{minipage}{18.0em}
This paper is partly based on a lecture by JF at the Coloquio de \'Algebras de Hopf, 
Grupos Cu\'anticos y Categor\'{\i}as Tensoriales at La Falda, C\'ordoba, in September 2009.
\\[2pt]
JF wishes to thank the organizers, and in particular N.\ Andruskiewitsch and 
J.M.\ Mombelli, for their hospitality at C\'ordoba and La Falda.
\end{minipage} &
\bP(0,83) \put(14,-60){\scalebox{.43}{\includegraphics{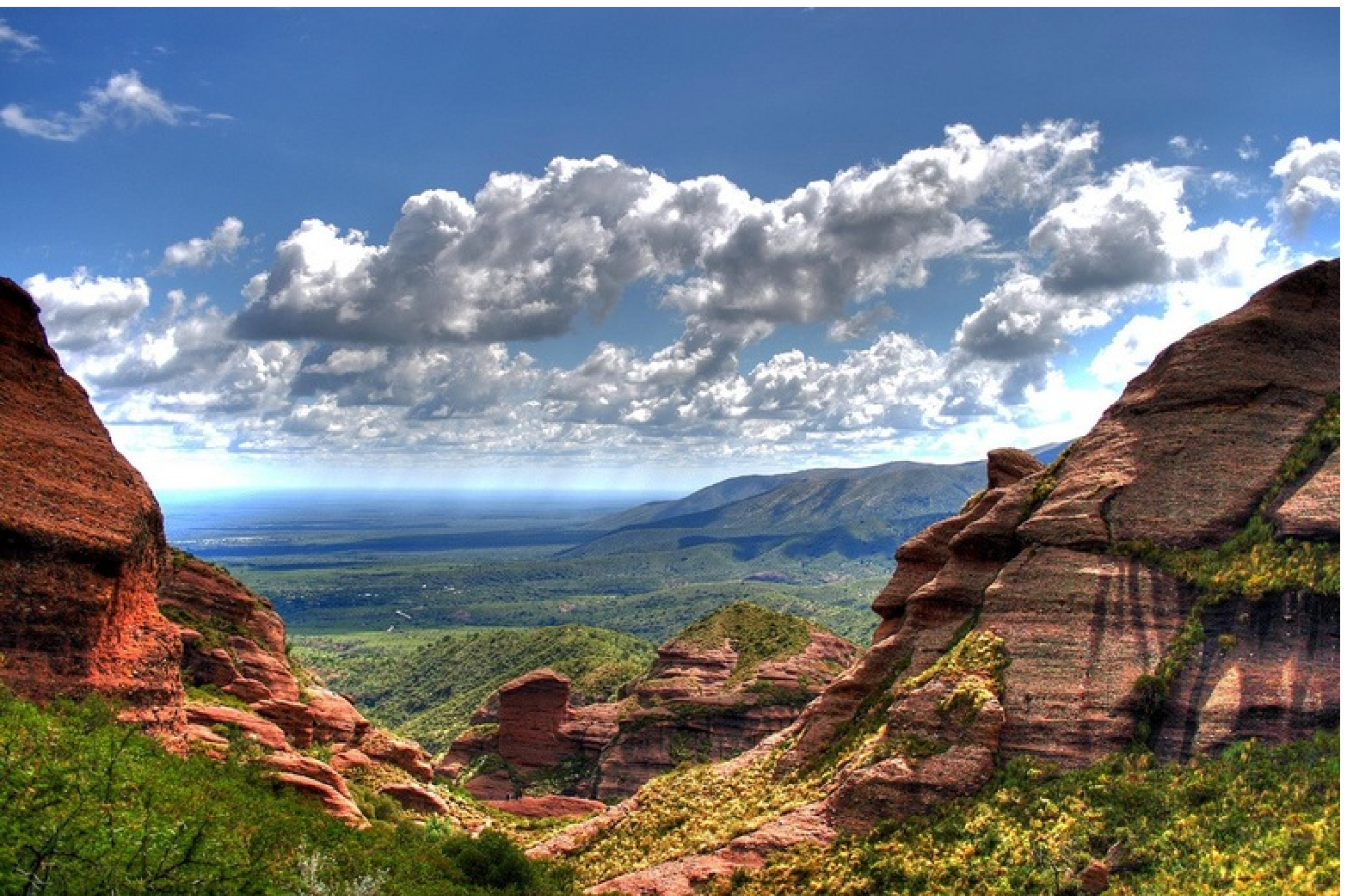}}} \eP
\end{tabular}

 \newpage

\section{Correlators and conformal blocks}\label{sec.corr.cb}

\subsection{CFT correlation functions}\label{subsec.corr}

A world sheet \X\ is characterized by its topology, i.e.\ the genus and the number of 
boundary components, and the number $m$ of marked points on \X\ (further aspects of
\X\ will be given below). Denote by $\mosp{}_\X$ the moduli space of conformal 
structures and locations of the marked points. As already indicated in the 
introduction, a correlation function, or \emph{correlator}, of a CFT is a function
  \be
  \CX:\quad \mosp{}_\X^{} \times \vec{\mathcal H} \,\to\, \complex \,,
  \ee
with $\vec{\mathcal H}$ an $m$-tuple of state spaces
  \be
  \vec{\mathcal H} = \mathcal H_1 \times\, \mathcal H_2 \times \dots \times \mathcal H_m
  \ee
assigned to the marked points. The state spaces are (\infdim) complex vector spaces,
and \CX\ is multilinear in $\vec{\mathcal H}$.
For our purposes we do not need to spell out all details of this description,
but the following information concerning the world sheet data and the vector spaces 
$\mathcal H_\alpha$ for $\alpha\eq1,2,...\,,m$ will be instrumental:

\Itemize

\itx
Each of the state spaces $\mathcal H_\alpha$ is a \rep\ space of either the chiral 
symmetry algebra $\V$ or of the tensor product $\V\otic\V$ of two copies of \V,
depending on whether the corresponding marked point lies on the boundary or in the
interior of the world sheet.

\itx
One framework for describing the chiral symmetry algebra \V\ in technical terms is by
the structure of a conformal vertex algebra, that is, an \infdim\ $\zet_{\ge0}$-graded 
complex vector space endowed with a product that depends on a formal parameter and 
with various additional structure. The precise form of this algebraic structure will, 
however, not concern us here; rather, it will be the properties of \,\C, rather than 
\V\ itself, that are relevant to us.\,%
 \footnote{~In fact there is a different approach to CFT, based on nets of von Neumann
 algebras instead of conformal vertex algebras, which is essentially equivalent at
 the categorical level. For some details and references see e.g.\ \cite{koRu3}.}
The crucial aspect for us is that there are notions of \V-module and of \V-module map 
(or intertwiner), which furnish the objects and morphisms, respectively, of the \rep\ 
category \repV\ of \V, and that one can define a tensor product of \V-modules 
\cite{hule3}, such that \repV\ is a monoidal category. Below it will be convenient to 
work with a strict monoidal category equivalent to \repV; we denote this category by \C. 

\itx
The \emph{world sheet} $\X \,{\equiv}\, (\X;\mbox{\boldmath$\tau$},(\vec p,\vec V))$ 
is a compact surface \X\ endowed with a conformal structure {\boldmath$\tau$}, 
i.e.\ an equivalence class, with respect to conformal rescalings, 
of (Euclidean) metrics, and with a finite collection $\vec p \eq \{p_\alpha\iN\X 
\,|\, \alpha\eq1,2,...\,,m\}$ of pairwise distinct points with labeling 
$\vec V \eq \{V_\alpha \,|\, \alpha\eq1,2,...\,,m\}$. \X\ can in particular have 
a non-empty boundary $\partial\X$, and it need not be orientable. 
The points $p_\alpha$ are referred to as \emph{insertion points}. Their labeling
$V_\alpha$ is given by either an object of \C\ -- if $p_\alpha$ lies on the 
boundary  $\partial\X$ -- or by a pair $(V_\alpha',V_\alpha'')$ of objects of 
\C\ --  if $p_\alpha$ lies in the bulk $\X{\setminus}\partial\X$.

\itx
Insertion points on $\partial\X$ partition the boundary $\partial\X$ into
segments. Each of these segments carries a decoration, which is called a 
\emph{boundary condition}. This decoration is again an object of a category,
namely of the category of $A$-modules for some algebra $A$ in \,\C\ (see the
dictionary at the end of section \ref{sec.main}).

\itx
Recalling that \C\ is the \rep\ category of \V, we see that for $p_\alpha\iN\partial\X$ 
the vector space $\mathcal H_\alpha$ is just the object $V_\alpha$ of \C\ regarded
concretely as a \V-module, while for $p_\alpha\iN\X{\setminus}\partial\X$ it is the
space $V_\alpha'\otic V_\alpha''$, regarded as a module over $\V\otic\V$. The 
construction of CFT correlators is compatible with direct sums; hence for 
rational CFTs, for which \C\ is semisimple, one may without loss of generality
assume that $V_\alpha$, respectively $V_\alpha'$ and $V_\alpha''$, are simple.

\end{itemize}

In physics terminology, an insertion point corresponds to the presence of a 
\emph{field insertion}. When the insertion point lies on the boundary of the 
world sheet, $p_\alpha\iN\partial\X$, one says that the field insertion ia a
\emph{boundary field}, while for $p_\alpha\iN\X{\setminus}\partial\X$ one speaks
of a \emph{bulk field}. 
The insertion points actually come with additional structure, which can be encoded
in the choice of a local coordinate system at each point. These data will be 
largely suppressed in the sequel, except in some of the pictures below, in which 
their presence is indicated either by drawing small oriented
arcs passing through the points, or by replacing them by small oriented intervals or 
circles (for boundary and bulk insertions, respectively). Another structure that
can be present on \X, and which for brevity will below be suppressed as well, are 
\emph{defect lines} which partition the world sheet in different regions that
support different \emph{phases} of the theory. Accordingly there is also another
type of field insertions: \emph{defect fields}, which can change the type of 
defect line. For these, and for various other details, too, we refer the reader to 
e.g.\ \cite{scfu4}, sections 2 and 4 of \cite{scfr2} and section 1 of \cite{ffrs5}.

As an illustration, the following pictures show two examples of world sheets
without defect lines: a closed one of genus two with two bulk insertion points, and 
one of genus one with three boundary components as well as with 
two bulk (at $p_1$ and $p_5$) and three boundary (at $p_2,\,p_3,\,,p_4$) insertions.

\bP(200,110)(-60,0) \put(-8,14) {\scalebox{.46}{\includegraphics{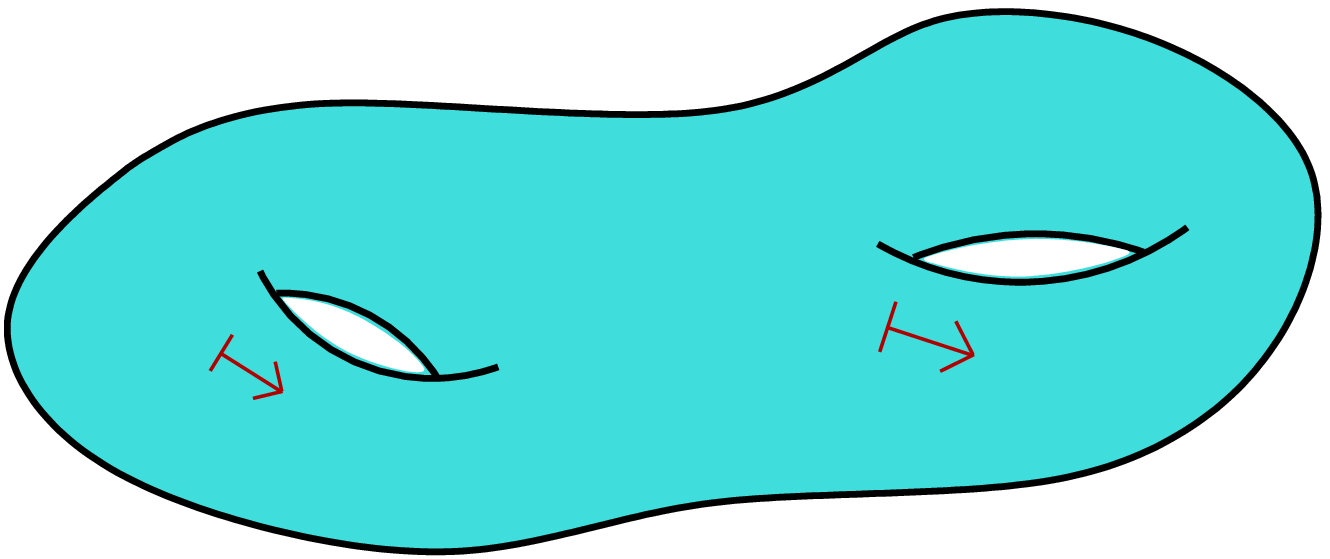}}} 
     \put(225,0){
     \put(0,0)    {\scalebox{.57}{\includegraphics{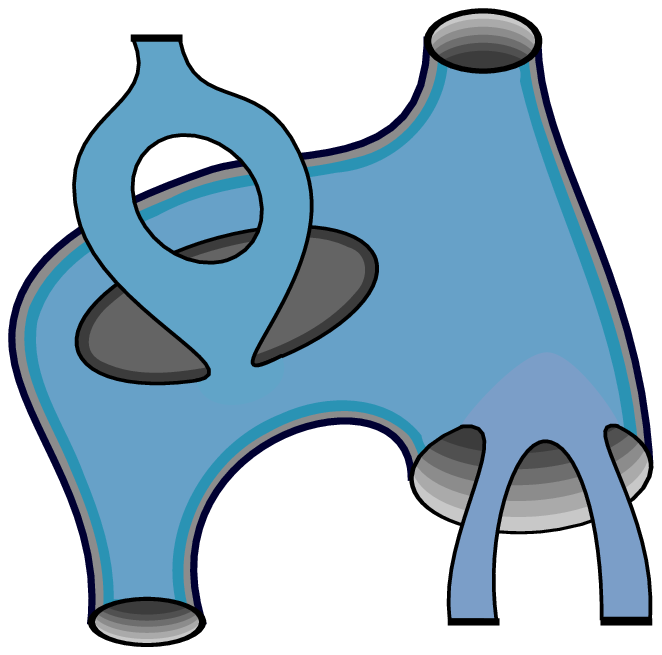}}} \setulen81
     \put(26.8,-8.7) {$\sse p_{\!1}^{}$} 
     \put(91.2,-3.9) {$\sse p_{\!2}^{}$} 
     \put(122.7,-3.9){$\sse p_{\!3}^{}$}
     \put(28.2,129.9){$\sse p_{\!4}^{}$} 
     \put(93,136.3){$\sse p_{\!5}^{}$} 
 }\eP


\subsection{Consistency conditions}\label{sec.coco}

The correlators of a CFT are subject to a system of consistency conditions.
There are two basically different types of such conditions. The first are the
\emph{Ward identities}, which implement compatibility with the symmetries of the
theory that are encoded in \V. They select, for any point on $\mosp_\X$, a subspace 
$\Bl_\X$ in the space of linear forms
  \be
  \Hom_\complex(\mathcal H_1 \otic \mathcal H_2 \otic\cdots\otic \mathcal H_m,\complex)
  \ee
on the tensor product of state spaces. The relevant subspace is obtained as the space 
of invariants with respect to a globally defined action of the symmetries. This action 
can be constructed from the algebra \V, which encodes the action of the symmetries 
locally in the vicinity of an insertion point, together with the geometrical data of 
the world sheet. The so obtained subspaces $\Bl_\X$, which for all rational CFTs are 
\findim, are 
called the vector spaces of \emph{conformal blocks}, or also of chiral blocks. 

Further, in the rational case the spaces of conformal blocks for all world sheets 
of a given topology and with given number and decorations of bulk and boundary 
insertions fit together into the total space of a finite-rank vector bundle over 
the moduli space $\mosp{}_\Xh$ of geometric data for the \emph{complex double}
\Xh\ of the world sheet \X. The surface \Xh\ is, by definition, the orientation 
bundle $\mathrm{Or}(\X)$ over \X, modulo an identification of the points over 
$\partial\X$:
  \be
  \Xh = \mathrm{Or}(\X) \,{\big/}
  _{\! p{\times}\{1\} \,\simeq\, p{\times}\{-1\},~ p\iN\partial\X} \,.
  \labl{def.Xh}
For instance, the double of a world sheet \X\ that is closed and orientable is the 
disconnected sum of two copies of \X\ with opposite orientation, the double 
of the disk $\mathrm D$ is a two-sphere, $\widehat{\mathrm D} \eq S^2$, and the
double of the Klein bottle is a two-torus, $\widehat{\kbottle}\eq\torus$.
 
The symmetries encoded in the conformal vertex algebra \V\ include in particular 
the conformal symmetry; accordingly, any \V-module carries a \rep\ of the Virasoro 
algebra. (Also, all of them have the same eigenvalue of the canonical central element 
of the Virasoro algebra; this value $c$ is called the central charge of \V.) This 
symmetry can be used to endow each bundle of conformal blocks with a projectively flat 
connection, the Knizhnik-Zamolodchikov connection. The monodromy of this connection 
furnishes a projective \rep\ $\rho_\Xh$ of the mapping class group ${\rm Map}(\Xh)$
of \Xh\ (the fundamental group of the moduli space $\mosp{}_\Xh$) on the fibers, i.e.\ 
on the spaces of conformal blocks (for details see e.g.\ \cite{frsh2,FRbe,looi9}). 
Another property of the system of conformal blocks of a rational CFT is the existence
of \emph{gluing isomorphisms} between (direct sums of) conformal blocks for surfaces
that are related by `cutting and gluing' (and thus generically have different topology);
see e.g.\ \cite[Sect.\,4]{looi9} or \cite[Sect.\,2.5.3]{fffs3}.

The bundles of conformal blocks, or also their (local) sections, are sometimes referred 
to as the correlators of the \emph{chiral CFT} that is associated to the chiral symmetry 
algebra \V. However, generically the bundles do not have global sections, so a chiral 
CFT behaves quite differently from a conventional quantum field theory, which has 
single-valued correlators. (Still, chiral CFT has direct applications in physics, 
for systems in which a chirality (handedness) is distinguished, e.g.\ in the 
description of universality classes of quantum Hall fluids.)

\medskip

The second type of consistency conditions obeyed by the correlators of the full CFT are 
twofold: first, the \emph{sewing constraints}, or factorization constraints, and second,
the \emph{modular invariance} constraints. These are algebraic equations which assert
that the image of a correlator under a gluing isomorphism is again a correlator (and 
thus incorporate the compatibility of correlators on world sheets that are related by 
cutting and gluing), respectively that the correlator for a world sheet \X\ is invariant 
under the action of the group ${\rm Map}_{\rm or}(\X)$ of oriented 
mapping classes of the world sheet, which is naturally embedded in ${\rm Map}(\Xh)$. 
In rational CFT the compatibility with cutting and gluing allows one to reconstruct 
all correlators by gluing from a small collection of special correlators. Invariance 
under ${\rm Map}_{\rm or}(\X)$ is referred to as \emph{modular} invariance, a term 
reflecting the special case that \X\ is a torus without insertions. In that case one 
has $\widehat{\mathrm T} \eq \mathrm T \,{\sqcup}\, {-}\mathrm T$, and the relevant 
\rep\ of ${\rm Map}(\widehat{\mathrm T})$ has the form $\rho_{\widehat{\mathrm T}} \eq
\rho_{\mathrm T}^{} \otic \rho_{\mathrm T}^*$ with $\rho_{\mathrm T}$ a \rep\ of the 
modular group $\slz \eq {\rm Map}_{\rm or}(\mathrm T)$. (Here ${-}\mathrm T$ denotes 
the torus with opposite orientation. The orientation reversal results in the 
complex conjugation in the second factor of $\rho_{\widehat{\mathrm T}}$.)

CFT defined on world sheets \X, with correlators satisfying, in addition to the
Ward identities, also all sewing constraints,
is referred to as \emph{full CFT}, as opposed to chiral CFT, which is defined
on closed oriented surfaces and deals with conformal blocks, which are solutions
to the Ward identities only. The problem to be addressed in the rest of the
present section and in the next section is how to obtain a full CFT on world sheets
\X\ from a corresponding chiral CFT on the complex doubles \Xh, and in particular,
how to select the correlator on any world sheet \X\ as an element of the space of 
conformal blocks on \Xh.


\subsection{Solution of the sewing and modular invariance constraints}\label{subsec.solu}

{}From now until the end of section \ref{sec.corr.cb} we restrict our attention to
\emph{rational} CFTs. Basically this means that the relevant \rep\ theory is 
semisimple and satisfies certain finiteness conditions. Technically, it can be 
formulated as a set of conditions on the vertex algebra \V\ \cite{huan29} (or
on the corresponding net of von Neumann algebras \cite{kalm}), and in terms of the 
decoration category \C\ it is characterized by \C\ being a \emph{modular tensor category}.

In short, a \mtc\ \C\ is a semisimple monoidal category which comes with a braiding,
a duality and a twist, which has finitely many simple objects up to isomorphism, for
which the tensor unit is simple, and
for which the braiding is maximally non-degenerate; for the precise definition,
see appendix \ref{app.mtc}. We denote the finite set of isomorphism classes of 
simple objects of \C\ by $\I$ and select a representative $\U_i$ for each class
$i\iN\I$. As representative for the class of the tensor unit $\one$ we take $\one$
itself, and denote its class by $0$, i.e.\ write $\one\eq\U_0$. As a \V-module, 
$\one\iN\C\eq\repV$ is just \V\ itself. 

A crucial feature of rational CFTs is that for solving the sewing constraints, and
thus for obtaining the correlators for all world sheets, of arbitrary topology and 
with any number and type of field insertions (as well as for other purposes), 
\emph{combinatorial} information about the theory is sufficient. Specifically, 
to account for the chiral symmetries it
is enough to treat the category \C\ as an abstract category (with suitable additional
structure), rather than as given concretely as the \rep\ category $\repV$ of the chiral
symmetry algebra. For instance, we can, and will, take \C\ to be \emph{strict} monoidal,
equivalent as a monoidal category (with additional structure) to the non-strict 
category $\repV$. Furthermore, one can replace the complex-analytic modular functor 
that is furnished by the system of conformal blocks by the equivalent 
\cite[Ch.\,4]{BAki} topological modular functor that is provided\,%
  \footnote{~See the discussion around formula \erf{tftc32} below.}
by the TFT functor \tftc\ that is associated to the \mtc\ \C. Hereby the fibers of the 
bundles of conformal blocks for \X\ are identified, as \rep\ spaces of mapping class 
groups,  with the state spaces $\tftc(\Xh)$ of the \C-decorated \threedim\ topological 
field theory \tftc.
In this description, the correlator \CX\ is just regarded as an element
  \be
  \CX \in \Bl_\X
  \ee
of such a \findim\ vector space.
As for the world sheet \X, it can be treated just as a topological manifold, endowed
with some additional structure to rigidify the situation. The information that is 
suppressed in this simplified setting can be restored completely when combining the 
results obtained in this setting with the explicit expressions for the conformal 
blocks as vector bundles over $\mosp{}_\Xh$. (The latter are \emph{not} fully known 
for all surfaces \Xh\ and all classes of RCFTs, but determining them is by all 
means an issue independent of the problems considered here.)

\medskip

Despite all these simplifications, to actually determine the solution of all sewing 
and modular invariance constraints is not an easy matter. After all, these constraints 
constitute infinitely many nonlinear equations, in particular one for each cutting or 
sewing, in infinitely many variables, namely the correlators for all world sheets. A 
traditional approach to these constraints has been to establish the
general solution to some specific small subset of constraints, most prominently,
to the requirement of modular invariance of the torus partition function (i.e., the 
correlator for a torus without insertions) $Z \eq \Cor(\torus,\tau,\emptyset)$,
see e.g.\ \cite{DIms,gann16}. The torus partition function is a vector in the 
space $\Bl_\torus$ of conformal blocks for the torus $\torus$. The terminology
`partition function' signifies the property of $Z$ that $\Bl_\torus$ has a basis 
  \footnote{~The complex conjugation of $\chii_j^{}$ reflects the opposite
  orientation of the second copy of $\torus$ in the
  double $\widehat{\torus} \eq \torus \,{\sqcup}\, {-}\torus$.} 
$\{\chii_i^{}\oti\chii_j^*\,|\,i,j\iN\I\}$ in which the expansion coefficients are 
non-negative integers and the coefficient with $i\eq j\eq 0$ is unity (this amounts 
to the basic physical requirement of uniqueness of the bulk vacuum state). Thus with
  \be
  Z = \sum_{i,j\in\I} \Z_{i,j} \, \chii_i^{}\oti\chii_j^*
  \labl{Zij}
one has
  \be
  \Z_{i,j}\in\zet_{\ge0} \qquand \Z_{0,0} = 1 \,.
  \labl{Z1}
Modular invariance of $Z$ amounts in this basis to the requirement
  \be
  [\Z,\rho_{\mathrm T}(\gamma)] = 0 ~~~ \mathrm{for~all} ~~ \gamma\iN\slz \,.
  \labl{Z2}
As conformal blocks depending on the modular parameter $\tau$ of the torus, the 
basis elements $\chii_i$ are given by the \emph{characters}
   \be
   \chii_i^{}(\tau) := \mathrm{tr}_{\U_i}^{}(\eE_{}^{2\pi\ii\tau(L_0-c/24)})
   \labl{def_chi}
of the irreducible modules of the chiral symmetry algebra \V.
(In \erf{def_chi}, $c$ is the central charge of the conformal vertex algebra \V\
and $L_0$ is the zero mode of the Virasoro algebra.)

The classification, for a given chiral algebra, of matrices $\Z\eq\big(\Z_{i,j}\big)$
obeying these conditions has produced remarkable results, starting with the discovery 
of an A-D-E-structure of the solution set for classes of models with simple objects
of sufficiently small categorical dimension.
It has not led to any deep insight in the structure of the solution set for general 
rational CFTs, though. And indeed this classification program had to be put into a 
different perspective by the observation that there exist spurious solutions, i.e.\ 
matrices satisfying \erf{Z1} and \erf{Z2} which are unphysical in the sense that they 
do not fit into a consistent collection of correlators that also satisfies all other 
sewing and modular invariance constraints. (For some details and references see 
section 3.4 of \cite{fuRs13}.) 

\medskip

Additional insight has been gained by the study of boundary conditions that 
preserve all chiral symmetries. In particular it has been realized that for consistency
of such boundary conditions, representations of the fusion rules by matrices with 
non-negative integral entries, so called NIM-reps, must exist \cite{bppz}. This 
requirement, and similar consistency conditions for unorientable surfaces, turn out to
be rather restrictive, see e.g.\ \cite{gann17,huss2}. These developments have 
triggered another approach to the solution of the sewing constraints, which we
call the \emph{TFT construction} of the correlators of rational CFTs.\,%
  \footnote{~The correlators on world sheets of genus 0 and 1 have also been obtained
  by direct use of the theory of vertex algebras \cite{huKo2,kong6}. The relation
  between these results and the TFT construction is discussed in \cite{koRu2}.}
This construction
associates to a world sheet \X\ a cobordism $\MX$ from $\emptyset$ to $\partial\MX$ 
such that $\partial\MX$ is homeomorphic to the complex double $\Xh$ (see \erf{def.Xh})
of the world sheet. The correlator for \X\ is then defined to be the element
  \be
  \CX = \tftc(\MX)\,1
  \labl{CX}
of the vector space $\Bl_\X \eq \tftc(\Xh)$ of conformal blocks.
It has been proven that the correlators produced by this construction satisfy
\emph{all} sewing and modular invariance constraints.

Besides the \mtc\ \C, the TFT construction requires one additional datum, namely
a certain \emph{Frobenius algebra} internal to \C. For each Morita class of such
algebras the construction gives one independent solution. The relevant algebras 
will be described in section \ref{sec.frob}.
The term TFT construction alludes to a tool that is crucial in the construction,
namely the \C-decorated \threedim\ \emph{topological field theory} (TFT) $\tftc$; 
this is \cite{TUra} a projective symmetric monoidal functor 
  \be
  \tftc:\quad \Cob_{3,2}^\C \to \Vect
  \labl{tftc32}
from a category $\Cob_{3,2}^\C$ of \threedim\ cobordisms to \findim\ complex vector 
spaces. Such a TFT associates to an oriented surface a vector space, called the TFT 
state space of the surface, and to a cobordism between surfaces a linear map between 
the corresponding state spaces. Both the surfaces and the cobordisms come with
additional data. In a \emph{\C-decorated} TFT these data include embedded arcs on
the surfaces and embedded ribbon graphs in the cobordisms (in such a way that for 
any world sheet \X\ the double \Xh\ is an object of $\Cob_{3,2}^\C$ and the 
cobordism $\MX$ is a morphism of $\Cob_{3,2}^\C$); the arcs are labeled
by objects of \,\C, and the ribbon graphs by objects and morphisms of \,\C. It is an
important result that to any (semisimple) \mtc\ there can indeed be associated a
TFT functor $\tftc$; for details we refer to \cite{TUra} and \cite[Ch.\,4]{BAki}; a 
brief summary can be found in \cite[Sect.\,2.5]{fffs3}. The linear maps obtained by 
the functor $\tftc$ furnish an invariant of oriented three-manifolds and of framed 
links in these manifolds \cite{retu2}.

In connection with the use of \tftc\ in the prescription \erf{CX} a word of caution
is in order. Namely, from a rational conformal vertex algebra \V\ and the modular 
tensor category $\C\eq\repV$ one obtains representations of mapping class groups in 
two distinct ways. The first is via the Knizhnik-Zamolodchikov connection on the bundles 
of conformal blocks; in this case the mapping class groups arise as the fundamental 
groups of the moduli spaces $\mosp{}_\Xh$. The second way is via three-dimensional 
TFT and the embedding of mapping class groups in cobordisms; this construction uses 
directly the modular tensor category \C, while the former is based on the vertex 
algebra \V\ and its conformal blocks. That the two constructions give equivalent 
representations is known for genus 0 and 1, 
but to establish equivalence model-independently and for any genus will require a 
considerably improved understanding of conformal blocks and in particular of their 
factorization properties for general rational vertex algebras.


\subsection{The TFT construction of RCFT correlators}\label{subsec.TFT}

The formula \erf{CX} gives the correlator $\CX$ as the invariant of a three-manifold 
\MX. This manifold, to which we will refer as the \emph{connecting manifold}, comes 
with a prescription for an embedded ribbon graph \GX; by abuse of notation we use the 
symbol \MX\ also to refer to the manifold including the graph \GX. In \erf{CX}, the 
three-manifold \MX\ is regarded as a cobordism from $\emptyset$ to the boundary 
$\partial\MX$. Thus taking the invariant, i.e.\ applying the TFT functor \tftc, 
yields a linear map from $\tftc(\emptyset)\eq\complex$ to $\tftc(\partial\MX) \eq 
\tftc(\Xh)$, which when applied to $1\iN\complex$ then gives the correlator -- a 
vector in $\tftc(\Xh)$, i.e.\ in the space of conformal blocks for the world sheet \X. 

The TFT construction is a prescription on how to obtain the connecting manifold $\MX$, 
including its embedded ribbon graph \GX, for any arbitrary world sheet \X. The details 
of this prescription are a bit lengthy. While many of the details will not concern us 
here, the following features are pertinent (for more information see e.g.\ 
\cite[App.\,B]{fjfrs} or \cite[App.\,A.4,A.5]{fuSs}):
\nxl2 \nxt 
The \findim\ vector space of which $\CX$ is an element is the space $\Bl_\X$ of 
conformal blocks (recall that $\Bl_\X$ is defined with the help of the complex double 
\Xh\ of \X). 
\nxl2 \nxt
The connecting manifold \MX\ is obtained by suitably `filling up' the double \Xh\
so as to get a three-manifold having \Xh\ as its boundary. A guideline for this 
filling-up procedure is that it should not introduce any new topological information;
indeed \MX\ has \X\ as a deformation retract, i.e.\ can be viewed as a fattening of 
the world sheet. (Moreover, in the extension from two to three dimensions there
should not, in physics terminology, be any additional dynamical information
involved; this is accounted for by taking the field theory relevant to \MX\ to be 
a \emph{topological} theory.) Concretely, as a manifold \MX\ is the total space of
an interval bundle over \X\ modulo an identification over $\partial\X$.
For instance, when \X\ is closed and orientable, then $\MX\eq X\Times [-1,1]$, the
connecting manifold $\mathscr M_{\mathrm D}$ for the disk is a three-ball, and the 
connecting manifold for the Klein bottle is a full torus.
\nxl2 \nxt
By construction, the world sheet \X\ is naturally embedded in the `equatorial plane' 
$\X\Times\{0\}$ of \MX. A crucial ingredient of the ribbon graph \GX\ in \MX\ is a 
dual triangulation \TX\ of the embedded world sheet. As an additional datum, beyond 
the \mtc\ \C\ and the
combinatorial data for \X,  the decoration of the ribbon graph \GX\ thereby involves 
the selection of an object $A$ of \C, whose role is to label each of the ribbons in 
the dual triangulation \TX. This object $A$ cannot be chosen arbitrarily; indeed, in 
order for the correlator \erf{CX} to be well-defined, $A$ is required to carry the 
structure of a \emph{simple \ssfa} in \C. (When allowing for unoriented world sheets,
$A$ must come with an additional structure, a braided analogue of an involution.)
Each of the trivalent vertices of the dual triangulation \TX\ is labeled
by either the product or coproduct morphism of $A$; the various properties of $A$
then ensure that the invariant $\tftc(\MX)$ is independent of the particular choice
of  dual triangulation of \X.


\subsection{Main results of the TFT construction}\label{sec.main}

The basic result of the TFT construction is the statement that the pair $(\C,A)$ is 
necessary and sufficient for obtaining a solution to the sewing constraints of 
rational conformal field theory. In more detail, one has the following
\nxl6
{\sc Theorem}:\nxl2
(i)~~The pair $(\C,A)$, with \C\ a \mtc\ and $A$ a simple \ssfa\ in \C, supplies
all required decoration data, and the TFT construction yields a family 
$\{\hspace*{.7pt}\CXCA\hspace*{.5pt}\}$ of correlators (on oriented world sheets 
of arbitrary topology and with arbitrary field insertions) that satisfy all sewing 
and modular invariance constraints of rational CFT.
\nxl2
(ii)~~Every solution to the sewing constraints of a non-degenerate rational CFT
based on the chiral data \C\ can be obtained via the TFT construction with a simple 
\ssfa\ $A$ in \C. Once a boundary condition of the CFT is selected, this algebra 
$A$ is determined uniquely up to isomorphism.
\nxl2
(iii)~\,Morita equivalent algebras give rise to equivalent families of correlators.

\vskip 8pt

We add a few remarks;
for more details the reader should consult the original literature.
\nxl2 \nxt
The assertion (i), first stated in \cite{fuRs} and \cite[Sect.\,5]{fuRs4}, was proved
in \cite{fjfrs}.
A crucial input of the proof is that every sewing operation can be obtained by a finite
sequence of finitely many distinct local moves. This allows one to reduce the infinitely 
many sewing constraints to three types of manipulations: action of (a generator of) the
relevant mapping class group, boundary factorization, and bulk factorization. These 
three types of constraints are treated in theorems 2.2, 2.9 and 2.13 of \cite{fjfrs}, 
respectively.
\nxl2 \nxt
Part (ii) was established in \cite{fjfrs2}. In the proof the boundary condition is 
taken to be the same on all segments of $\partial\X$; different choices of boundary 
condition give rise to Morita equivalent algebras.
The qualification `non-degenerate' refers to properties of the vacuum state and of
the two-point correlators of boundary fields on the disk and of bulk fields on the 
sphere, all of which are expected to hold in a rational CFT.
\nxl2 \nxt
Part (iii) was claimed in \cite{fuRs}. The precise meaning of equivalence -- namely 
equality of correlators up to a factor $\gamma^{-\chi(\X)/2}$ with $\chi(\X)$ the
Euler characteristic of \X\ and $\gamma$ the quotient of the dimensions of the two
Morita equivalent algebras involved -- has been formulated in section 3.3 of 
\cite{ffrs5}, where also a proof of the statement is presented. Note that in a
(semisimple) \mtc\ a Frobenius structure can be transported along Morita contexts.
\nxl2 \nxt
As they stand, parts (i) and (iii) apply to oriented world sheets. There is a 
version for unoriented (in particular, unorientable) world sheets as well (for 
part (i), see theorems 2.4, 2.10 and 2.14 of \cite{fjfrs}), but it requires a few 
modifications. In particular, as already mentioned, the algebra $A$ must possess 
an additional structure $\sigma\iN\End(A)$, which is a braided version of an 
involution \cite{fuRs8} ($(A,\sigma)$ is then called a \emph{Jandl algebra}), and 
the notion of Morita equivalence gets generalized accordingly \cite[def.\,13]{fuRs11}.  

\medskip

Instrumental in all these findings is another fundamental result: All relevant
concepts from rational CFT have a precise mathematical formalization that is
entirely expressible in terms of the data \C\ and $A$. These may be collected
in the form of a dictionary between physical concepts and mathematical structures.
Some entries of this dictionary are the following:\\[-2.1em]
\begin{center}
{\small \begin{tabular}{ll}
~\nxl5 \hline ~\nxl{-12}
\multicolumn1c{physical concept} & \multicolumn1c{mathematical structure}
\nxl2 \hline \nxl{-7}%
phase of a CFT       & symmetric special Frobenius algebra $A$ in \C
\nxl3
boundary condition   & $A$-module $M \iN \CA$
\nxl3
defect line separating phases $A$ and $B$  & $A$-$B$-bimodule $X \iN \CAB$
\nxl3
boundary field changing the boundary       & module morphism $\Psi\iN\HomA(M\oti U,M')$
       \\ \multicolumn1r{condition from $M$ to $M'$}
\nxl3
bulk fields in region with phase $A$       & bimodule morphism
                              $\Phi\iN\HomAA(U\,{\otimes}{}^+_{}\!A\,{\otimes}{}^-_{}V,A)$
\nxl3
defect field changing the defect type      & bimodule morphism
                              $\Theta\iN\HomAB(U\,{\otimes}{}^+_{}\!X\,{\otimes}^-_{}V,X')$
       \\ \multicolumn1r{from $X$ to $X'$}
\nxl1
CFT on unoriented world sheets & $A$ a Jandl algebra
\nxl3
non-chiral internal symmetry & element of Picard group \PicCAA\ \cite{ffrs5}
\nxl5\hline\end{tabular} }
\end{center}

\vskip9pt

\noindent
In the expressions for bulk and defect fields, the notation $\otimes^\pm$ refers 
to bimodule structures of the corresponding objects which are obtained by 
combining the product of the algebra(s) and suitable braidings, e.g.
  \be \bearll&
  U \,{\otimes^+} A := (\,U \oti A,\,
    (\id_U\oti m) \cir (c_{U,A}^{~-1} \oti \id_A),\, \id_U\oti m\,)
  \nxm8 {\rm and}~~&
  U \,{\otimes^-} A := (\,U \oti A,\,
    (\id_U\oti m) \cir (c_{\!A,U}^{} \oti \id_A),\, \id_U\oti m\,) \,,
  \eear \labl{braid_ind}
and analogously in the other cases (see e.g.\ \cite[eqs.\,(2.17,18)]{fuRs10}).
\\
One may note that the structures appearing in the list above naturally fit into a 
bi-category, with the phases, defect lines and defect fields being, 
respectively, the objects, 1-mor\-phisms and 2-morphisms.


\subsection{Frobenius algebras in \C} \label{sec.frob}

As the algebra $A$ plays a central role in the TFT construction, let us pause to
collect some information about the relevant class of algebras. We find it convenient 
to represent the relevant structural morphisms graphically. (For an introduction to 
the graphical calculus for strict monoidal categories see e.g.\ \cite{joSt5,MAji,KAss} 
or section 2.1 of \cite{fuRs4}. Note in particular that the positions of the
various pieces in such pictures are relevant only up to suitable isotopy.)
We draw the pictures such that the domain of a morphism is at the bottom and the
codomain is at the top.

The various properties of $A$ are defined as follows.
\nxl2 \nxt
A (unital, associative) \emph{algebra}, or monoid, in a monoidal category \C\ 
consists of an object $A$ together with morphisms $\,m \eq 
\bP(16,0) \put(2,-6){\scalebox{.17}{\includegraphics{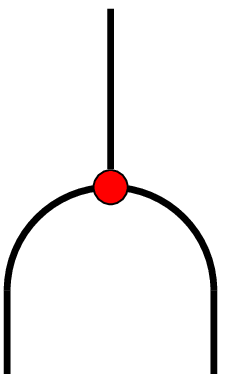}}} \eP
\iN\Hom(A\Oti A,A)$ (the product, or multiplication, morphism) and $\,\eta \eq
\bP(7,0) \put(2,-2){\scalebox{.20}{\includegraphics{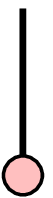}}} \eP
\iN\Hom(\one,A)$ (the unit morphism) such that the associativity and unit properties
  \Eqpic{p1} {290}{15} { \put(0,-4){ \setUlen15
  \put(0,0)   {\scalebox{.33}{\includegraphics{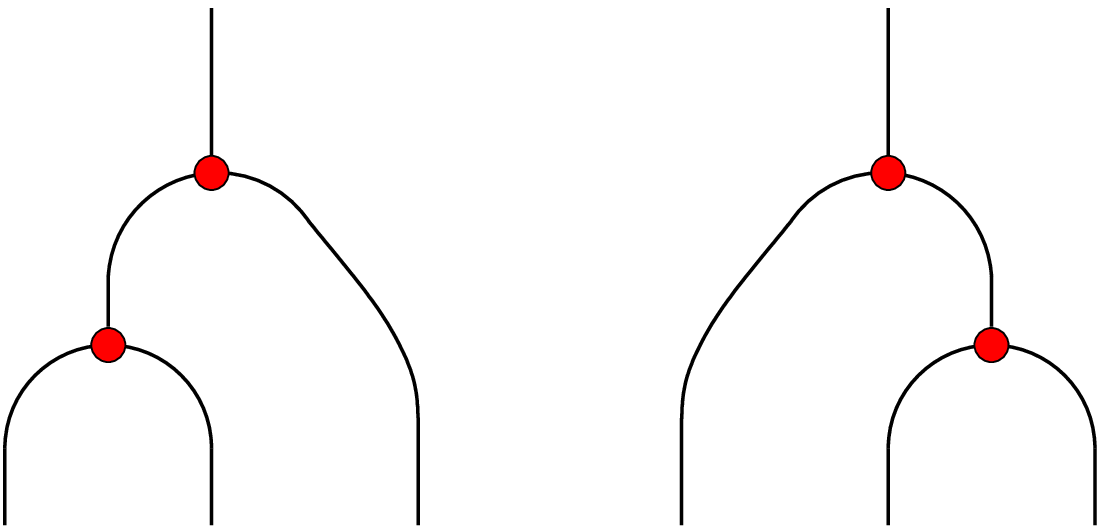}}} \put(32,14) {$ = $}
  \put(93,16) {and}
  \put(130,0) {\scalebox{.33}{\includegraphics{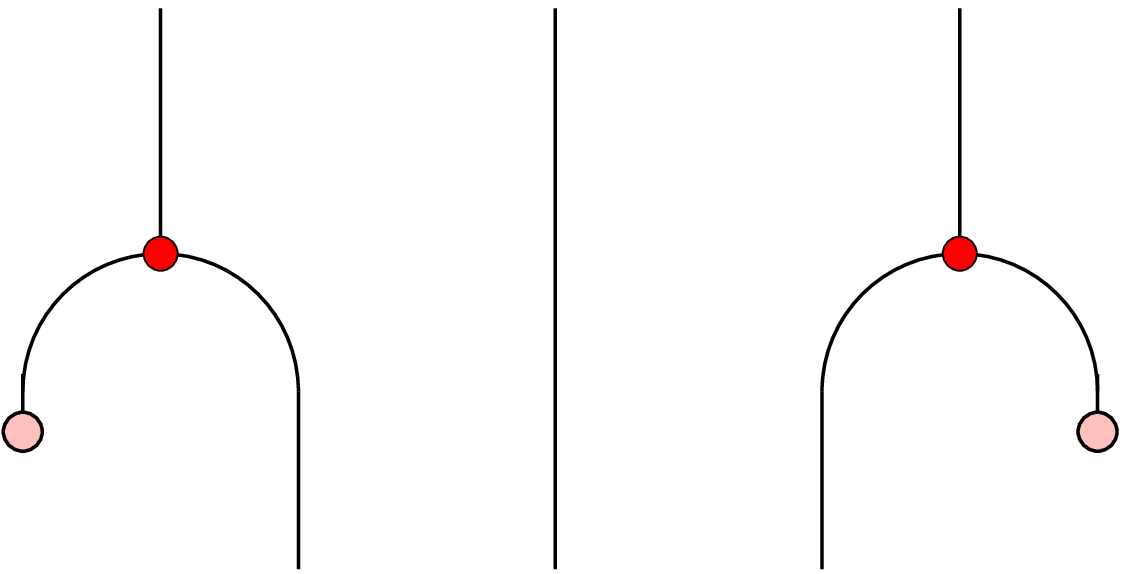}}}
  \put(154,16){$ = $} \put(171,16){$ = $}
  } }
are satisfied.
\nxl2 \nxt
Likewise, a \emph{coalgebra} in \C\ is an object $C$ together with coproduct and 
counit morphisms $\,\Delta \eq
\bP(16,0) \put(2,-7.5){\scalebox{.17}{\includegraphics{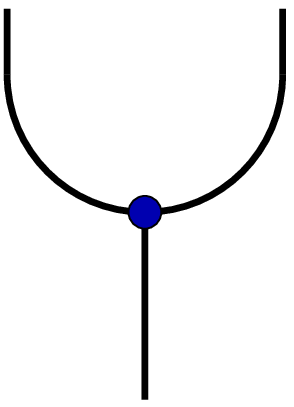}}} \eP
\iN\Hom(A,A\Oti A)$ and $\,\eps \eq
\bP(7,0) \put(2,-4.2){\scalebox{.17}{\includegraphics{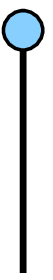}}} \eP
\iN\Hom(A,\one)$ such that
  \Eqpic{p2} {290}{15} { \put(0,-4){ \setUlen15
  \put(0,0)   {\scalebox{.33}{\includegraphics{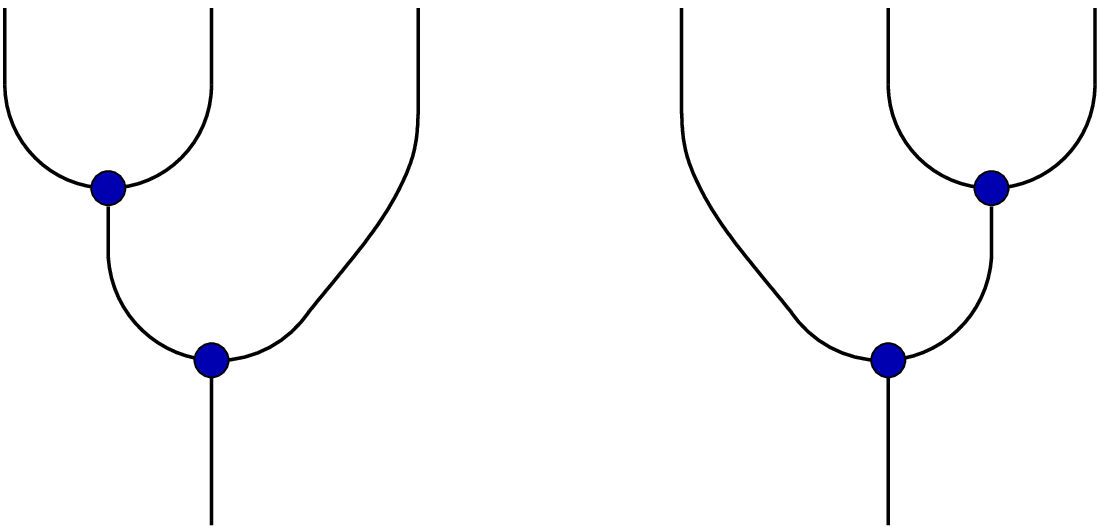}}} \put(32,14) {$ = $}
  \put(93,16) {and}
  \put(130,0) {\scalebox{.33}{\includegraphics{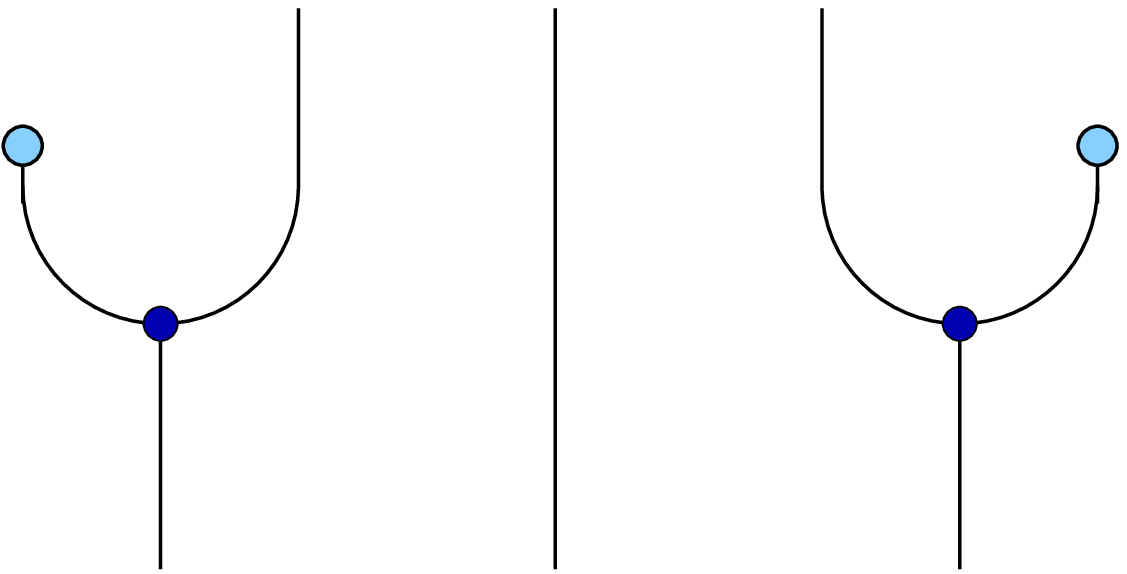}}}
  \put(153,13){$ = $} \put(172,13){$ = $}
  } }

\noindent \nxt
A \emph{Frobenius algebra} in \C\ is a quintuple $(A,m,\eta,\Delta,\eps)$ such that
$(A,m,\eta)$ is an algebra in \C, $(A,\Delta,\eps)$ is a coalgebra, and the two
structures are connected by the requirement that the coproduct is a bimodule morphism,
i.e.\ that
  \Eqpic{p3} {180}{21} { \put(0,-4){ \setUlen15
    \put(0,0)   {\scalebox{.33}{\includegraphics{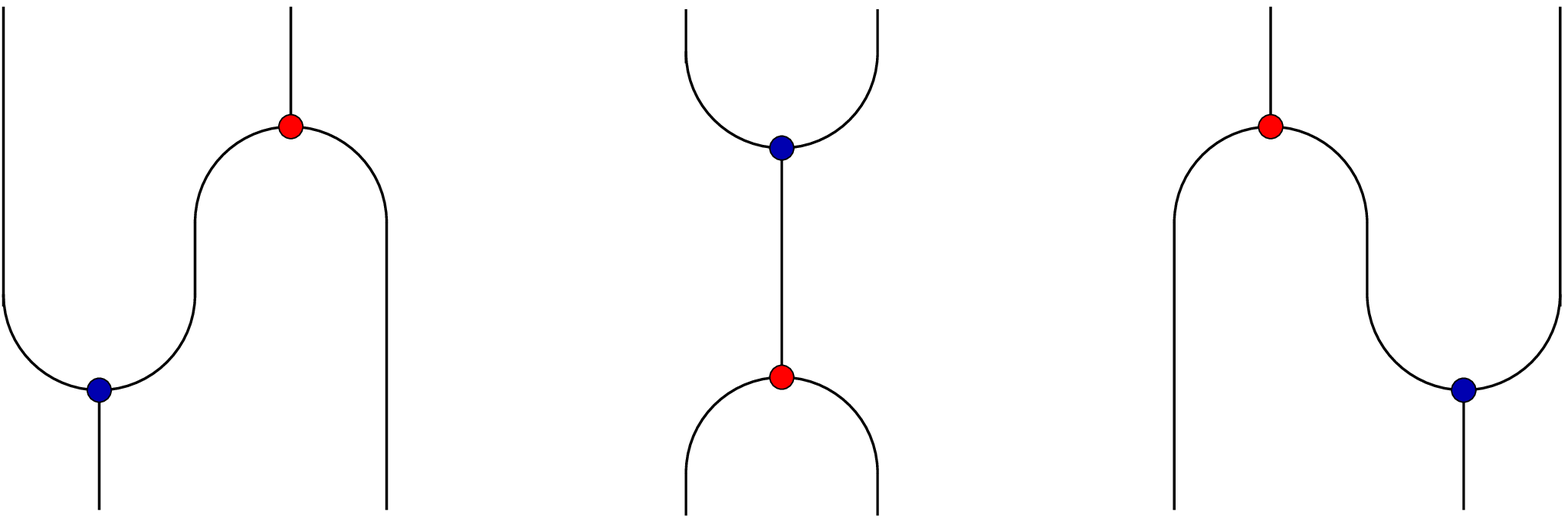}}}
    \put(43,19) {$ = $} \put(81,19) {$ = $}
  } }
\nxt
A \emph{symmetric} algebra in a sovereign monoidal category \C\ (that is,
a monoidal category endowed with a left and a right duality which coincide as
functors) is an algebra for which the equality
  \Eqpic{p4} {110}{27} {\setUlen15
    \put(0,3)   {\put(0,0){\scalebox{.33}{\includegraphics{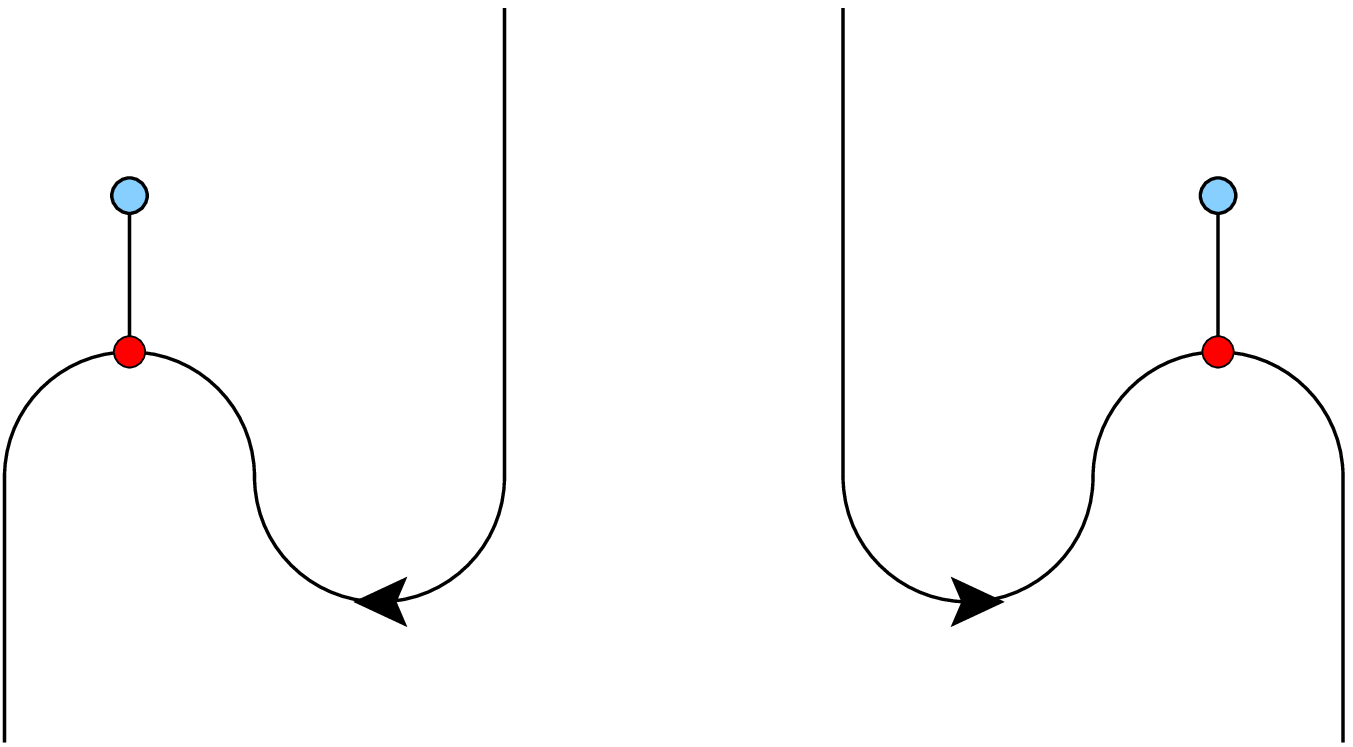}}}
    \put(40,22) {$ = $}
    \put(-3,-6) {$\sse A $} \put(29.5,49) {$\sse A^{\!\vee} $} }
  }
of morphisms in $\Hom(A,A^\vee)$ holds. If the algebra is also Frobenius, then both of 
these are \emph{iso}morphisms. Note that the equality only makes sense if the left and 
right dual objects $A^\vee$ and $^{\vee\!}A$ are equal, as is ensured if \C\ is sovereign.
\nxl2 \nxt
A \emph{special} Frobenius algebra in a rigid monoidal category \C\ is a Frobenius
algebra for which the relations 
  \Eqpic{p5} {190}{22} { \put(0,-3){ \setUlen15
    \put(0,7)  {\scalebox{.33}{\includegraphics{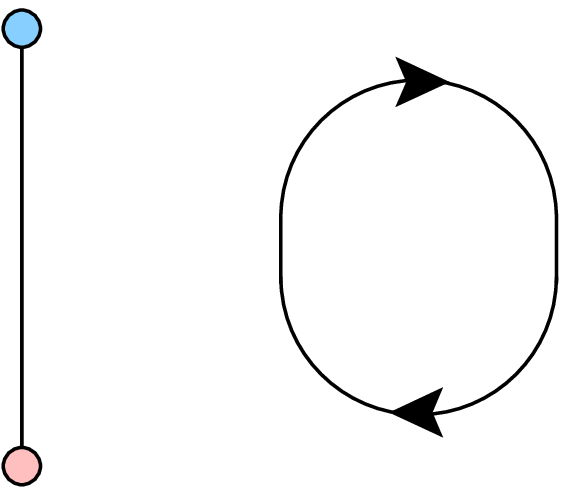}}} \put(6.4,20.5){$=$}
    \put(41,20.5) {$\ne0$} \put(71,20.5) {and} \put(33,0){
    \put(70,0) {\scalebox{.33}{\includegraphics{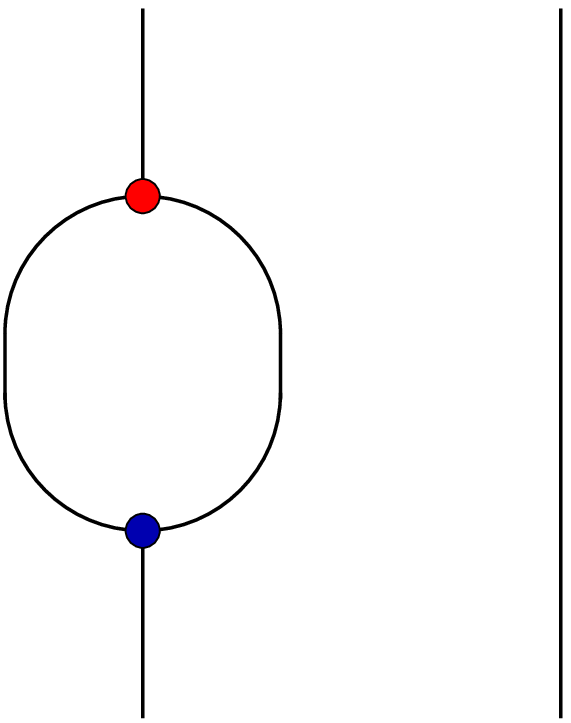}}} \put(93,21) {$=$} }
  } }
are satisfied, i.e.\ the product is left-inverse to the coproduct and, up to a factor 
$\dim(A)\nE0$, the counit is left-inverse to the unit. (Sometimes this property is
instead called normalized special, with specialness just meaning that the two
equalities hold up to non-zero constants; if $A$ is symmetric Frobenius, the product
of these constants is unity.)
\nxl2 \nxt
A \emph{simple} Frobenius algebra is one that is simple as an $A$-bimodule.

\medskip

The following is a (non-exhaustive) list of examples of simple symmetric special 
Frobenius algebras:
\nxl2
(a)~\,In any (strict) monoidal category, the tensor unit, with all structural morphisms
being unit morphisms, $\,A \eq (\one,\id_\one,\id_\one,\id_\one,\id_\one,\id_\one)$.  
\nxl2
(b)~\,In any rigid monoidal category \C, and for any object $U$ of \,\C, the 
following algebra:
 \\
$ A \eq (U^\vee{\otimes}\,U,\id_{U^\vee}{\otimes}\tilde d_U{\otimes}\id_U,
\tilde b_U,\id_{U^\vee}{\otimes}b_U{\otimes}\id_U,d_U)$.
Any such algebra is Morita equivalent to the one in (a).
\nxl2
(c)~\,For \C\ the \mtc\ of integrable modules over the affine Lie algebra
$\widehat{\mathfrak{sl}}(2)$ at positive integral value $k$ of the level, the
following objects, coming in an A-D-E pattern, are symmetric special Frobenius algebras
(we label the simple objects by their highest $\mathfrak{sl}(2)$-weight, i.e.\ as 
$\U_0,\,\U_1,\,...\,,\U_k$):
  \begin{eqnarray}
  && A({\rm A}) = \U_0 \equiv \one ~~~(k\iN\zet_{>0}) \,, \qquad\quad
     A({\rm D}) = \U_0 \oplus \U_k ~~~(k\iN2\zet_{>0}) \,,  \nonumber \nxl5
  && A({\rm E}_6) = \U_0 \oplus \U_6 ~~~(k\eq10) \,, \qquad\quad
     A({\rm E}_7) = \U_0 \oplus \U_8 \oplus \U_{16} ~~~(k\eq16) \,,  \nonumber \nxl5
  && A({\rm E}_8) = \U_0 \oplus \U_{10} \oplus \U_{18} \oplus \U_{28} ~~~(k\eq28) \,.
  \label{ADE}
  \end{eqnarray}
These exhaust the set of Morita classes of simple \ssfa s in this category \cite{kios}.
\nxl2
(d)~\,For $K\,{\le}\,\PicC$ a subgroup of the Picard group (i.e., the group of 
isomorphism classes 
of invertible objects) of \C, subject to the requirement that the three-cocycle on 
$\PicC$ that is furnished by the associativity constraint of \,\C, when restricted to
the subgroup $K$ is the coboundary of a two-cocycle $\omega$ on $K$, there is an algebra 
$A(K,\omega) \,{\cong}\, \bigoplus_{g\in K} S_g$, with cohomologous two-cocycles 
$\omega$ giving isomorphic algebras \cite{fuRs9}. Algebras of this type are called 
\emph{Schellekens algebras}; in the physics literature the corresponding CFT models 
are known as \emph{simple current constructions} \cite{scya6,fusS6}.
\nxl2
(e)~\,In the $m$-fold Deligne tensor product $\C^{\boxtimes m}$ of \C, there is an 
algebra structure on the object $A \eq \bigoplus_{i_1,i_2,...,i_m\iN\I\,} 
(\U_{i_1}\Times \U_{i_2}\Times \dots\Times \U_{i_m})^{\oplus N_{i_1i_2...i_m}} 
\in \C^{\boxtimes m}$, with multiplicities given by $N_{i_1i_2...i_m} \eq
\dimc\Hom_\C(\U_{i_1}\oti \U_{i_2}\oti \dots\oti \U_{i_m},\one)$. Algebras of this 
type correspond to so-called permutation modular invariants \cite{bfrs2}.


\subsection{A digest of further results}\label{subsec.digest}

The TFT construction supplies universal formulas for coefficients of correlation 
functions, valid for all RCFTs. The coefficients of $\Cor(X)$ with respect to a 
basis in the appropriate space $\Bl_\X$ of conformal blocks are expressible through 
the structural data for the \mtc\ \C\ and the \ssfa\ $A$.

For some world sheets of particular interest, the construction gives the
following results.

\Itemize

\itx
A basic correlator is the torus partition function $Z\eq\Cor(\torus,\tau,\emptyset)$
\erf{Zij}. The TFT construction gives $Z$ as the invariant of a ribbon graph in 
the three-manifold $\torus{\times} [-1,1]$, and expressing it as in \erf{Zij} in 
the standard basis of characters yields the coefficients $\Z_{i,j}$ as invariants 
of ribbon graphs in the closed three-manifold $S^2{\times}S^1 $. Choosing the 
simplest dual triangulation of the world sheet $\mathrm T$, these graphs in the
connecting manifold $\torus{\times} [-1,1]$ look as 
follows (see (5.24) and (5.30) of \cite{fuRs4}):
  \begin{eqnarray}
  \bP(320,116)(-62,-17) 
    \put(-64,42)  {$\Cor(\torus;\emptyset)\,\hat=$}
    \put(0,0)     {\scalebox{.30}{\includegraphics{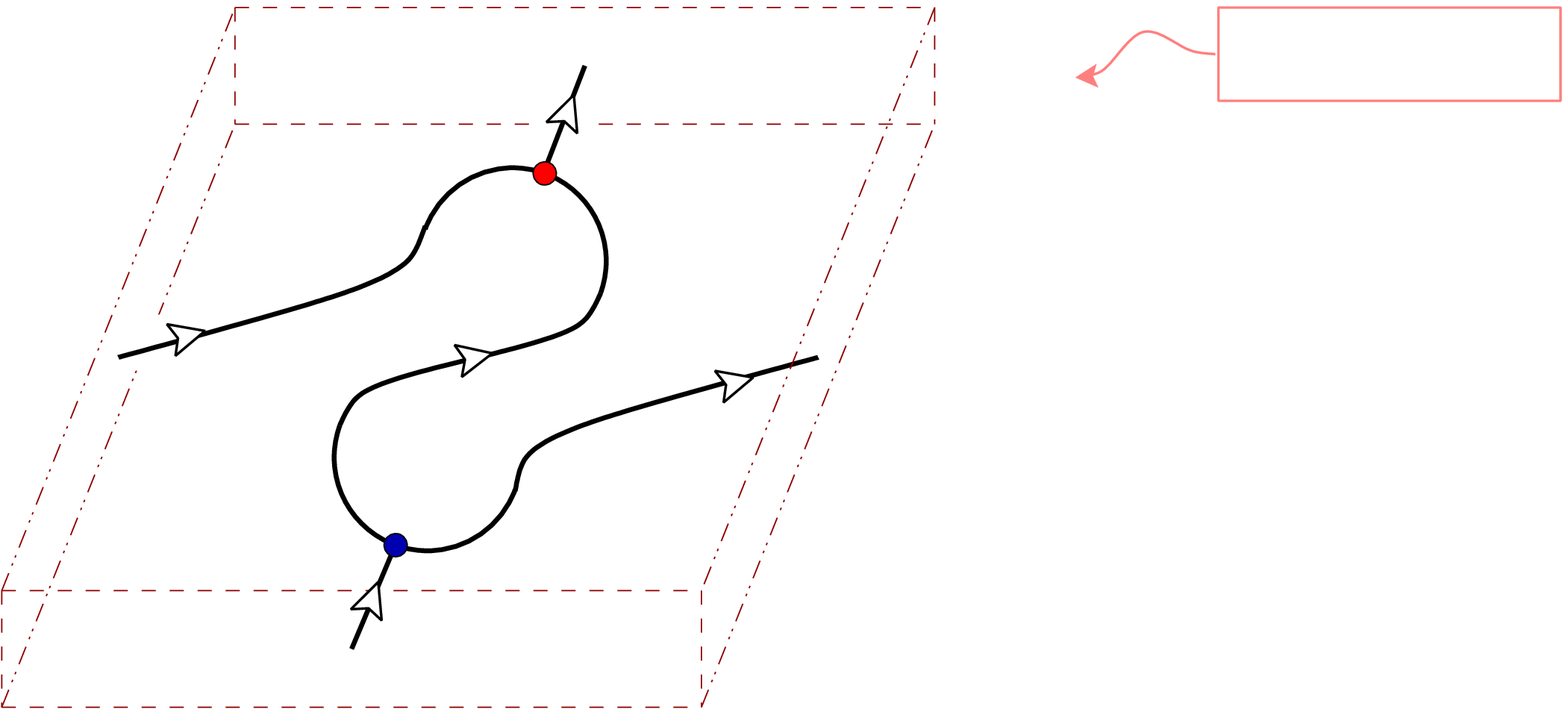}}} \setulen78
    \put(183.2,94.5) {$ \scs \torus{\times} [-1,1] $} \eP
  \bP(100,79)(0,2)
    \put(-49,61)  {$\Z_{i,j}\,\hat=$}
    \put(0,0)     {\scalebox{.30}{\includegraphics{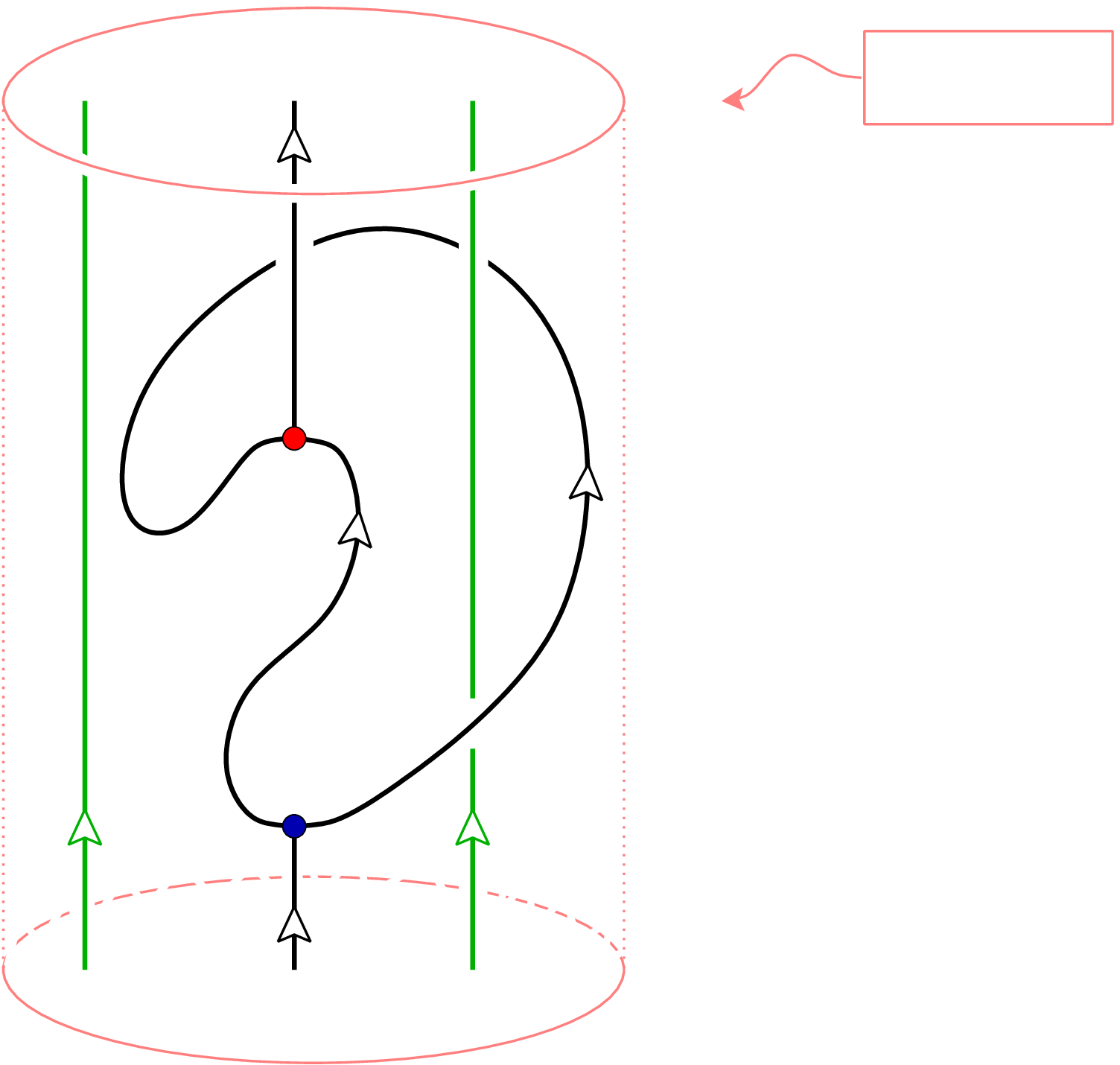}}} \setulen78
    \put(13.5,14.1) {$\scs i $} \put(44.5,14.1) {$\scs A $}
    \put(71.9,14.8) {$\scs j $} \put(29,58)     {$\scs A $}
    \put(78,89)     {$\scs A $} \put(45,106)    {$\scs A $}
    \put(130,143.2) {$ \scs S^2{\times}S^1 $} \eP
  \nonumber\\[-1.7em]~ \label{pics.Z}
  \end{eqnarray}
In plain formulas, the resulting expression for the numbers $\Z_{i,j}$ looks still 
a little complicated; in the special case that both $\dimc\Hom(\U_i,A)$ and
$\dimc\Hom(\U_i\oti\U_j,\U_k)$ for all $i,j,k\iN\I$ are either 0 or 1,
it reduces to \cite[eq.\,(5.85)]{fuRs4}
  \be
  \Z_{i,j}^{} = \tftc\Big(\,
  \bPo \put(1.2,-18){ \put(0,0) {\scalebox{.1}{\includegraphics{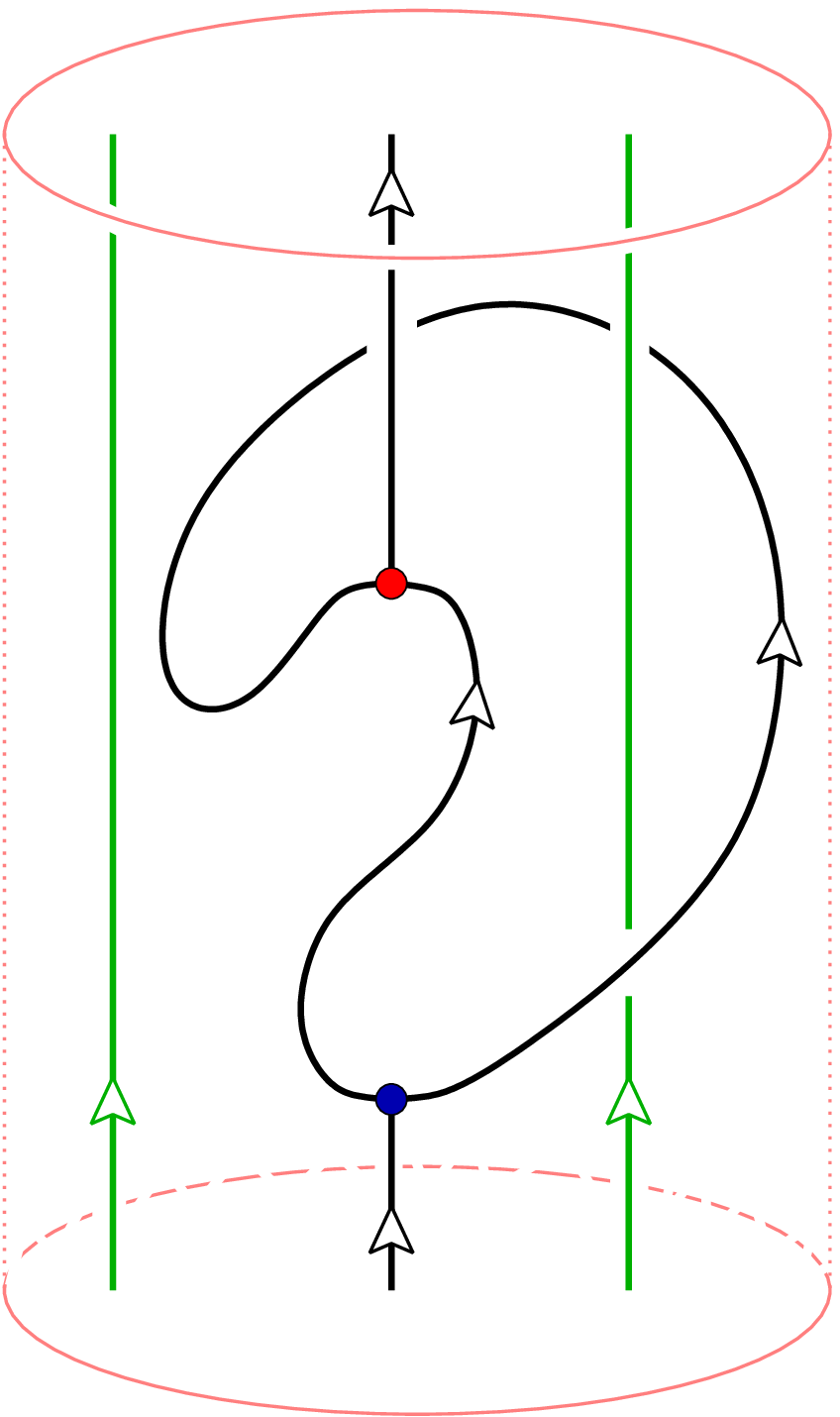}}}
  \setulen26 \put(13.5,13.1){$\scs i $} \put(44.5,12.1){$\sss A $}
  \put(71.9,14.8){$\scs j $} } \eP
  \hsp{2.39} \Big)^{}_{\!(\!S^2_{}{\times}S^1_{}\!)}
  = \sum_{a,b,c\in\mathcal J}\! {m_{bc}}^{\!a}\, \Delta_a^{\,cb}\,
  \sum_{k\in\I} \Gs cbj{\bar\imath}ak\, \Rs {}cba\,
  \Frac{\theta_k}{\theta_j}\, \Fs cbj{\bar\imath}ka \,,  
  \ee
with $\mathcal J\eq\{i\iN\I\,|\,\dimc\Hom(\U_i,A)\eq1\}$, $m_{ij}^{\!k}$ and 
$\Delta_i^{\,jk}$ the structure constants of the product and the coproduct,
respectively, $\theta_i$ the twist eigenvalues, and {\sf F} and {\sf R}
coefficients of the associativity constraint and of the braiding with respect
to standard bases of morphism spaces (and ${\sf G} \eq {\sf F}^{-1}$; for
detailed definitions see section 2.2 of \cite{fuRs4}).

\itx
Inserting the data for the algebras of the $\widehat{\mathfrak{sl}}(2)$ A-D-E 
classification \erf{ADE}, one recovers the well-known expressions \cite{caiz2}
for the torus partition functions of the $\widehat{\mathfrak{sl}}(2)$ theories.
One also shows (Thm\,5.1 of \cite{fuRs4}) that the partition function is modular 
invariant, which in terms of the matrix $\Z$ means $[\gamma,\Z] \eq 0$ for 
$\gamma\iN\slz$, that $\Z_{0,0} \eq 1$ iff $A$ is simple, and that the coefficients 
$\Z_{i,j}$ are nonnegative integers. In fact, 
  \be
  \Z_{i,j} = \dim_\complex^{}\HomAA(\U_i\otip\! A\otim \U_{\!j},A) \,.
  \ee
One also shows that the torus partition functions obtained for the opposite algebra 
and the direct sum and tensor product of algebras are $\Z^{A^{\rm opp}}_{} \eq 
\big(\Z^A\big)^{\!\rm t}_{} $, $\Z^{A\oplus B}_{} \eq \Z^A \,{+}\, \Z^B$, and
${\widetilde{\Z}}^{A\otimes B}_{} \eq {\widetilde{\Z}}^A\, {\widetilde{\Z}}^B$
with ${\widetilde{\Z}}_{i,j} \eq \Z_{i,\bar j}$.
Also recall that any correlator on an oriented world sheet, and hence in particular 
the matrix $\Z\eq\Z(A)$, depends only on the Morita class of $A$.

\itx
As an example of a correlator for a non-orientable world sheet we give the partition 
function on the Klein bottle \kbottle. Recall that for non-orientable world sheets 
one needs a Jandl algebra $(A,\sigma)$ rather than just a Frobenius algebra $A$ as 
an input. One finds \cite[Sect.\,3.4]{fuRs8} that the correlator and the expansion 
coefficients $K_j$ in the basis $\{\chii_j^{}\,|\,j\iN\I\}$ of conformal blocks of 
the torus $\torus \eq \widehat{\kbottle}$ are given by
  \begin{eqnarray}
  \bP(300,103)(-53,3) 
    \put(-72,49)  {$\Cor(\kbottle;\emptyset)~\hat=$}
    \put(0,0)     {\scalebox{.30}{\includegraphics{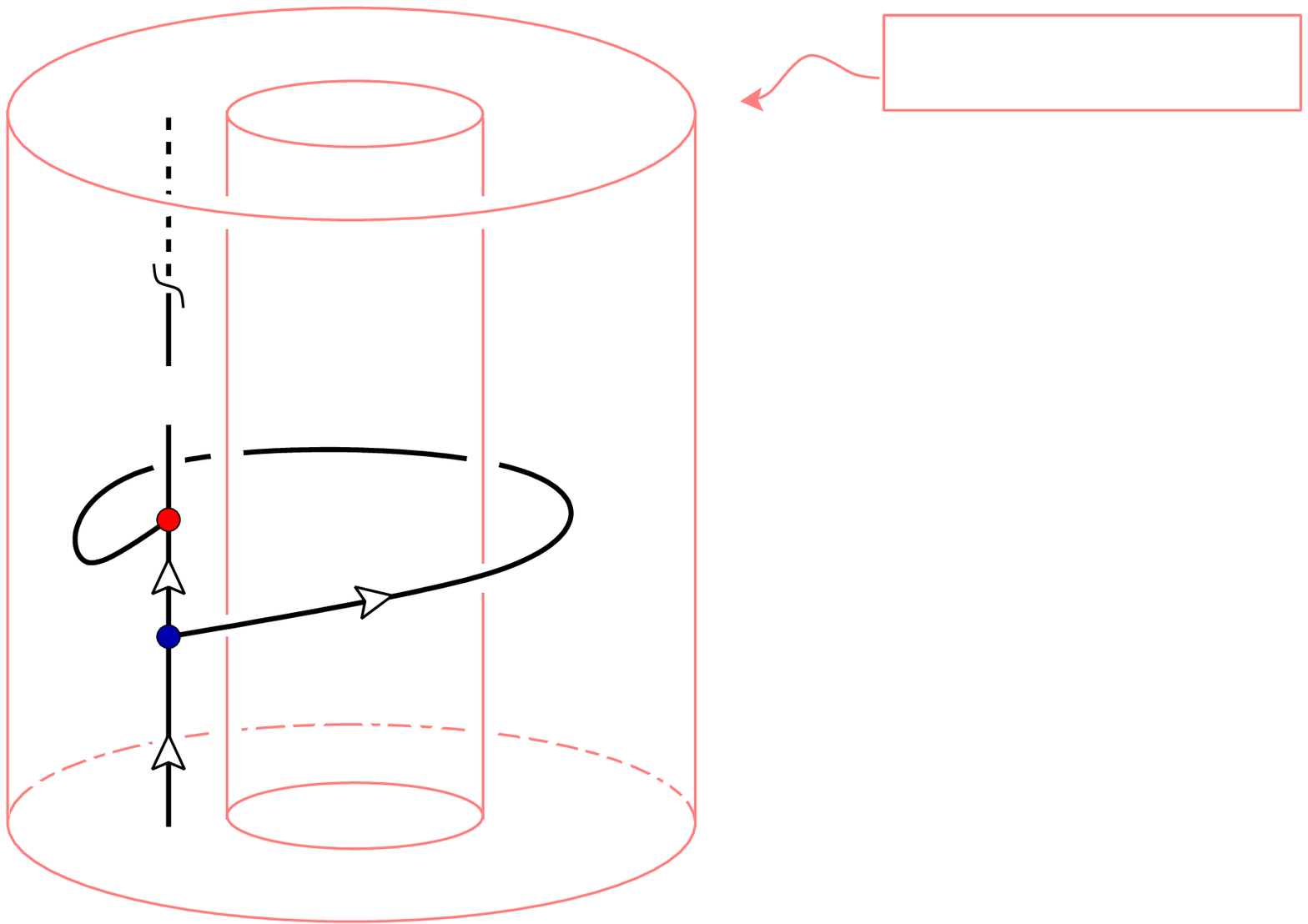}}} \setulen78
    \put(15.3,49.1) {$\scs A $}   \put(22.9,77) {$\scs \sigma $}
    \put(134,125.1) {$\scs I{\times}S^1{\times}I{/}{\sim} $}
    \put(131,108)   {$\scs (r,\phi)_{\rm top} $}
    \put(131,96)    {$\scs {\sim}\,(\frac1r,-\phi)_{\rm bottom} $} \eP
  \bP(100,79)(0,3)
    \put(-47,49)  {$K_j~\hat=$}
    \put(0,0)     {\scalebox{.30}{\includegraphics{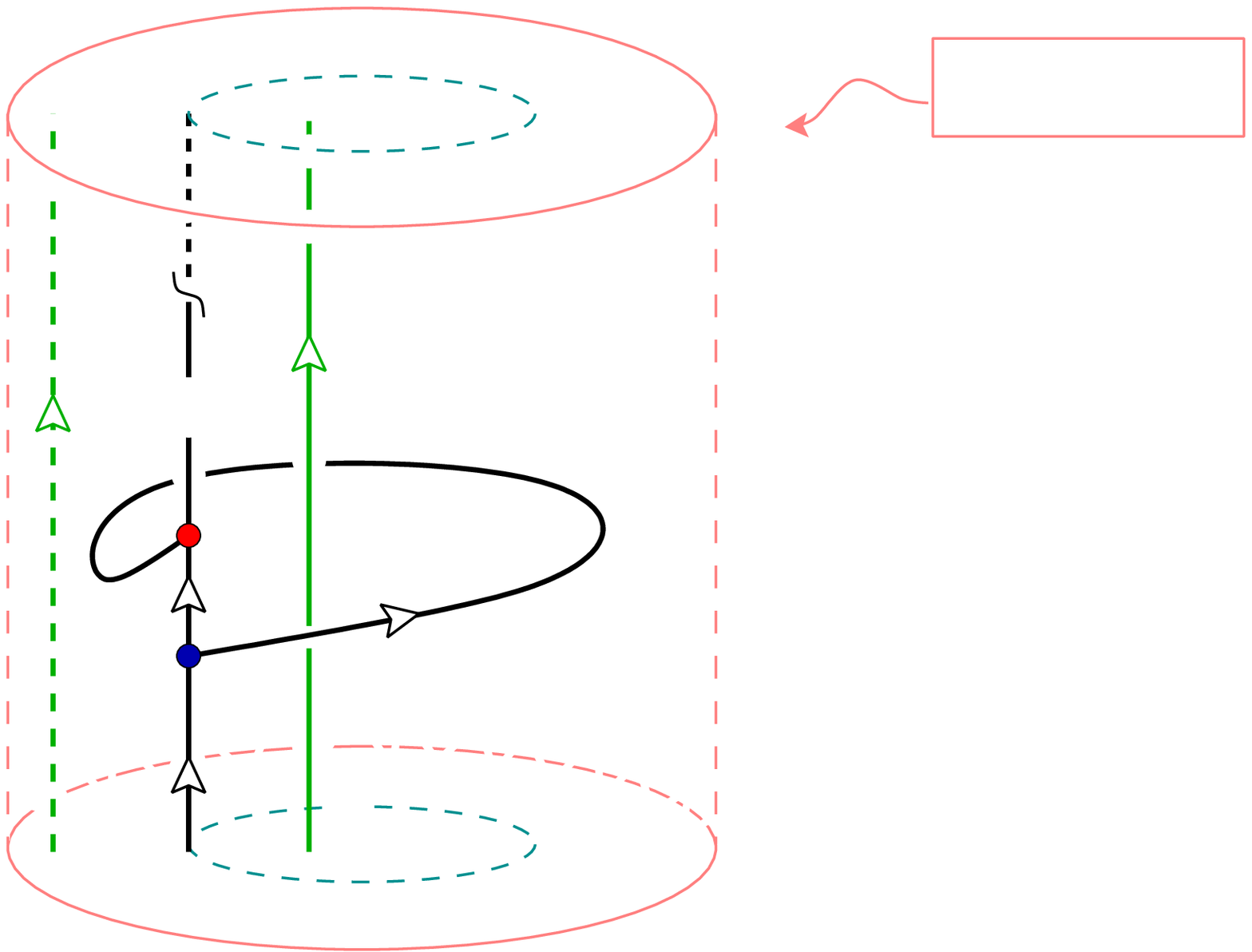}}} \setulen78
    \put(9.3,15.4)  {$\scs j$} \put(17.6,48.9) {$\scs A$}
    \put(25.2,76.6) {$\scs \sigma$}    \put(45.7,14.4) {$\scs j$}
    \put(137.1,121.7){$\scs S^2{\times}I\!{/}{\sim}$} \eP
  \nonumber\\[-1.7em]~ \label{pics.K}
  \end{eqnarray}
The coefficients $K_j$ obtained this way satisfy \cite[thm.\,3.7]{fuRs8}
$K_\ja \eq K_\jb,~ K_\ja+\Z_{\ja\ja} \iN 2\zet$ and 
$|K_\ja| \,{\le}\, \frac12\,\Z_{\ja\ja}$\,.

\itx
There is only one Jandl structure on the Frobenius algebra $A\eq\one$, namely
$\sigma\eq\id_\one$. In this particular case the formulas \erf{pics.Z} for 
$\Z\eq\Z(A)$ and \erf{pics.K} for $K\eq K(A,\sigma)$ reduce to
  \be
  \Z_{\imath,\jmath}(\one) =\,\delta_{\imath,\jb}^{} \qquand
  K_\jmath(\one,\id_\one) = \left\{\!\begin{array}{ll} \mbox{\sc F$\!$S}(\U_\jmath) 
  & {\rm if}\ \jmath\eq\jb \\[2pt] 0 & {\rm else} \eear\right.
  \ee
(as first obtained in \cite{fffs3}), with {\sc F$\!$S} denoting the
Frobenius-Schur indicator (see e.g.\ \cite{fffs3,natA4,ngsc2}).

\itx
Finally, as an example for a world sheet with non-empty boundary, consider the disk 
${\rm D}$ with one bulk field insertion $\vhi\eq\vhi_{i,j}$ and with boundary 
condition $M$. One finds \cite[eq.\,(4.20)]{fuRs10} that the correlator is non-zero
only if $j\eq\overline i$, in which case the coefficient of $\Cor({\rm D};\vhi;{M})$
in an expansion in a standard (\onedim) basis of the space of two-point blocks on the
sphere it is given by
  \be \bearl ~\nxl5
  \bP(280,87) \setulen60
  \put(143.5,0)  {\scalebox{.228}{\includegraphics{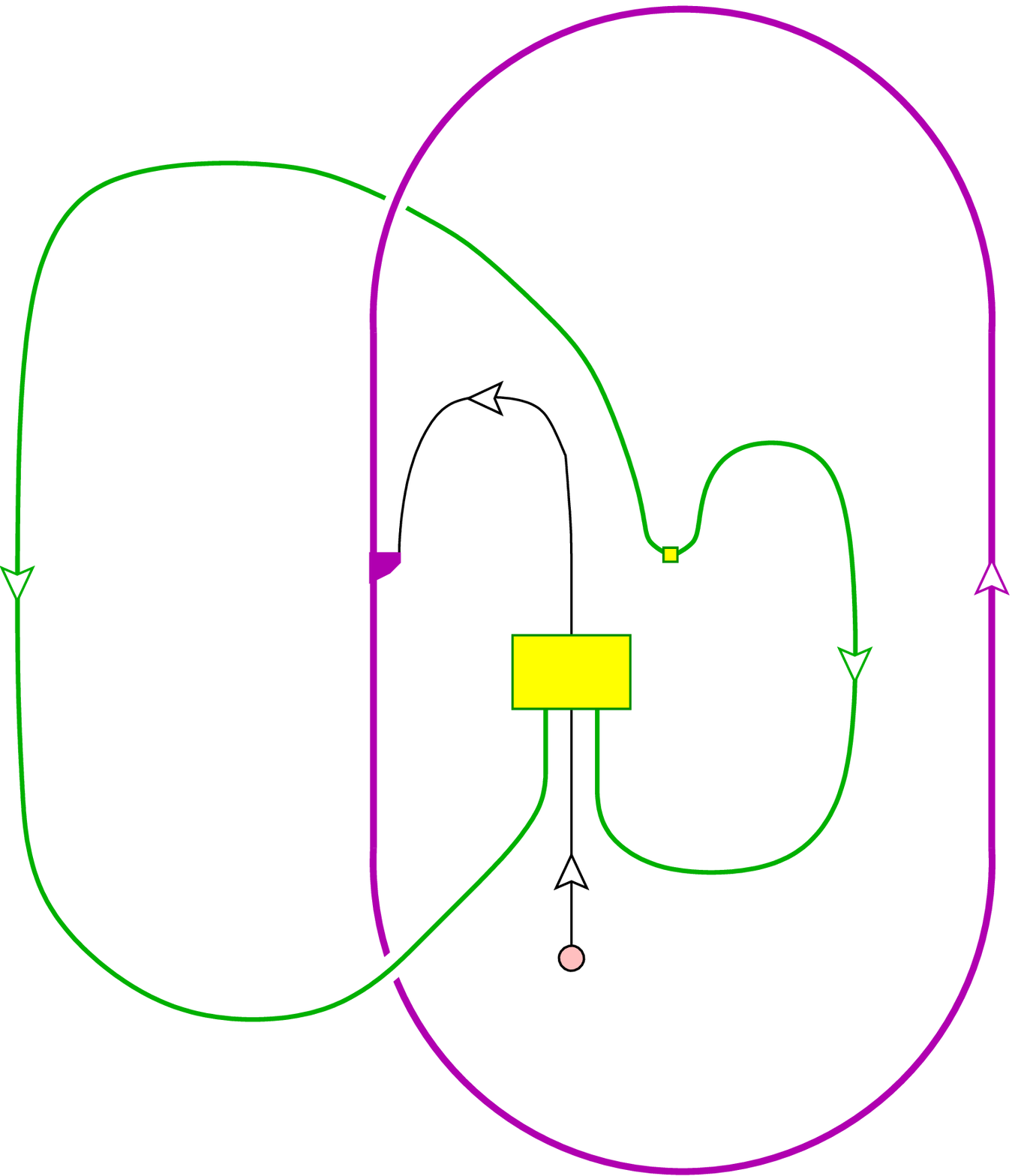}}}
   \put(65,0){
  \put(178.5,58) {$\scs \overline{i} $}
  \put(72,123)   {$\scs i $} 
  \put(167,78.3) {$\scs \vhi $} 
  \put(159.2,101){$\scs A $}
  \put(241,134)  {$\scs M $}
  } \eP
  \eear \ee
\end{itemize}

The study of these and other correlators has also motivated the investigation
of properties of Frobenius algebras and their \rep\ theory.
Here is a list of selected results that arose in this manner:

\Itemize

\itx
For $A$ a special Frobenius algebra in a rigid monoidal category \C, every
$A$-module is a submodule of an induced (free) module $(A\oti U,m\oti \id_U)$
for some object $U$ of \C\ \cite[Lemma\,4.8(ii)]{ffrs}.

\itx
For $A$ a simple symmetric special Frobenius algebra in a \mtc\ 
\C, every simple $A$-bimodule is a sub-bimodule of a braided-induced bimodule 
$U\Otip\! A\,\Otim{V}$ (see \erf{braid_ind}) for some simple objects $U,V$ of 
\C\ \cite[Prop.\,4.7]{ffrs5}.
        
\itx
For $A$ a symmetric special Frobenius algebra in a \mtc\ the complexified
Grothendieck ring $\kO(\CAA)\,{\otimes_\zet}\,\complex$ of the category of 
$A$-bimodule is isomorphic as a \complex-algebra
to $\bigoplus_{i,j\in I}\End_\complex\big(\HomAA(A\otip U_i,A\otim U_j)$
\cite{fuRs12}.

\itx
For $A$ a symmetric special Frobenius algebra in a \mtc\ \C\ there is an exact sequence 
  \be
  1 \,\to\, {\rm Inn}(A) \,\to\ {\rm Aut}(A) \,\to\, \PicCAA
  \ee
of groups \cite[Prop.7\,]{fuRs11}. One can also describe conditions under which there
exists a Morita equivalent Frobenius algebra $A'$ for which the corresponding 
homomorphism from ${\rm Aut}(A)$ to $\PicCAA$ is surjective \cite{bfrs}.

\itx
For $A$ a simple symmetric special Frobenius algebra in a \mtc\ \C, the 
\complex-vector space $\bigoplus_{\ia\in\I} \Homaa{\U_\ia\otip\! A\otim \U_\ib}A$ can 
be endowed with a natural structure of a semisimple unital commutative associative 
algebra $\mathcal A$. The simple modules of this algebra are in bijection with the 
simple $A$-modules, i.e.\ with the elementary boundary conditions of the full CFT 
defined by \C\ and $A$. If $A$ is Morita equivalent to $\one$, then $\mathcal A$ is 
naturally isomorphic to the complexified Grothendieck ring of \C\ \cite{fuSs}.
     
\itx
Every simple algebra in a \mtc\ is Morita equivalent to one for which 
$\dimc \Hom(\one,A) \eq 1$ \cite{ostr}. This can be used to show 
\cite[Prop.\,3.6]{fuRs9} that the number of Morita classes of simple symmetric 
special Frobenius algebras in a modular tensor category is finite. As an immediate 
consequence, associated to any chiral RCFT there are only finitely many different 
full RCFTs, in particular only finitely many different physical modular invariant
torus partition functions.
\end{itemize}


\section{Verlinde-like relations in non-rational CFT}\label{sec.verl}

\subsection{Beyond rational CFT}

The results for RCFT correlators reported in section \ref{sec.corr.cb} may be summarized 
by the statement that for \emph{rational} CFT, the transition from chiral to full CFT is 
fully under control and can be treated in an entirely model-independent manner. In 
particular all correlators can be constructed combinatorially\,%
  \footnote{~To go beyond the combinatorial framework, one must promote the geometric
  category of topological world sheets to a category of world sheets with metric,
  and similarly for the relevant algebraic category of vector spaces. Confidence that
  this can be achieved comes from the result \cite{BAki} that the notions of a
  (\C-decorated) topological modular functor and of a (\C-decorated)
  complex-analytic modular functor are equivalent. An intermediate step might consist
  of the construction of cohomology classes on the moduli spaces of curves using
  combinatorial models for the moduli space like in \cite{costK5}.}
as elements of the appropriate spaces of conformal blocks of the chiral theory. This 
does not mean that a complete understanding 
of rational CFT has been achieved, but the issues that still need to be resolved
are confined to the realm of chiral CFT.

In contrast, attempts to understand more general CFTs, like Liouville theory or
logarithmic CFTs, at a similar level are still under development. In the present and 
the next section we discuss two features that are expected to be relevant for making 
progress in this direction: relations between fusion rules and modular transformation 
properties of characters, and a generalization of the notion of \mtc\ to chiral 
symmetry structures with non-semisimple \rep\ categories which is based on the 
presence of a Hopf algebra in \C\ that is obtained as a coend.

\medskip

The first of these observations has been made in a specific class of models which are
collectively termed \emph{logarithmic CFTs} (or sometimes rational logarithmic CFTs), 
and which have been intensively studied in the physics literature. The proper 
categorical setting remains, however, unclear even for these models. For instance, 
in spite of much recent progress (see e.g.\ \cite{hulz3,huan28,miya13}), still some 
mysteries need to be unraveled for the tensor structure of logarithmic CFTs.
{}From the physics perspective, one is tempted to argue that certain features 
valid in RCFT are still present in the non-rational case, and that they imply specific 
properties of the vertex algebra and its relevant \rep\ category \C. Concretely, the 
existence of a sensible notion of operator product expansion is expected to imply that 
\C\ is monoidal, for obtaining reasonable monodromy properties of conformal blocks \C\ 
should be braided, the nondegeneracy of two-point blocks on the disk and on the sphere
can be taken as an indication that \C\ is rigid, while the scaling symmetry points to
the possible presence of a twist (balancing). But as any such reasoning is basically 
heuristic, none of these properties can be taken for granted. For instance, it has been 
argued, in the context of vertex algebras \cite{miya13}, that rigidity must be relaxed.

For the purposes of the present section we adopt the following setting: we assume 
that the category \C\ of chiral data is a braided \emph{finite tensor category}, 
that is \cite{etos}, a \ko-linear abelian braided rigid monoidal category for 
which every object $U$ is of finite length and has a projective cover $P(U)$, which 
has finitely many simple objects up to isomorphism, and for which the tensor unit 
$\one$ is simple. Here $\ko$ is an algebraically closed field; in the CFT context, 
$\ko \eq \complex$. For finite tensor categories the tensor product functor 
$\otimes$ is exact in both arguments. 
We continue to denote the set of isomorphism classes of simple objects by $\I$.

\medskip

While it is not known whether the framework of finite tensor categories will 
ultimately suffice to understand all aspects of logarithmic CFTs,
in any case the relevant categories \C\ will no longer be semisimple, so that 
in particular various arguments used in the TFT construction no longer apply.
Still one may hope that \threedim\ \tft, or at least invariants of links
in three-manifolds, will continue to provide tools for studying such CFTs. It is
thus a significant result that such invariants exist \cite{lyub6,kaRad3,henni,muNa},
and that even \rep s of mapping class groups can be constructed \cite{lyub8,kerl5}.
However, these invariants and mapping class group \rep s do not really fit together,
and actually it is reasonable to expect that for a full treatment one must work
with \emph{extended} TFTs related to higher categorical structures (see e.g.\ 
\cite{luri4}) and allow the TFT functor to take values in complexes of vector 
spaces. But such generalizations have yet to be established.


\subsection{The fusion rules of a semisimple modular tensor category}

By the \emph{fusion rules}, or fusion algebra, of a chiral CFT, one means the 
complexified Grothendieck ring
  \be
  \F := \kc\otimes_\zet^{}\complex \,.
  \labl{fusalg}
Recall that the Grothendieck group of an abelian category \C\ 
is the abelian group \kc\ generated by isomorphism classes $[U]$ of objects of \C\
modulo the relations $[W] \eq [U] \,{+}\, [V]$ for any short exact sequence
$\,0\To U\To W\To V\To 0\,$ in \C.

To set the observations about fusion rules for logarithmic CFTs into context, it is 
worth to describe first in some detail the analogous results for (semisimple) rational 
CFTs. The Grothendieck group of a \mtc\ \C\ inherits various properties from \C.
It is a commutative unital ring, with product
  \be
  [U]\fus[V] := [U\oti V]
  \ee
and unit element $[\one]$, and evaluation at $[\one]$ defines an involution
$[U] \,{\mapsto}\, [U^\vee]$. \kc\ has a distinguished basis 
$\{\hspace*{.2pt} [\U_i] \hspace*{.2pt}\}_{i\in\I}^{}$ 
in which the structure constants $\N ijk$, obeying
  \be
  [\U_i] \fus [\U_j] = \sum_{k\in\I}\N ijk\, [\U_k] \,,
  \ee
are non-negative integers, and as a consequence of semisimplicity these structure
constants are given by $\N ijk\eq \dimc\Hom(\U_i\oti\U_j,\U_k)$.

It follows that the fusion algebra \erf{fusalg} of a \mtc\ is a commutative, 
associative, unital \complex-algebra for which evaluation at the unit is an 
involutive automorphism. Together, these properties imply that \F\ is a 
\emph{semisimple} algebra.  As a consequence, \F\ has another distinguished basis 
$\{ e_l \}$, consisting of primitive orthogonal idempotents, such that 
$e_l\fus e_{l'} \eq \delta_{l,l'}\,e_l$.
The transformation between the two distinguished bases $\{ [\U_i] \}$ and 
$\{ e_l \}$ furnishes a unitary matrix $\So$ 
such that $[\U_i] \eq \sum_{l} (\Soinv)_{0l}^{}\SO_{il}\,e_l$ with 
$(\Soinv)_{0l}^{} \nE 0$. The structure constants $\N ijk$ can then be written as
  \be
  \N ijk = \sum_l \frac {\SO_{il}\,\SO_{jl}\,\SOstar_{lk}} {\SO_{0l}} \,,
  \labl{furu.diag}
a relation known as the \emph{diagonalization of the fusion rules}.
 
One of the distinctive features of \mtcs\ is the fact that the so defined
matrix $\So$ coincides with a multiple of the matrix $\soo$ \erf{def.s} whose 
invertibility enters the definition of a \mtc:\,%
 \footnote{~To be precise, the equality holds for a specific ordering of the 
 basis $\{ e_l \}$ of idempotents. Note that for any semisimple \complex-algebra
 the sets of primitive orthogonal idempotents and of irreducible \rep s are in
 bijection, albeit not in natural bijection. Implicit in \erf{verl1} is the 
 statement that for \F\ one specific such bijection is distinguished; in this
 bijection the braiding enters as an additional structure.}
  \be
  \So = \Soo \equiv \SO_{00}\,\soo \,.
  \labl{verl1}
This equality has been established, from three related points of view, in 1989 in
\cite{witt50}, \cite{mose3} and \cite{card9}. The proof can be formulated entirely in
categorical terms, as a series of identities between elements of the morphism space
$\End(\one)\eq \complex\,\id_\one$ (which we identify with \complex):
  \begin{eqnarray}
  \bP(390,65)(62,0) \setlength\unitlength{1.8pt}
   \put(-5,-2){
    \put(20,17)    {$\dsty\frac{\sOO_{i,k}}{\sOO_{0,k}}\,\sOO_{j,k}\hspace*{.5em}
                    =\hspace*{.5em}\dsty\frac{\sOO_{i,k}}{\sOO_{0,k}}$}
    \put(76,0)     {\scalebox{.225}{\includegraphics{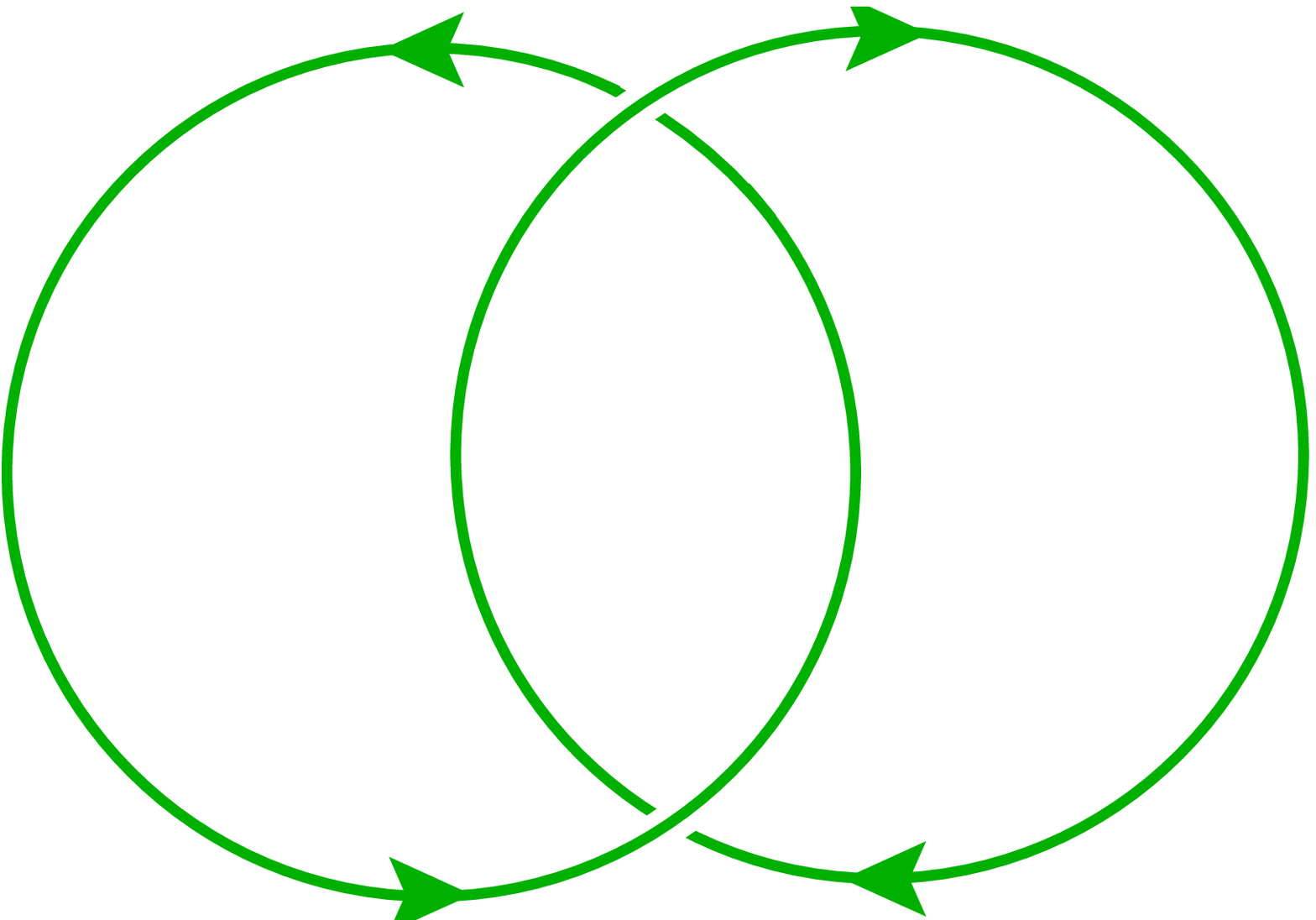}}}
    \put(92.3,17.5){$\sss j$}
    \put(115.5,17.5){$\sss k$}
    \put(145,17)   {$=$} 
    \put(163,0)    {\scalebox{.225}{\includegraphics{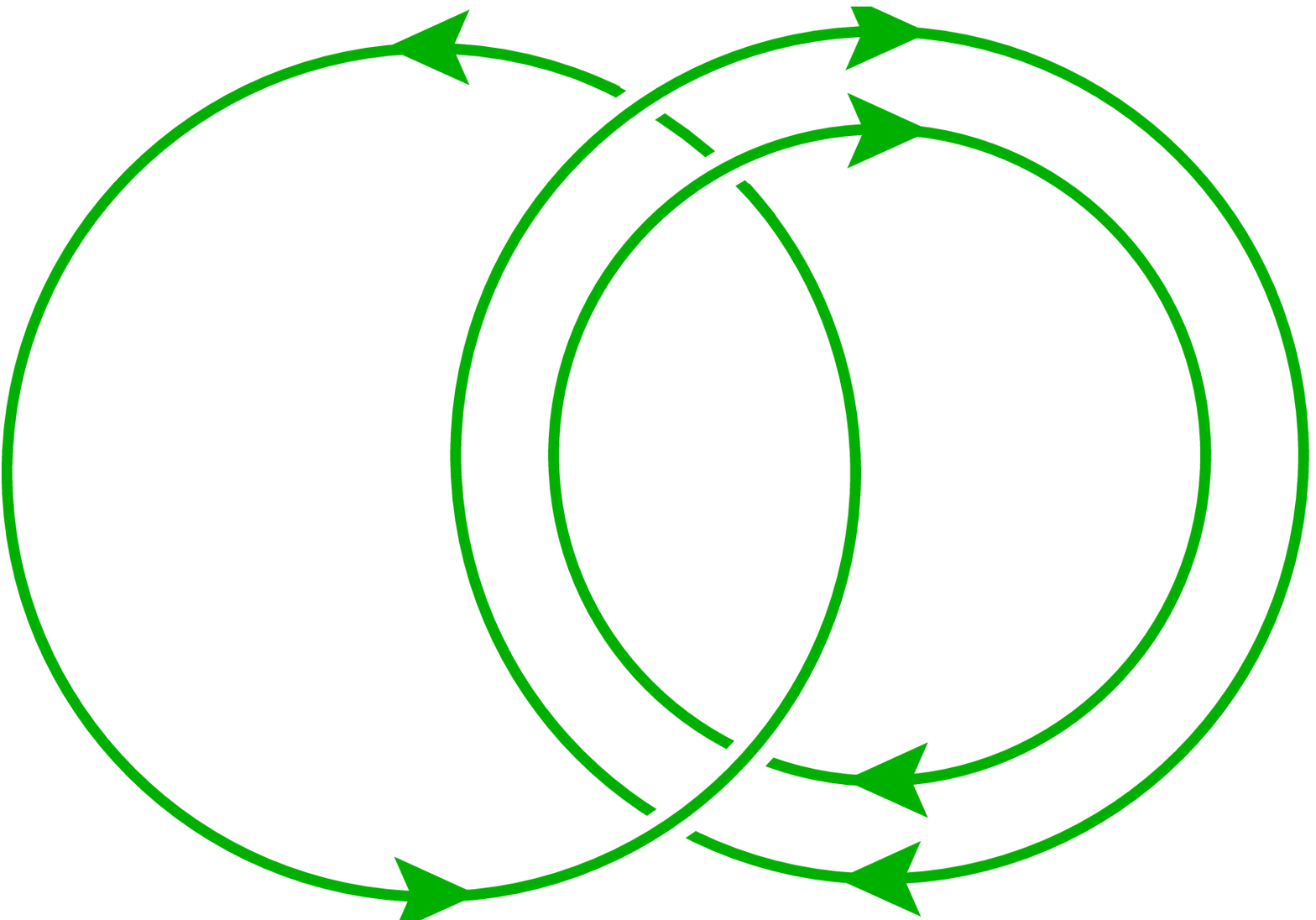}}}
    \put(180,17,5) {$\sss j$}
    \put(189,17.5) {$\sss i$}
    \put(202,17.5) {$\sss k$}
  }
  \put(-33,1){\bPo 
    \put(48,-38)   {$=~\,\dsty\sum_p\sum_\alpha$}
    \put(88,-55)   {\scalebox{.225}{\includegraphics{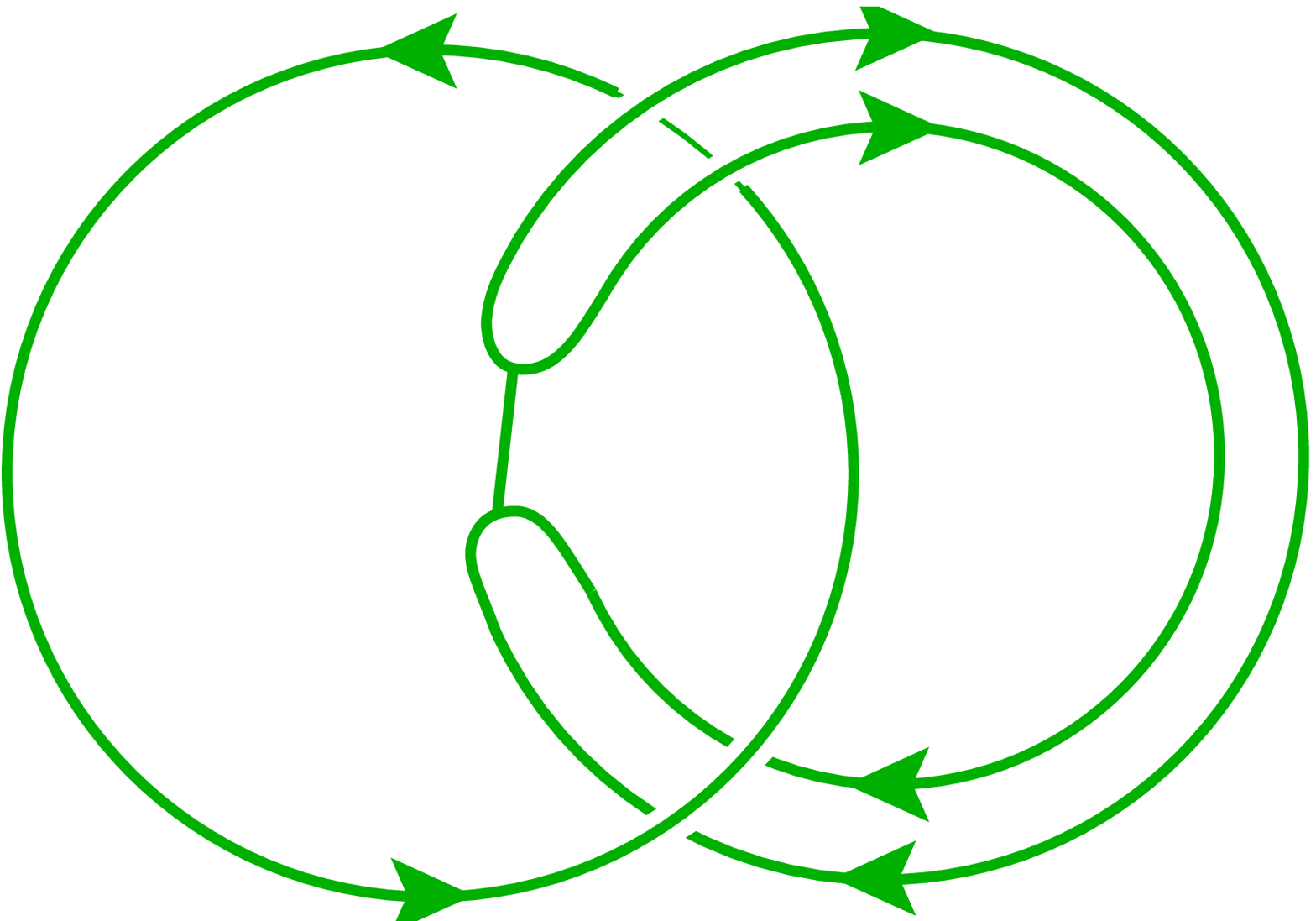}}} 
    \put(106.2,-35){$\sss p$}
    \put(114.5,-32.4){$\sss \overline\alpha$}
    \put(112.7,-37){$\sss \alpha$}
    \put(127.5,-37){$\sss k$}
    \put(134.5,-32){$\sss i^{\!\vee}$}
    \put(147.5,-32){$\sss j^{\!\vee}$}
    \put(160,-38)  {$=~\,\dsty\sum_p\sum_\alpha $}
    \put(198,-55)  {\scalebox{.225}{\includegraphics{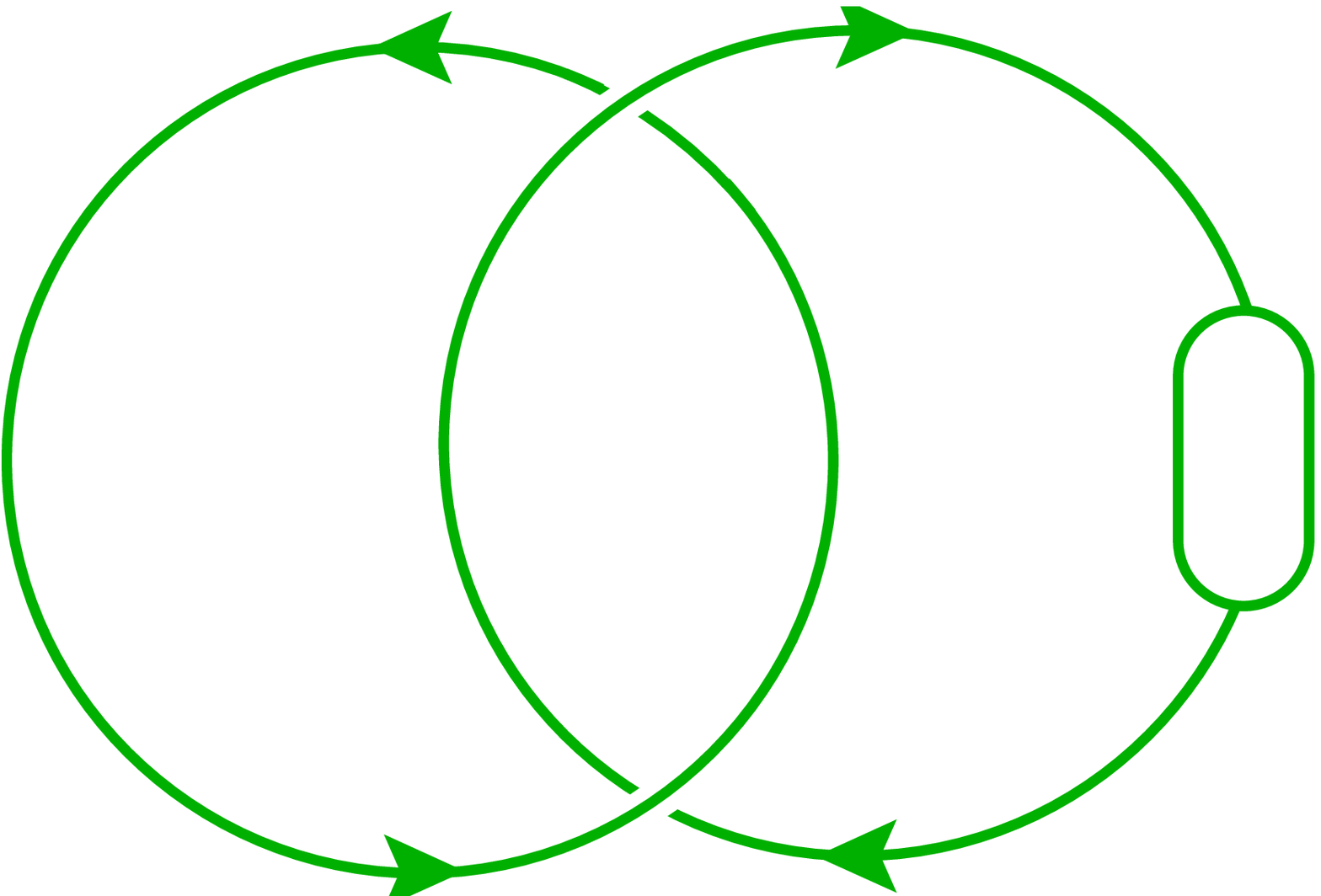}}}
    \put(214,-37)  {$\sss p$}
    \put(231.5,-37){$\sss k$}
    \put(246.5,-34){$\sss i^{\!\vee}$}
    \put(258.5,-34){$\sss j^{\!\vee}$}
    \put(254.8,-46){$\sss \overline\alpha$}
    \put(255.4,-26.8){$\sss \alpha$}
    \put(271,-38)  {$=~~\dsty\sum_p \N ij{\,p} \sOO_{p,k}$}
  \eP } \eP
  \nonumber \\[6.6em]~
  \end{eqnarray}
Here the first two equalities follow directly from the definition of $s$
(together with the fact that simple objects are absolutely simple), the third 
uses semisimplicity to express $\id_{\U_j}\oti\id_{\U_i}$ through a choice of 
bases in the spaces $\Hom(\U_i\oti\U_j,\U_p)$ and their dual bases in 
$\Hom(\U_p,\U_i\oti\U_j)$, the forth holds by the defining properties of 
duality morphisms and braiding, and the last is a property of dual bases.

An independent, much stronger statement, can be established when one not just works 
with \C\ as a category, but includes information from its concrete realization as 
the \rep\ category of a rational conformal vertex algebra \V. Namely, one 
can then show that the matrix $\Soo$ also coincides with the matrix
$\Schi$ that implements the modular transformation $\tau\,{\mapsto}\,{-}\frac1\tau$
on the space spanned by the characters $\chii_{\U_i}$ \erf{def_chi} of the simple 
\V-modules $\U_i$. Thus the diagonalization of the fusion rules can be
rephrased as the \emph{Verlinde formula}
  \be
  \N ijk = \sum_l \frac
  {S^{\sss\chi}_{il}\, S^{\sss\chi}_{jl}\, S^{\chi\,*}_{\,lk}} {S^{\sss\chi}_{0l}} \,.
  \labl{verl2}
The Verlinde formula was conjectured in 1988 \cite{verl2}. It was established, around 
1995, for the particular case of WZW models, a class of models for which the vertex 
algebra \V\ can be 
constructed from an affine Lie algebra. For these models the proof can be obtained with 
methods from algebraic geometry (see e.g.\ \cite{tsuY,falt3,beau8,sorg}). A proof for 
all rational conformal vertex algebras was achieved in 2004 \cite{huan21,huan24}.


\subsection{Verlinde-like formulas for logarithmic minimal models}\label{verl.L1p}

One reason why the findings above, which all apply to \emph{rational} CFT, are of 
interest to us here is that Verlinde-like relations have also been found in non-rational 
CFT, namely for the specific class of so-called \emph{logarithmic minimal models} 
$\mathcal L_{1,p}$, with $p\iN\zet_{\ge2}$. For these models, there is a corresponding 
vertex algebra $\Vep\eq\V(\mathcal L_{1,p})$ that is quite well 
understood \cite{kaus,caFl,adMi3,naTs2}. In particular, \Vep\ obeys the so-called 
$C_2$-cofiniteness condition, and as a consequence the category $\Cep\eq\Rep(\Vep)$ 
of grading-restricted generalized \Vep-modules is a braided finite tensor category 
(see \cite{naTs2} and section 5.2 of \cite{huan29}).
Specifically, the number of isomorphism classes of simple objects of \Cep\ is $2p$; 
there are two isomorphism classes of simple objects which are projective, while the 
isomorphism
classes of non-projective simple objects come in $p\,{-}\,1$ pairs, with non-trivial 
extensions existing within each pair, but not among members of different pairs 
\cite{fgst2,naTs2}. We write the label set $\I$ as a corresponding disjoint union
  \be
  \begin{array}{lll}\dsty
  \I = \bigsqcup_{a=1}^{p+1}\I_a \quad & {\rm with}~~ &
  \I_a = \{ i_a^+, i_a^- \} ~~~{\rm for}~~ a\eq1,2,...\,,p{-}1 \\ & {\rm and} &
  \I_p = \{ i_p \} \,, \quad \I_{p+1} = \{ i_{p+1} \} 
  \eear \labl{Ia}
of $p\,{-}\,1$ two-element subsets and two one-element subsets, with $\U_i$ projective
for $i \iN \I_p\,{\cup}\,\I_{p+1}$ and $\U_i$ non-projective 
for $i \iN \I \,{\setminus} (\I_p\,{\cup}\,\I_{p+1})$.

Under the modular transformation $\tau\,{\mapsto}\,{-}1/\tau$ the characters 
$\chii(\tau)$ of \Vep-modules acquire explicit 
factors of the modular parameter $\tau$ or, in other words, logarithms of the parameter 
$q\eq\eE^{2\pi\ii\tau}$ in which the characters are power series. There is thus no 
longer an \slz-\rep\ on the ($2p$-dimensional) span of \Vep-characters. However, there 
is a $3p{-}1$-dimensional \slz-\rep\ $\wrho$ on the space spanned by the $2p$ simple 
characters $\chii_i(\tau)$ together with $p\,{-}\,1$ \emph{pseudo-characters}
$\pchi_a(\tau)$. The quantities $\pchi_a$ are known explicitly as functions of $\tau$,
see e.g.\ \cite{floh5,flga} and
\cite[Sect.\,2.2]{fgst}: they are of the form of a product of $\,\ii\,\tau$
times a power series in $q$. It is expected \cite{fgst} that this $3p{-}1$-dimensional 
space coincides with the space of zero-point conformal blocks on the torus.
For any $C_2$-cofinite vertex algebra this space of conformal blocks can be 
constructed with the help of certain symmetric linear functions on the endomorphism 
rings of projective modules \cite{miya8,arik3} and can be shown \cite{miya8,flga} to 
carry a \findim\ \slz-\rep. 

We denote by $\Sxchi$ the \rep\ matrix for the modular transformation 
$\tau\,{\mapsto}\,{-}\frac1\tau$ in the \rep\ $\wrho$. Under such a transformation 
a pseudo-character is mapped to a linear combination of ordinary characters
\cite{fgst,gaTi}, i.e.\ $\Sxchi$ has the block-diagonal form
  \be
  \wrho\big(\zzmatrixS\big) \equiv \Sxchi 
  = \left( \!\! \begin{array}{cc} S^{\sss\chi\chi_{\phantom|}}_{{\phantom:}} \!\!&\!\!
  S^{\sss\chi\pchi_{\phantom|}}_{{\phantom:}} \\ S^{\sss\pchi\chi_{\phantom|}}_{{\phantom:}}
  \!\!& 0 ~ \eear \!\!\right) .
  \labl{Shat}

Further, introducing label sets $A_a\,{:=}\,\{+,-,\nul\}$ for $a\eq1,2,...\,,p{-}1$
and $A_p \eq A_{p+1} \,{:=}\,\{\nul\}$ and labeling the characters and pseudo-characters
in such a way that
  \be
  \chii_{(a\alpha)} = \left\{ \bearll
  \chii_{\!i_a^\alpha} & {\rm for}~ a\eq 1,2,...\,,p{-}1\,,~\alpha \eq +,- \,, \\~\\[-.98em]
  \pchi_a & {\rm for}~ a\eq 1,2,...\,,p{-}1\,,~\alpha \eq \nul \,, \\~\\[-.98em]
  \chii_{i_a^{}}  & {\rm for}~ a\eq p, p{+}1\,,~\alpha \eq \nul \,, \eear \right.
  \ee
as well as $\one \eq \U_{(1+)} \,{\equiv}\, \U_{i_1^+}$,
one shows \cite[Prop.\,1.1.2]{gaTi} by direct calculation that the complex numbers
  \be
  \hsp{-.7}\begin{array}{r}\dsty
  N_{(a\alpha),(b\beta)}^{~~~~~~(c\gamma)} = \sum_{d=1}^{p+1} \sum_{\delta\in A_d}
  \frac {\sxchi_{(c\gamma),(d\delta)}} {{(\sxchi_{(1\nul),(d\nul)})}^2} \Big(
    \sxchi_{(1\nul),(d\nul)} \sxchi_{(a\alpha),(d\nul)} \sxchi_{(b\beta),(d\delta)}
  + \sxchi_{(1\nul),(d\nul)} \sxchi_{(a\alpha),(d\delta)} \sxchi_{(b\beta),(d\nul)}
  \hsp{.4}\\~\\[-1.6em]
  -\, \sxchi_{(1\nul),(d\delta)} \sxchi_{(a\alpha),(d\nul)} \sxchi_{(b\beta),(d\nul)}
  \Big) \eear
  \labl{gaTi.verl}
are integers, and that those for which all three labels $(a\alpha),(b\beta),(c\gamma)$
correspond to ordinary characters rather than pseudo-characters are nonnegative and 
coincide with the Jordan-H\"older multiplicities for the corresponding tensor products 
of simple objects of \C. (For labels outside this range, the integers 
$N_{(a\alpha),(b\beta)}^{~~~~~~(c\gamma)}$ can be negative, though.)
The close similarity of \erf{gaTi.verl} with the Verlinde formula \erf{verl2}
suggests to call it a \emph{Verlinde-like formula}.

The Jordan-H\"older multiplicities obtained this way constitute the structure constants
$\N ijk$ of the fusion algebra $\F \eq \kO(\Cep)\,{\otimes_\zet^{}}\,\complex$ of \Cep. 
This algebra is non-semisimple, so that the fusion rules can no longer be diagonalized. 
However, there is another type of Verlinde-like relation: one can show \cite{garu2,peRr2} 
that the matrices $N_i$ with ${(N_i)}_j^{~k} \eq \N ijk$) can \emph{simultaneously} be 
brought to Jordan form $N_i^{\rm J}$ according to 
  \be
  N_i = Q\,N_i^{\rm J}\,Q^{-1} \,,
  \ee
with the entries of the similarity matrix $Q$ as well as of all matrices $N_i^{\rm J}$ 
being expressible through the entries of $\Sxchi$. While the concrete expressions are 
somewhat unwieldy and up to now a conceptual understanding of this similarity 
transformation is lacking, the very existence of the transformation is a 
quite non-trivial observation. 

Besides the fusion algebra \F, of dimension $2p$, there is also an algebra of
dimension $4p{-}2$ that arises from the tensor product of \Cep. Namely, the (conjectured) 
tensor product of any two simple objects of \Cep\ decomposes as a direct sum of simples 
and of projective covers of simples. As a consequence the tensor product closes among 
the union of (direct sums of) all simple objects $\U_i$ and their projective covers 
$P_i\eq P(\U_i)$, and one can thus extract a $4p{-}2$-dimensional \emph{extended 
fusion algebra} $\widetilde{\F}$ spanned by the set $\{[\U_i],\,[P_i]\,|\,i\iN\I\}$. 
$\widetilde{\F}$ is a commutative unital associative algebra \cite{peRr}. Moreover, 
similarly as for \F, the matrices $\widetilde N_\ell$ made out of the structure constants of 
$\widetilde{\F}$ can simultaneously be brought to Jordan form by a similarity transformation, 
  \be
  \widetilde N_\ell = \widetilde Q\,\widetilde N_\ell^{\rm J}\, \widetilde Q^{-1} \,.
  \ee
And again the entries of both $\widetilde Q$ and $\widetilde N_\ell^{\rm J}$ can be 
expressed through the entries of $\Sxchi$ \cite{rasm22}, although, again, a deeper 
understanding of the resulting formulas is lacking. Thus there is again a
Verlinde-like relationship.

Finally, even though \slz-transformations do not close on the linear span of characters 
of \Vep-modules, it is nevertheless possible to extract an interesting $2p$-dimensional
\slz-\rep\ $\frho$ from the modular transformations of characters. This is achieved
either by separating a so-called automorphy factor from the transformation formulas 
\cite{fhst}, or by decomposing $\wrho$ as the `pointwise' product (i.e.\ such that
$\wrho(\gamma) \eq \wrhp'(\gamma)\,\wrhp''(\gamma)$ for any $\gamma\iN\slz\,$) of two 
commuting $(3p{-}1)$-dimensional \slz-\rep s $\wrhp'$ and $\wrhp''$, one of which
restricts to the span of characters \cite{fgst}.
Moreover, this description gives rise to yet another Verlinde-like formula: by a 
similarity transformation $\mathcal Q$ the fusion matrices $N_i$ (formed, as 
described above, from the subset $\{\N ijk\}$ of the Jordan-H\"older multiplicities 
\erf{gaTi.verl}) are simultaneously brought to \emph{block-diagonal} form 
$N_i^{\rm b.d.}$, i.e.
  \be
  N_i^{} = \mathcal Q\, N_i^{\rm b.d.} \mathcal Q^{-1} \,,
  \ee
and both  $\mathcal Q$ and the 
$N_i^{\rm b.d.}$ have simple expressions in terms of the entries of $\mathcal S$.
But again the explicit expressions obtained this way are not too illuminating.
On the other hand, the block structure of the matrices $N_i^{\rm b.d.}$ is given
by two $1{\times}1$-blocks and $p{-}1$ $2{\times}2$-blocks, which precisely
matches the structure of nontrivial extensions among simple objects of \C.


\subsection{Hopf algebras for the $\mathcal L_{1,p}$ models}\label{hopf.L1p}

By inspection, the Perron-Frobenius dimensions of the simple objects
(i.e.\ the Perron-Frobenius eigenvalues of the fusion matrices $N_i$) of \Cep\
with $p$ prime coincide with the dimensions of the simple modules over the
$p$-restricted enveloping algebra of $\mathfrak{sl}(2,\mathbb F_p)$
(see e.g.\ \cite{hump3}), and the same holds for their respective projective 
covers. In view of the relationship between modular \rep s and quantum groups 
at roots of unity, this may be taken as an indication of the existence of a 
suitable quantum group with a \rep\ category that is equivalent to \Cep\ at 
least as an abelian category.

Such a Hopf algebra indeed exists for any $p\iN\zet_{\ge2}$, namely the 
$2p^3$-dimensional restricted quantum group \uqz\ at the value $q\eq\eE^{\pi\ii/p}$ 
of the deformation parameter: there is an equivalence
  \be
  \Rep(\Vep) \,\simeq\, \uqz\Mod
  \labl{RepVep=Repuqz}
of abelian categories \cite{fgst,fgst2,naTs2}. 

The two \slz-\rep\ $\wrho$ and $\frho$ described in the previous subsection also
arise naturally in the study of \uqz. First, it is known \cite{lyma,kerl2} that the 
center of \uqz, which has dimension $3p{-}1$, carries a \rep\ of \slz, obtainable by 
composing the Frobenius and Drinfeld maps (see \erf{def_fmap} and \erf{def_fQ_p}
below) between \uqz\ and its dual, and this \rep\ can be shown to be isomorphic with 
$\wrho$ \cite{fgst}. Second, $\frho$ is obtained through the multiplicative Jor\-dan 
decomposition of the ribbon element of \uqz\ \cite{fgst}.

\medskip

In view of these results there is little doubt that the quantum group \uqz\ is 
intimately related with the logarithmic minimal models $\mathcal L_{1,p}$. Let us 
therefore briefly summarize a few main features of \uqz. As an associative algebra, 
\uqz\ is freely generated by three elements $\{E,F,H\}$ modulo the relations
  \be \bearl
  E^p = 0 = F^p \,, \qquad K^{2p} = \one \,, \qquad
  K\,E\,K^{-1} = q^2\,E \,, \qquad K\,F\,K^{-1} = q^{-2}\,F
  \nxm8
  {\rm and} \qquad
  E\,F - F\,E = \dsty\frac{K-K^{-1}}{q-q^{-1}} \,.
  \eear \ee
The coproduct $\Delta$ and counit $\eps$ act on the generators as
  \be \bearl
  \Delta(E) = \one \oti E + E \oti K \,, \qquad
  \Delta(F) = K^{-1} \oti F + F \,, \qquad \Delta(K) = K \oti K \qquad{\rm and}
  \nxm6
  \eps(E) = 0 = \eps(F) \,, \qquad \eps(K) = 1 \,,
  \eear \ee
and this extends uniquely to all of \uqz\ by requiring that it endows \uqz\ 
with the structure of a bialgebra. This bialgebra structure becomes a Hopf algebra 
by defining the antipode $\apo$ to act as
  \be
  \apo(E) = -\,E\,K^{-1} \,, \qquad \apo(F) = -\,K\,F \qquand \apo(K) = K^{-1} \,.
  \ee

\medskip

The structure of \uqz\Mod\ as abelian category, which appears in the equivalence 
\erf{RepVep=Repuqz}, has been established with the help of the reduced form 
(basic algebra) of \uqz\ as an algebra. This reduced form has the structure
  \be
  \mathrm{Red}(\uqz) \,=\, \complex \,\oplus\, \complex \,\oplus \Big[
  \mathscr P\big( {\raisebox{.2mm}{\small$\bullet$}\hspace*{1.2pt}
  \raisebox{-.87mm}{$\stackrel{\dsty \,\Longrightarrow}\Longleftarrow$}\,
  \raisebox{.2mm}{\small$\bullet$} } \big) /_{\!\dsty \sim}^{} \Big]^{\!\oplus(p{-}1)}_{} ,
  \labl{basic.alg}
where the algebra in the square brackets, which is 8-dimensional, is the path algebra
of the indicated quiver modulo certain relations \cite{sute2,xiao6,arik}. This direct
sum decomposition shows in particular that the $2p$ isomorphism classes of simple 
\uqz-modules can be partitioned into $p\,{-}1$ pairs $\I_a\eq\{i_a^+,i_a^-\}$ 
precisely as in \erf{Ia} (together with two isolated classes) such 
that non-trivial extensions only exist between the two members of each pair $\I_a$.

There is also an analogue of the pseudo-characters $\pchi$ of \Vep: for each of the
$p\,{-}1$ decomposable projective modules $\mathrm P_a \,{:=}\, \bigoplus_{j\in\I_a}\! 
P_j$ one independent pseudo-character is obtained \cite{fgst4,semi9,gaTi}
by inserting a specific linear endomorphism of $\mathrm P_a$ into the trace.
Such linear maps can be non-zero because of the isomorphisms
  \be
  \soc(P_i) \,\cong\, S_i \,\cong\, \mathrm{Top}(P_i)
  \ee
for all $i\iN\I$. Note, however, that no non-zero map of this form is a morphism
of \uqz-modules, and it is in fact not yet known how to obtain such linear maps
in a systematic manner from Hopf algebra theory and the R-matrix.

\medskip

On the other hand, for $p\,{>}\,2$ the equivalence \erf{RepVep=Repuqz} does not 
extend to an equivalence between $\Rep(\Vep)$ and $\uqz\Mod$ as \emph{monoidal} 
categories. Indeed, whereas the monoidal category $\Rep(\Vep)$ can be endowed with a 
braiding, the category \uqz\Mod\ cannot, as \uqz\ is not quasitriangular.
\uqz\ is, however, `almost' quasitriangular, in the following sense: adjoining
a square root of the generator $K$ to \uqz\ one obtains a Hopf algebra $\widetilde U$ 
which contains \uqz\ as a Hopf subalgebra and which \emph{is} quasitriangular, with an 
explicitly known R-matrix $R \iN \widetilde U\oti\widetilde U$; and the corresponding 
monodromy matrix $Q \eq R_{21} \,{\cdot}\, R$ is not only an element of 
$\widetilde U\oti\widetilde U$, but even of $\uqz\oti\uqz$. This is also reflected in 
the structure of tensor products of indecomposable \uqz-modules: most of them satisfy 
$V\oti V'\,{\cong}\,V'\oti V$, while those which do not obey this relation involve 
pairs $U,U'$ of indecomposable \uqz-modules such that $U\oti V \,{\cong}\, V\oti U'$ and
$U'\oti V \,{\cong}\, V\oti U$ for any of the former modules $V$ and such that $U\,{\oplus}\,U'$, but 
not $U$ or $U'$ individually, lifts to an $\widetilde U$-module \cite{sute2,koSai}.


\subsection{Factorizable ribbon Hopf algebras}\label{sec.fac.rib}

It is tempting to expect that there are further interesting CFT models whose category 
\C\ of chiral data is equivalent, as an abelian category, to $H\Mod$ for some
non-semisimple \findim\ Hopf algebra $H$. For instance, in \cite{fgst4} candidates 
for such Hopf algebras are constructed for the case of the logarithmic minimal 
models $\mathcal L_{p,p'}$, with $p,p'\iN\zet_{\ge2}$ coprime and $p\,{<}\,p'$.
Just like for the subclass of $\mathcal L_{1,p}$ models such an equivalence will
usually not extend to an equivalence of monoidal categories. In particular,
for any \findim\ Hopf (or, more generally, quasi-Hopf) algebra $H$, the category
$H\Mod$ is a finite tensor category \cite{etos}, whereas the results of 
\cite{garW,rasm21,woodS} indicate that already for $\mathcal L_{2,3}$ the category \C\ 
is no longer a finite tensor category, as e.g.\ the tensor product of \C\ is not exact.  

Still, the structures observed in logarithmic CFT models and in suitable classes
of Hopf algebras are sufficiently close to vindicate a deeper study of such
Hopf algebras. In this regard it is gratifying that a Verlinde-like formula
has been established \cite{coWe6} for a particular subalgebra, namely the 
Higman ideal, of any \findim\ factorizable ribbon Hopf algebra over an
algebraically closed field \ko\ of characteristic zero.

\medskip

We will describe this Verlinde-like formula in section \ref{ssec.cowe}. To appreciate
its status, some information about the following aspects of \findim\ factorizable 
ribbon Hopf algebras $H \eq (H,m,\eta,\Delta,\eps,\apo)$ will be needed:

\Itemize

\itx
Chains of subalgebras in the center of $H$ and in the space of central forms 
in $\Hs\eq \Homk(H,\ko)$; 

\itx
the notions of quasitriangular, ribbon, and factorizable Hopf algebras;

\itx
the Drinfeld and Frobenius maps between $H$ and $\Hs$.

\end{itemize}

We first recall that the \emph{character} $\chii_M$ of a $H$-module $M\eq(M,\rho)$ is 
the map that assigns to $x\iN H$ the number $\tilde d_M \cir (\rho \oti \id_{M^*}) 
\cir (x \oti b_M)$. In graphical description, the character $\chii_M$ is given by
  \Eqpic{def_char} {343} {18} { \put(0,-6){  
  \put(51,36)    {$ \chii_M ~=~
                 \tilde d_M \cir (\rho \oti \id_{M^\vee_{}}) \cir (\idH\oti b_M) ~= $}
  \put(255,3){  {\Includepichopf{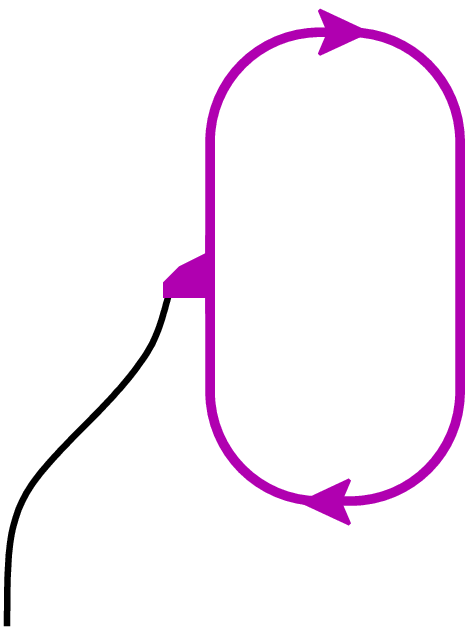}}
  \put(-3.1,-7) {$\sse H $}
  \put(15.9,60) {$\sse M $}
  \put(24.4,37.8) {$\sse \rho $}
  } } }
(The graphical description used here and below has the advantage that most results
translate directly from \Vectk\ to more general 
ribbon categories, and thus to the setting of section \ref{sec.coend}.)

The relevant chain of ideals in the center $Z(H)$ is
  \be
  Z_0(H) \,\subseteq\, \Hig(H) \,\subseteq\, \Rey(H) \,\subseteq\, Z(H) \,,
  \labl{chainH}
which features the \emph{Reynolds ideal} $\Rey(H) \eq \soc(H) \,{\cap}\, Z(H)$ of $H$,
the \emph{Higman ideal} (or projective center), i.e.\ the image of the trace map 
$\tau\colon H\To H$ (with respect to some pair of dual bases of $H$), and the span 
$Z_0(H)$ of those central primitive idempotents $e$ 
for which $He$ is a simple left $H$-module. For the space
  \be
  C(H) := \{ x \iN \Hs \,|\, x \cir m \eq x \cir m \cir c_{H,H} \}
  \ee
of \emph{central forms} (or class functions, or symmetric linear functions)
there is a similar chain of subspaces
  \be
  C_0(H) \,\subseteq\, I(H) \,\subseteq\, R(H) \,\subseteq\, C(H)
  \labl{chainHs}
with $R(H)$ the span of characters of all $H$-modules, $I(H)$ the span of 
characters of all projective $H$-modules, and $C_0(H)$ the span of characters 
of all simple projective $H$-modules. 
\erf{chainHs} is obtained from the chain \erf{chainH} of ideals in $H$ via
a symmetrizing form, and $H$ is semisimple iff any of the inclusions in these chains 
is an equality \cite[Prop.\,2.1\,\&\,Cor.\,2.3]{coWe6}. (In the example of $H\eq\uqz$,
the dimensions of the respective spaces are $2 < p\,{+}\,1 < 2p < 3p\,{-}\,1$.)

\medskip

A \emph{quasitriangular} Hopf algebra $H$ is a Hopf algebra with an
\emph{$R$-matrix}, i.e.\ with an invertible element $R\iN H\oti H$ satisfying
$\Delta^{\!\rm op} \eq \mathrm{ad}_R \cir \Delta$ and
  \be
  (\Delta \oti \idH)(R) = R_{13}\,R_{23} \qquand
  (\idH \oti \Delta)(R) = R_{13}\,R_{12} \,.
  \ee
These properties in turn imply
  \be
  R_{12}\,R_{13}\,R_{23} = R_{23}\,R_{13}\,R_{12} \qquand
  (\eps\oti\idH) \cir R = \eta = (\idH\oti\eps) \cir R
  \ee
as well as, by the invertibility of the antipode,
  \be
  (\apo\oti\idH) \cir R = R^{-1} = (\idH\oti\apo^{-1}) \cir R \qquand
  (\apo\oti\apo) \cir R = R \,.
  \ee
Here we made the canonical identification $H \,{\equiv}\, \Homk(\ko,H)$ 
analogously as $\Hs \,{=} \, \Homk(H,\ko)$, so that in
particular $R\iN\Hom(\ko,H\oti H)$. We describe $R$ pictorially as
  \begin{eqnarray}
  \eqpic{def_R}{100}{18} { \put(0,-24){
  \put(-1,26)    {$ R ~= $}
  \put(37,8)  {\Includepichopfsm{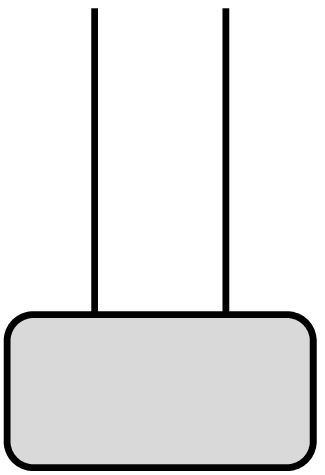}}
  \put(74,26)    {$ \in H\oti H\,. $}
  } }
  \label{def_R} \\[-.2em]~ \nonumber
  \end{eqnarray}
{}From $R$ one forms the \emph{monodromy matrix}
  \be
  Q := R_{21} \,{\cdot}\, R \equiv (m\oti m)\cir (\idH\oti c_{H,H}^{}\oti\idH)
  \cir [ (c_{H,H}^{}\cir R) \oti R] \,\in H\oti H \,.
  \labl{def.Q}

A \emph{ribbon} Hopf algebra $H$ is a quasitriangular Hopf algebra with a
\emph{ribbon element}, i.e.\ a central element $v\iN Z(H)$ such that $\apo \cir v \eq v$ 
and $\eps \cir v \eq 1$ as well as $\Delta \cir v \eq (v\oti v) \,{\cdot}\, Q^{-1}$.
The defining properties of the ribbon element imply that $v^2 \eq m\cir(u\oti 
(\apo\cir u))$ with $u\iN H$ the \emph{canonical element} $u \,{:=}\, m \cir 
(\apo\oti\idH) \cir R_{21}$. The product $b \,{:=}\, m \cir (v^{-1}\oti u)$ is called 
the \emph{balancing element} of the ribbon Hopf algebra $H$; $b$ is group-like and 
satisfies $\apo^2 \eq \mathrm{ad}_b$. A ribbon Hopf algebra is called \emph{factorizable} 
iff the \emph{Drinfeld map}, i.e.\ the morphism $\drin$ from \Hs\ to $H$ defined as
  \Eqpic{def_fQ_p} {290} {52} {
  \put(-3,52) {$\drin ~:=~ (d_H\oti \idH) \circ (\idHv\oti Q) ~\equiv $}
  \put(188,2) { {\Includepichopf{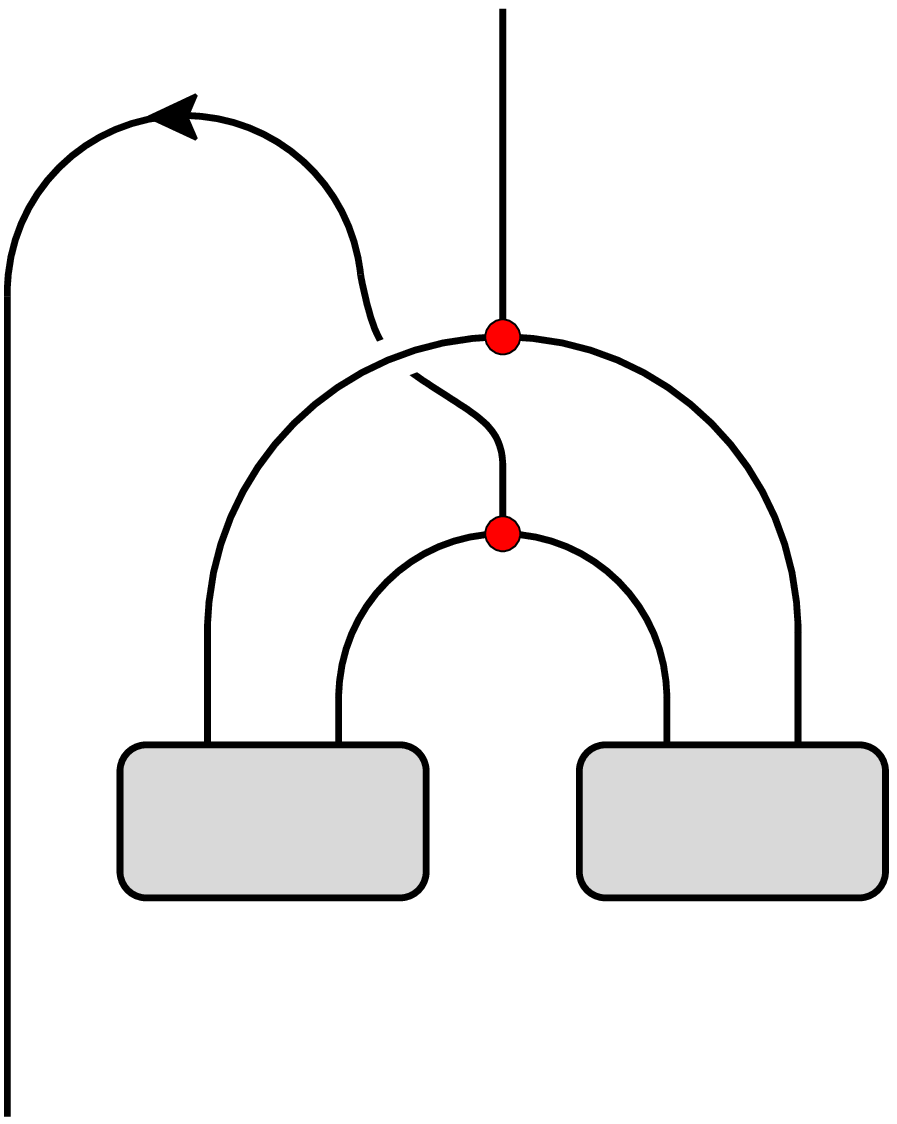}}
  \put(-3.2,-7) {$\sse \Hs $}
  \put(23.9,29.5) {$ R $}
  \put(52.1,125)  {$\sse H $}
  \put(75.1,29.5) {$ R $}
  } }
is invertible (as a linear map) or, equivalently \cite{fgst}, iff the monodromy matrix
\erf{def.Q} can be written as $Q \eq \sum_\ell h_\ell \oti k_\ell$
with $\{h_\ell\}$ and $\{k_\ell\}$ two bases of $H$.

\medskip

A left integral of $H$ is a morphism $ \Lambda \iN \Hom(\one,H) $ such that
$m \cir (\idA \oti \Lambda) \eq \Lambda \cir \eps$, and a right integral of $\Hs$
(or right cointegral of $H$) is a morphism $ \lambda \iN \Hom(H,\one) $ such that
$ (\lambda \oti \idH) \cir \Delta \eq \eta \cir \lambda $. Pictorially,
  \Eqpic{def_l_int} {350} {17} { \put(0,-7){
  \put(58,29)    {$ = $}
  \put(0,0) { {\Includepichopf{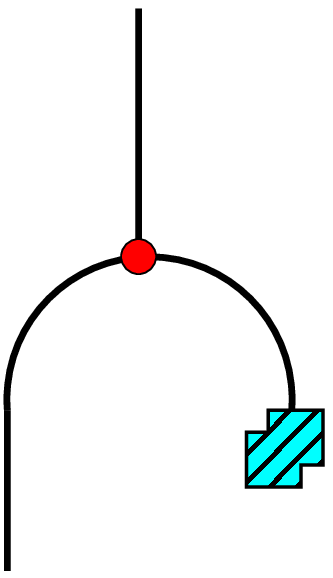}}
  \put(37.5,12)  {$\sse\Lambda$}
 }
  \put(84,0) { {\Includepichopf{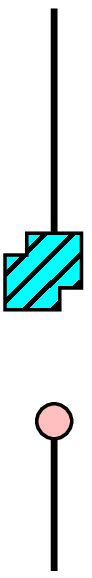}}
  \put(11.4,32)  {$\sse\Lambda$}
  \put(9.7,15)   {$\sse\eps$}
 }
 \put(158,28) {and}
 \put(230,0) { 
  \put(57,29)    {$ = $}
  \put(0,0) { {\Includepichopf{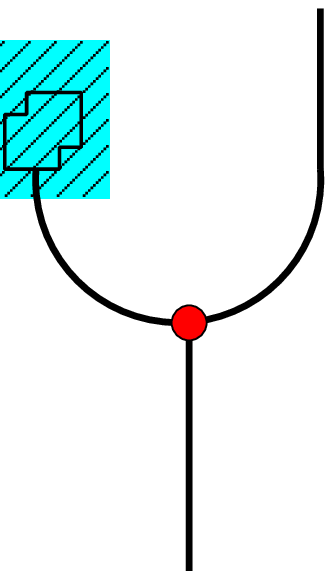}}
  \put(-6.4,46)  {$\sse\lambda$}
 }
  \put(84,0) { {\Includepichopf{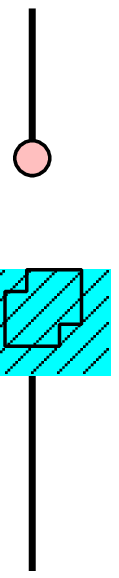}}
  \put(10.9,28)  {$\sse\lambda$}
  \put(7.5,45)   {$\sse\eta$}
 } } } }
respectively.
Given a left integral $\Lambda \iN H$ and a right integral $\lambda \iN \Hs$,
the \emph{Frobenius map} \fmap\ and its inverse $\fmapi\,{\equiv}\,\fmap^{-1}$  
are the morphisms
  \be
  \fmap:~~ H \to \Hs: \quad h \,\mapsto\, \lambda \,{\leftharpoonup}\, \apo(h)  \qquand
  \fmapi:~~ \Hs \to H: \quad p \,\mapsto\, \Lambda \,{\leftharpoonup}\, p \,,
  \ee
i.e.
  \be
  \eqpic{def_fmap} {420} {105} {
  \put(3,40)  {$ \fmap ~= $}
  \put(35,0) {  {\Includepichopf{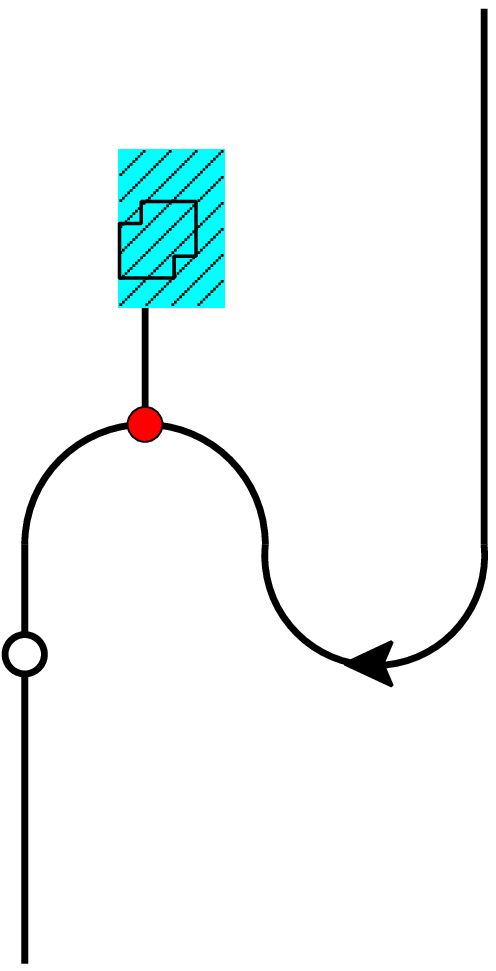}}
  \put(-1.1,-8) {$\sse H $}
  \put(5.9,32.4)  {$\sse \apo $}
  \put(22.7,79) {$\sse \lambda $}
  \put(48.4,109){$\sse \Hs $}
  \put(75,40)  {$ \in\,\Hom(H,\Hs) $}
  } 
 \put(226,0){
  \put(6,40)  {$ \fmapi ~= $}
  \put(47,0) {  {\Includepichopf{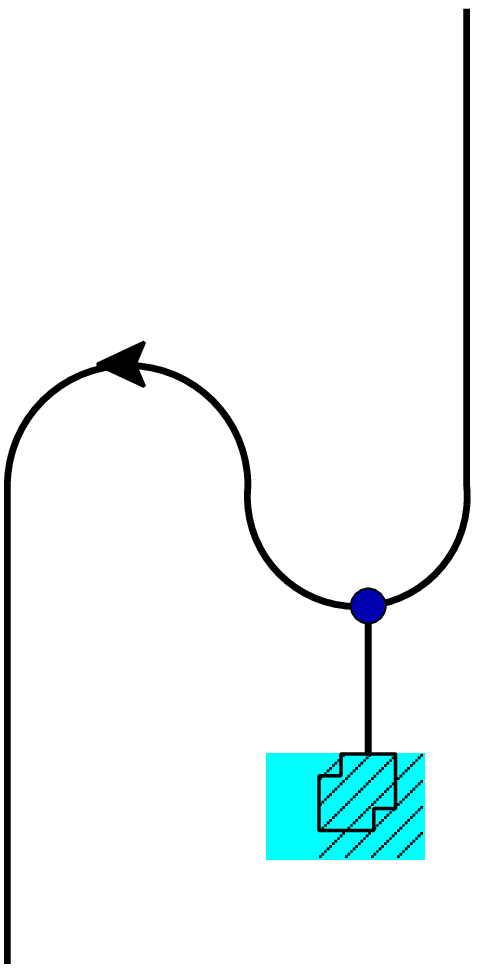}}
  \put(-3.2,-8) {$\sse \Hs $}
  \put(40.4,10.2){$\sse \Lambda $}
  \put(48.1,109){$\sse H $}
  \put(75,40)   {$ \in\,\Hom(\Hs,H) $}
 } } }
  \labl{def_fmap}
(For a graphical proof that these two morphisms are indeed each others' inverses see
appendix \ref{app.fmap}.)
The Frobenius map is a morphism of left $H$-modules (and of right \Hs-modules):
  \Eqpic{fmap_modmorph} {360} {45} {
  \put(16,0)   {\Includepichopf{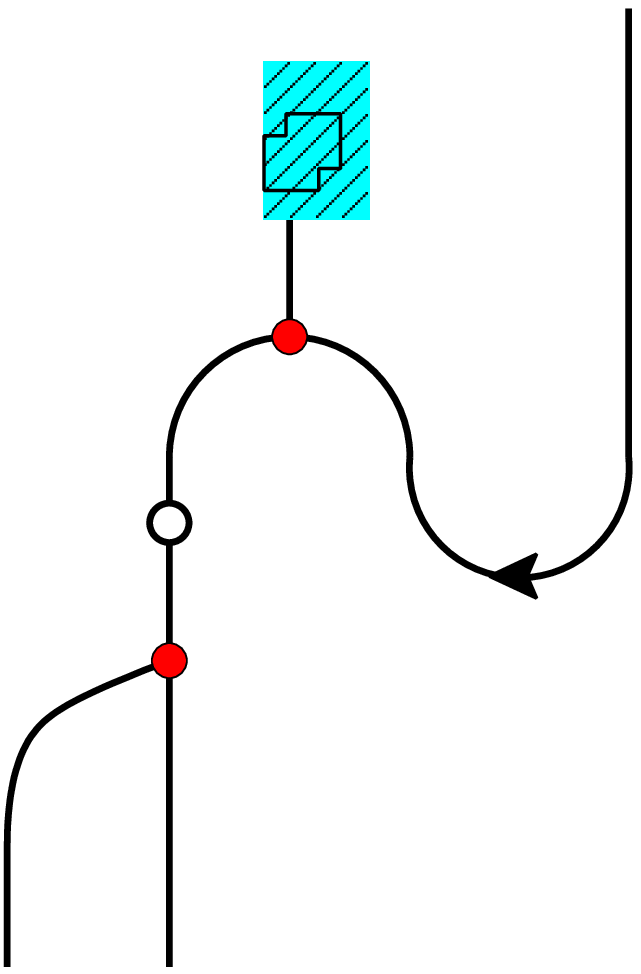}}
  \put(39.4,87){$\sse\lambda$}
  \put(107,49) {$ = $}
  \put(130,0)  {\Includepichopf{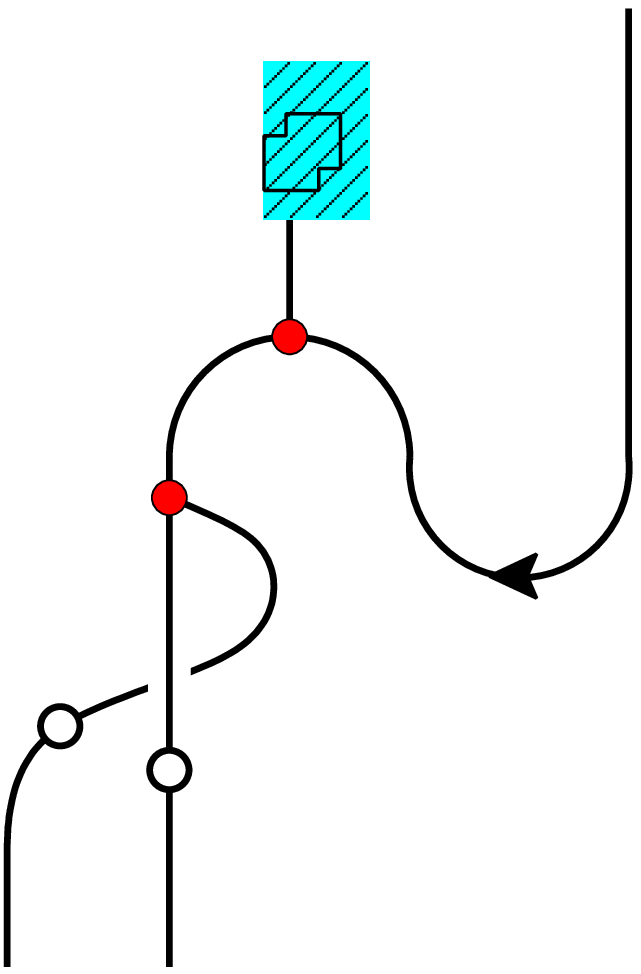}}
  \put(218,49) {$ = $}
  \put(240,0)  {\Includepichopf{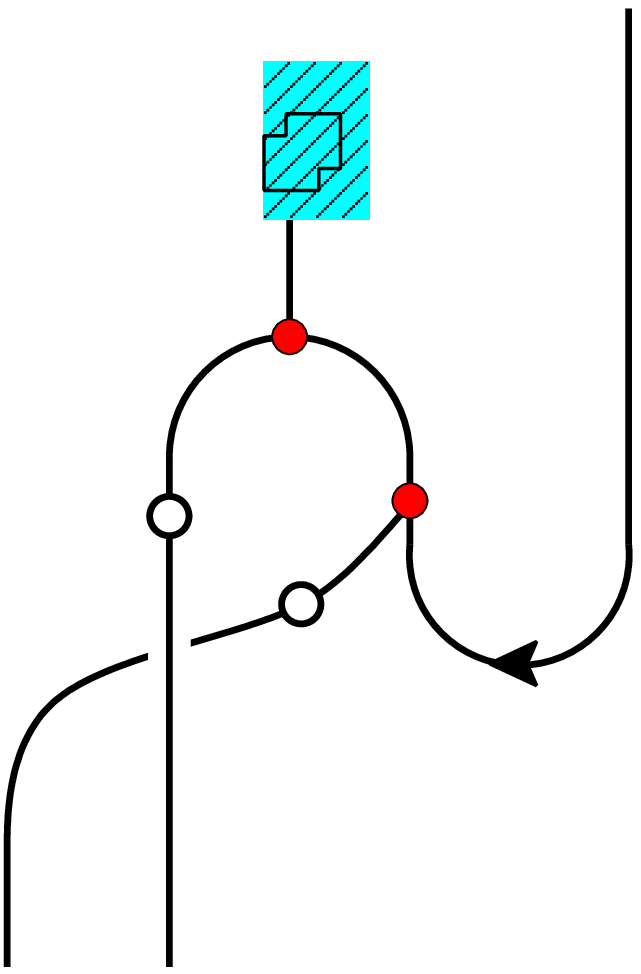}}
  }

For any  \findim\ factorizable ribbon Hopf algebra $H$ over an algebraically closed 
field \ko\ of characteristic zero, the Frobenius map and Drinfeld map furnish an 
algebra isomorphism $Z(H) \,{\cong}\, C(H)$, as well as isomorphisms between the 
respective other members in the two chains \erf{chainH} and \erf{chainHs} \cite{coWe6}.


\subsection{A Verlinde-like formula for the Higman ideal}\label{ssec.cowe}
 
We now describe the result of \cite{coWe6} announced in section \ref{sec.fac.rib}. 
Recall first that for $H$ a \findim\ ribbon Hopf algebra, one defines the 
\emph{$s$-matrix} of $H$ as the square matrix whose entries are obtained by 
composing characters and the Drinfeld map $\drin$ as follows:
  \Eqpic{def_S}  {360} {56} {
  \put(0,62) {$s_{i,j} ~=~ \chii_j \circ (\drin(\chii_i)) ~=$}
  \put(137,0) {\Includepichopfsm{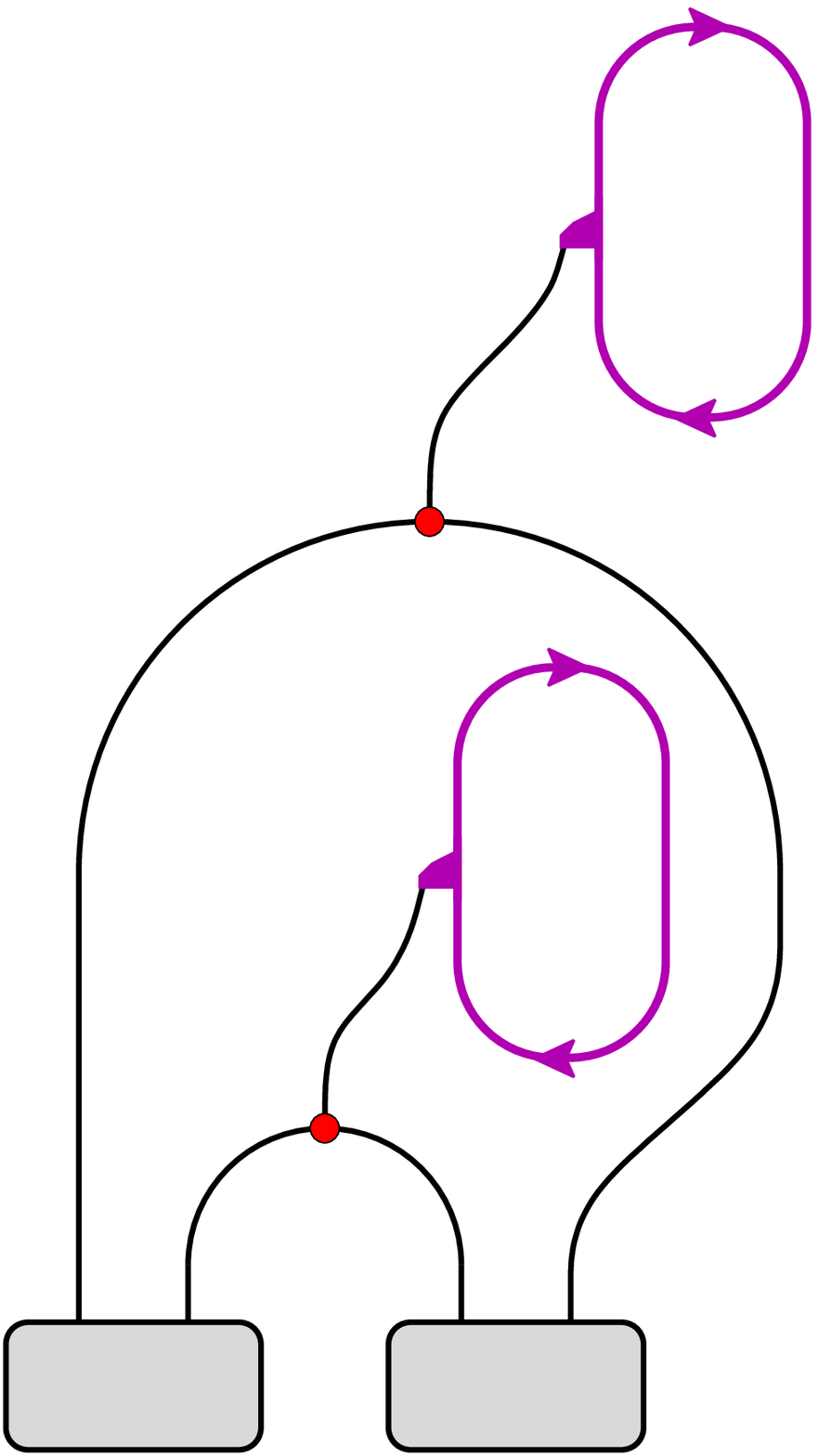}}
  \put(100,0) { \setulen70
  \put(99.2,87)  {$\sse \U_i$}
  \put(117.8,173){$\sse \U_j$}
 }
  \put(236,62)    {$ = $}
  \put(258,22) { {\Includepichopfsm{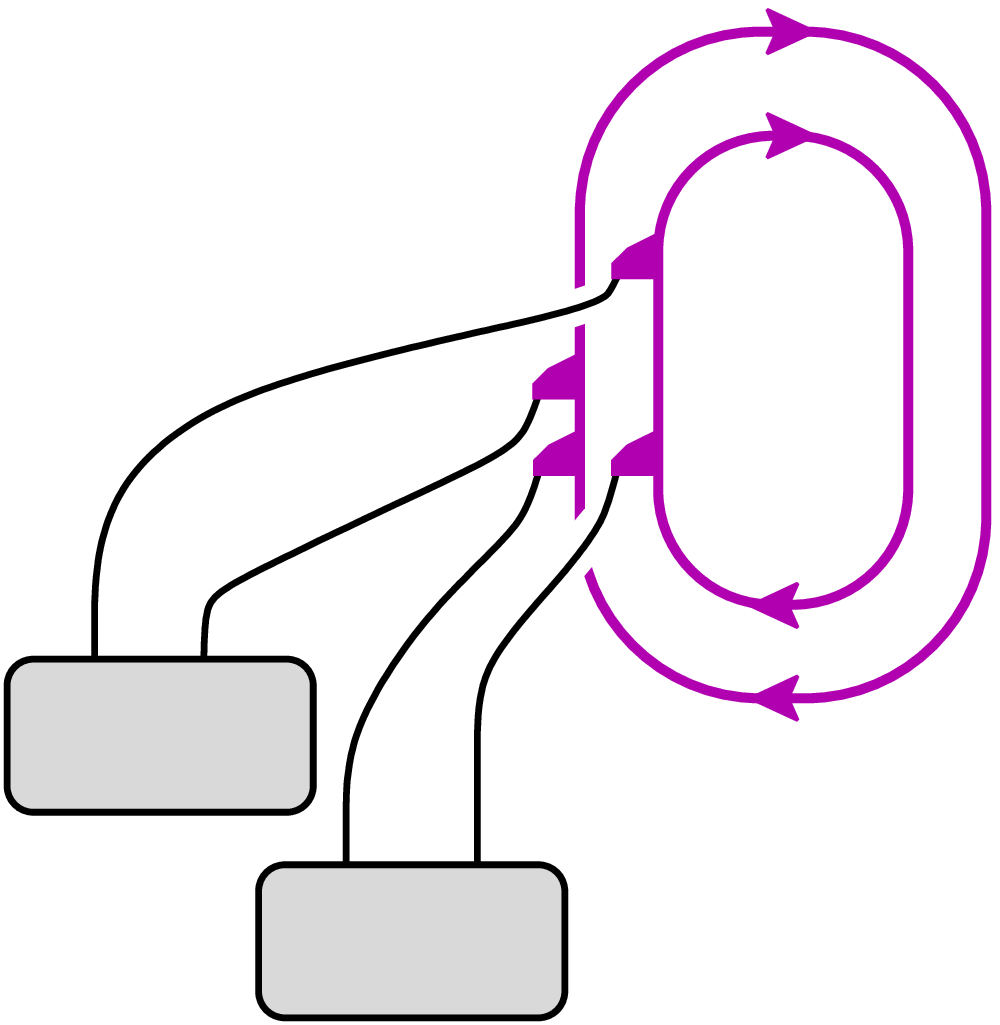}} \setulen70
  \put(54.1,99)  {$\sse \U_i$}
  \put(73.7,70)  {$\sse \U_j$}
 } }
The composition of the flip map with the action $(\rho_{\U_i}{\otimes}\rho_{\U_j})(R)$ 
of $R$ provides a non-trivial braiding on $H\Mod$. The $s$-matrix given by \erf{def_S} 
is then precisely the matrix $\soo$ defined in formula \erf{def.s}.

The result of \cite{coWe6} can be regarded as a specific generalization of the formula 
\erf{furu.diag} (the diagonalization of the fusion rules). It is a statement about the 
Higman ideal $\Hig(H)$ which reduces to \erf{furu.diag} (together with the identification
\erf{verl1}) for $H$ semisimple, i.e.\ for $\Hig(H) \eq H$. To cover the non-semisimple 
case we need two generalizations $\Sb \,{\equiv}\, \check s^{\dsty\circ\!\!\circ}
_{\phantom;}$ and $\Sl \,{\equiv}\, \hat s^{\dsty\circ\!\!\circ}_{\phantom;}$ of 
the $s$-matrix. They are defined by
  \be
  \Sb_{i,j} := \chii_j \circ (\Drinh(\chii_i))  \qquand
  \Sl_{i,j} := (\fhmap(\apo\cir e_j)) \circ (\Drinh(\chii_i))
  \ee
for $i,j\iN\I$, respectively, and involve, besides the primitive idempotents $e_j\iN H$ 
(satisfying $P_j \eq H\,e_j$), the \emph{modified Frobenius map} $\fhmap\colon\, H \To \Hs$ 
and \emph{modified Drin\-feld map} $\Drinh\colon\, \Hs \To H$ which are defined by
  \be
  \fhmap\colon~ h\,\mapsto\, (\lambda\,{\leftharpoonup}\,b) \,{\leftharpoonup}\,
  \apo^{-1}(h) \qquand \Drinh\colon~ p \,\mapsto\, \drin(p\,{\leftharpoonup}\,\uiv) \,,
  \ee
where $b$ is the balancing element of $H$. ($\Drinh$ is an algebra isomorphism 
$C(H) \,{\stackrel\cong\longrightarrow}\, Z(H)$, and the modifications of Drinfeld and 
Frobenius maps cancel, $\Drinh \circ \fhmap \eq \drin \circ \fmap$.) Graphically,
  \Eqpic{def_So_} {400} {87} { 
  \put(13,88)   {$\Sb_{i,j} ~= $}
  \put(44,0) { {\Includepichopf{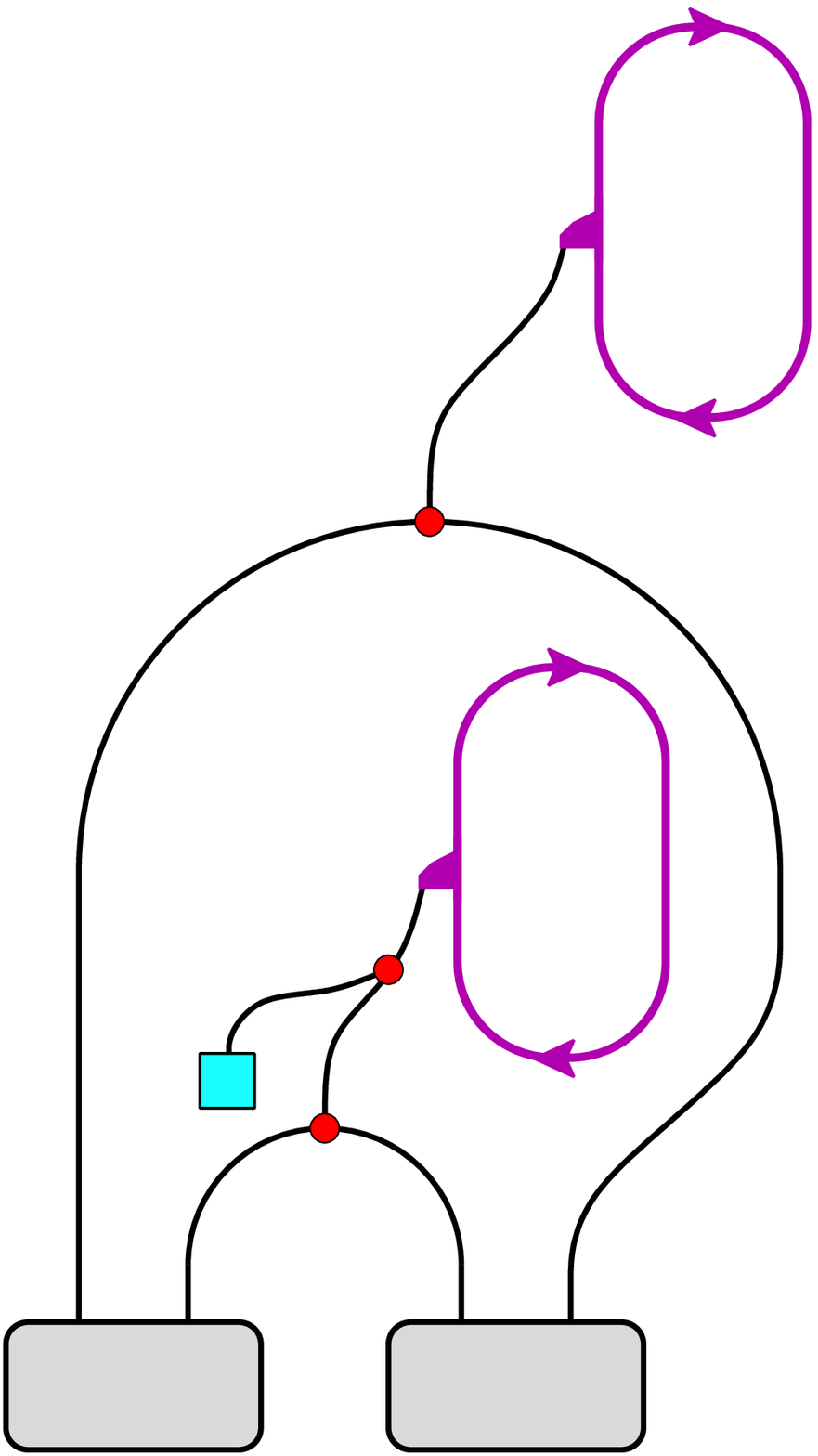}}
  \put(17.4,53)  {$\sse \uiv $}
  \put(51.2,87)  {$\sse \U_i $}
  \put(69.1,174) {$\sse \U_j $}
  \put(12.9,5.6) {$ R $}
  \put(63.7,5.6) {$ R $}
 }
  \put(230,0){
  \put(0,88)   {$\Sl_{i,j} ~=$}
  \put(30,0) { {\Includepichopf{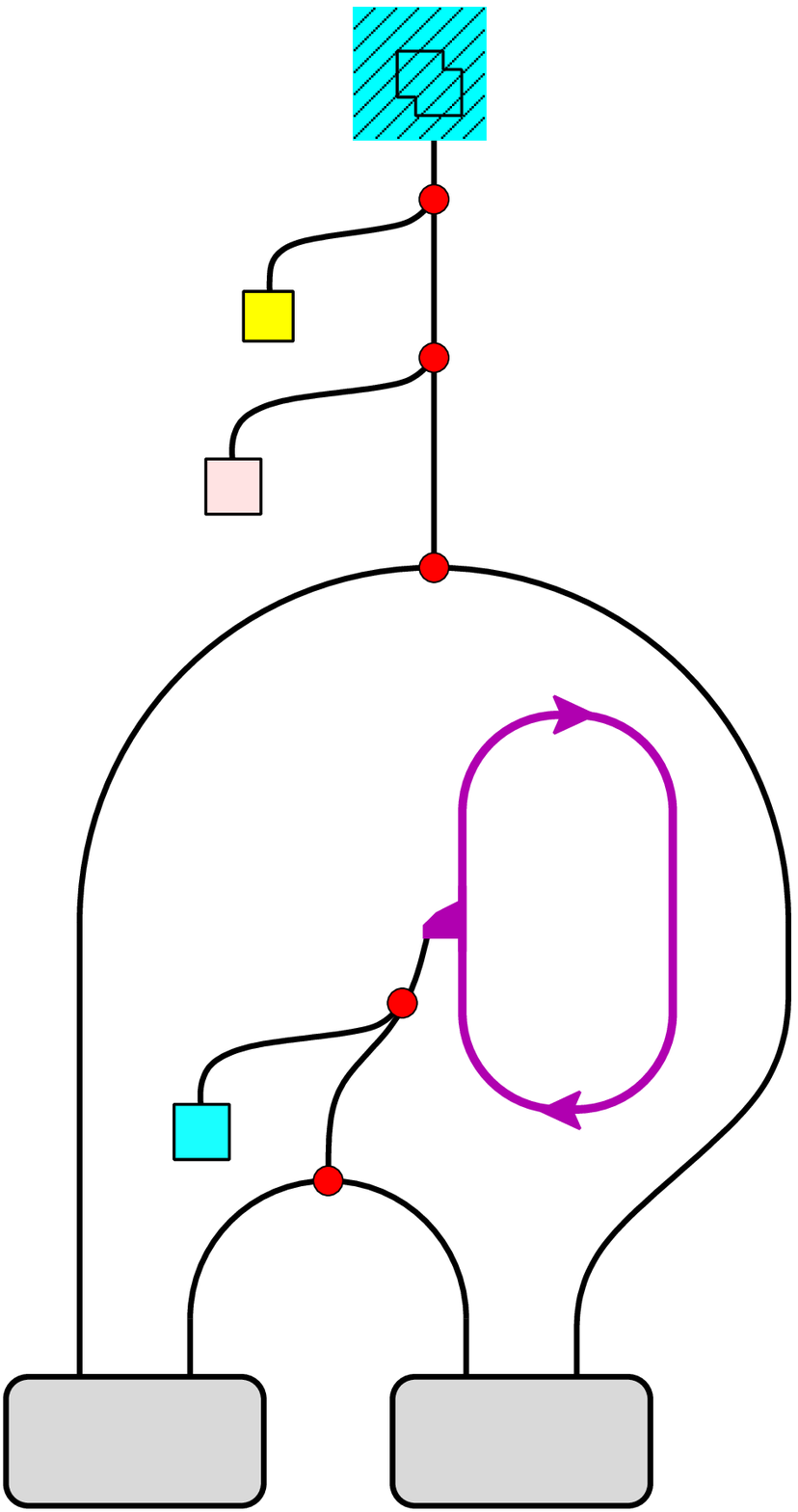}}
  \put(18.3,132) {$\sse e_j $}
  \put(13.3,53.5){$\sse \uiv $}
  \put(26.2,153.7){$\sse b $}
  \put(47.7,179.4){$\sse \lambda $}
  \put(50.8,87)  {$\sse \U_i $}
  \put(12.9,5.6) {$ R $}
  \put(63.7,5.6) {$ R $}
  } } }

\medskip

We now consider the following two bases for the Higman ideal $\Hig(H)$: the images 
$ \{ \Drinh(\chii_{P_j}) \}$ under the modified Drinfeld map of the characters of 
the indecomposable projective $H$-modules $P_j\eq P(\U_j)$, and the images
$\{ \tau(e_j) \}$ under the trace map $\tau$ of the primitive orthogonal idempotents 
$e_j$ that correspond to the $P_j$. It can be shown \cite[Prop.\,3.11]{coWe6} that 
the matrix $\SF$ that transforms the latter basis to the former satisfies
  \be
  \SF = (\mathrm M(C))^{-1}_{}\, \mathrm M(C\,\Sl) \,,
  \labl{SF}
where $C$ is the Cartan matrix (with entries $C_{i,j} \eq \chii_{P_i} \cir e_j$) of 
$H$ and $\mathrm M(A)$ stands for the $\kap\,{\times}\,\kap$ major minor of a matrix 
$A$, with 
  \be
  \kap := \dim_\ko(\Hig(H)) \,.
  \ee

The tensor product of a projective $H$-module with any module is again 
projective. Hence for $i,j,k\in\{1,2,...\,,\kap\}$ one can define nonnegative integers 
$\Nh ijk$ by
  \be
  \chii_{S_i}^{} \, \chii_{P_j}^{} = \sum_{k=1}^m \Nh kji\, \chii_{P_k}^{} \,.
  \labl{Nh}
Consider now the matrices $\NNh i$ that are given by $\NNh i\,{:=}\,(\Nh jki)$.
As shown in \cite[Thm.\,3.14]{coWe6}, the basis transformation matrix $\SF$ 
simultaneously diagonalizes these matrices $\NNh i$,
and the entries of the resulting diagonal matrices are, up to factors of 
dimensions $d_k \eq \dim(\U_k)$, given by the entries of the matrix $\Sb$:
  \be
  \SF^{-1}_{}\,\NNh i\, \SF = \big(\mathrm{diag}(d_j^{-1}\,\Sb_{i,j}
  \big)_{j=1,...,\kap} \,.
  \ee

This is the Verlinde-like relation announced above. Regarding \erf{Nh} as (the
appropriate analogue of) fusion rules for the Higman ideal $\Hig(H)$, it asserts
that the matrix $\SF$ given by \erf{SF} diagonalizes the fusion rules for $\Hig(H)$.


\section{Hopf algebras and coends}\label{sec.coend}

\subsection{Hopf algebras in braided monoidal categories}\label{ssec.hopf.in.cat}

One idea to comprehend CFT beyond the rational case is to generalize the notion of 
modular tensor category, as defined in appendix \ref{app.mtc}, in such a way that
semisimplicity of the category is not required.
As it turns out, such a generalization can be achieved with the help of 
a Hopf algebra that, for a large class of rigid braided monoidal categories, can be 
defined as the coend of a suitable functor. We will thus need some background
information about coends and about Hopf algebras in braided categories.

\medskip

A \emph{bialgebra} in a braided monoidal category \C\ is an object $H$ of \C\
endowed with a product $m$, unit $\eta$, coproduct $\Delta$ and counit $\eps$
such that $(A,m.\eta)$ is an algebra and $(A,\Delta,\eps)$ is a coalgebra (compare
section \ref{sec.frob}), and such that the coproduct and counit are (unital) algebra
morphisms, i.e.
  \Eqpic{p6} {370} {27} {
  \put(0,0) {\scalebox{.38}{\includegraphics{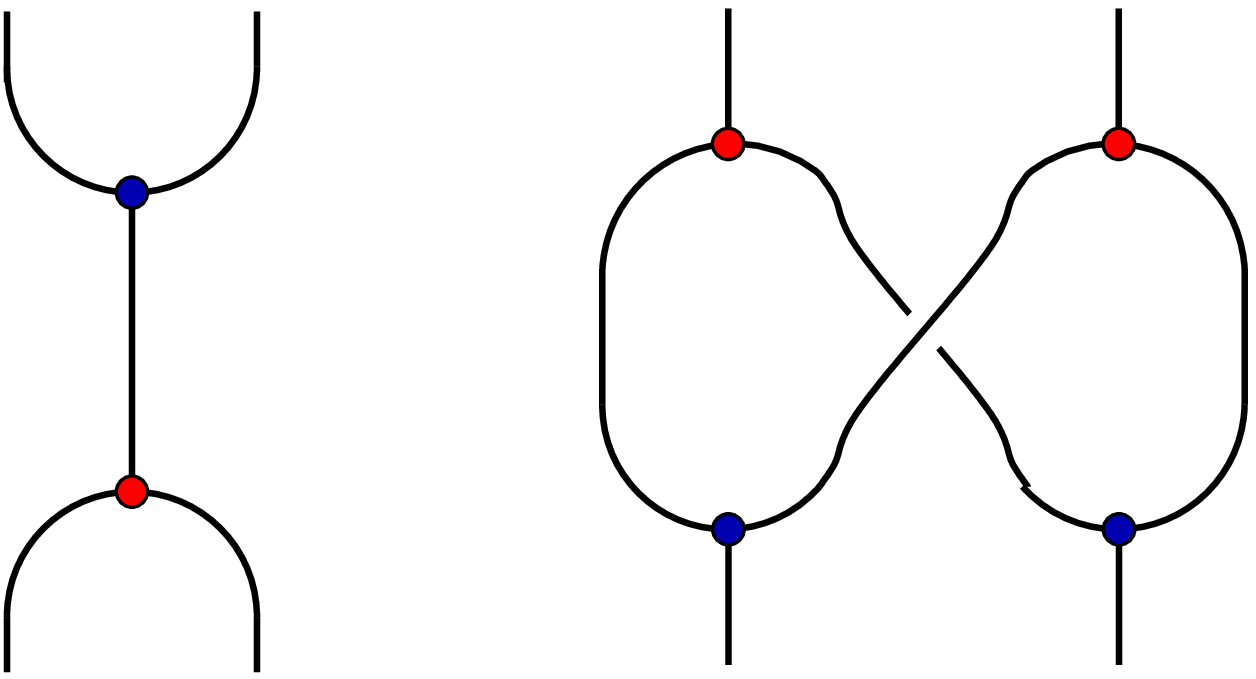}}} \setUlen13
              \put(30,26)  {$ = $}
  \put(142,26) {and}
  \put(190,9) { \scalebox{.41}{\includegraphics{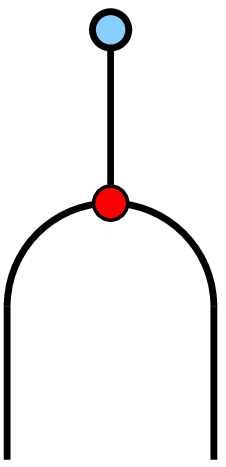}}
             \put(17,20){$ = $}
             \put(38,0) {\scalebox{.41}{\includegraphics{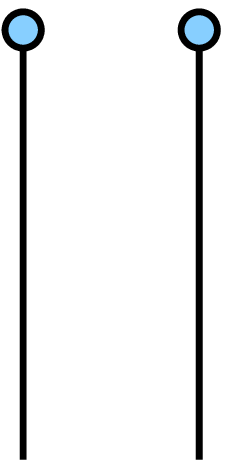}}} 
  } }
A \emph{Hopf algebra} $(H,m,\eta,\Delta,\eps,\apo)$ in \C\ is a bialgebra together
with an antipode, i.e.\ an endomorphism $\apo$ of $H$ satisfying
  \Eqpic{antihom_apo} {150} {34} {
  \put(0,0)    {\Includepichopf{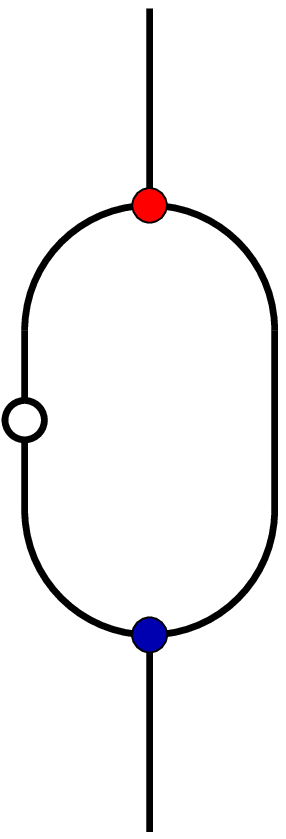}}
  \put(-7.7,42.5){$\sse \apo $}
  \put(54,43)   {$ = $}
  \put(86,0)   {\Includepichopf{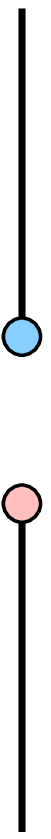}}
  \put(112,43)  {$ = $}
  \put(144,0)  {\Includepichopf{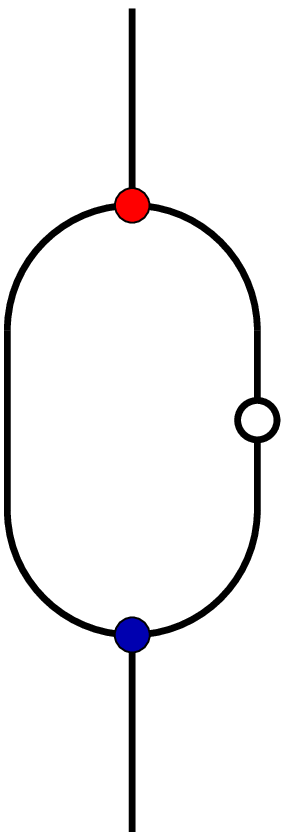}}
  }

Just like in the case of Frobenius algebras, the use of graphical calculus is not 
only convenient for visualizing definitions, but also for giving proofs. 
For instance, it can be checked graphically that the antipode is an algebra-
and coalgebra-antihomomorphism.
As another simple illustration, that the left coadjoint action, i.e.\ 
  \Eqpic{def_lads} {258} {36} {
  \put(53,-6) { {\Includepichopfsm{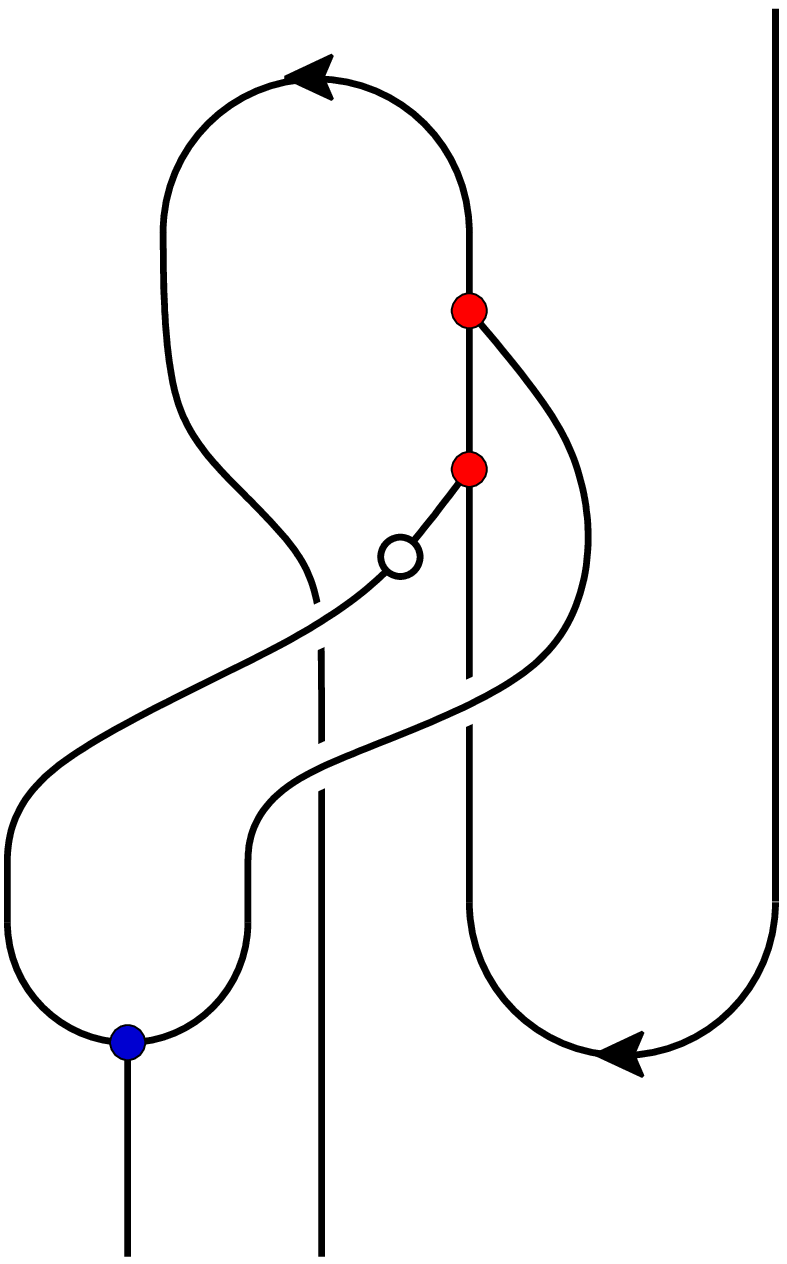}}
  \setulen 70
  \put(111,61)  {$ \in \Hom(H\oti H^\vee,H^\vee) $}
 } }
endows the dual $H^\vee$ with the structure of a left module over $H$ is seen by
the following sequence of equalities:
  \be
\bP(400,171) \setulen70
   \put(-45,25){
  \put(0,0)   { \Includepichopfsm{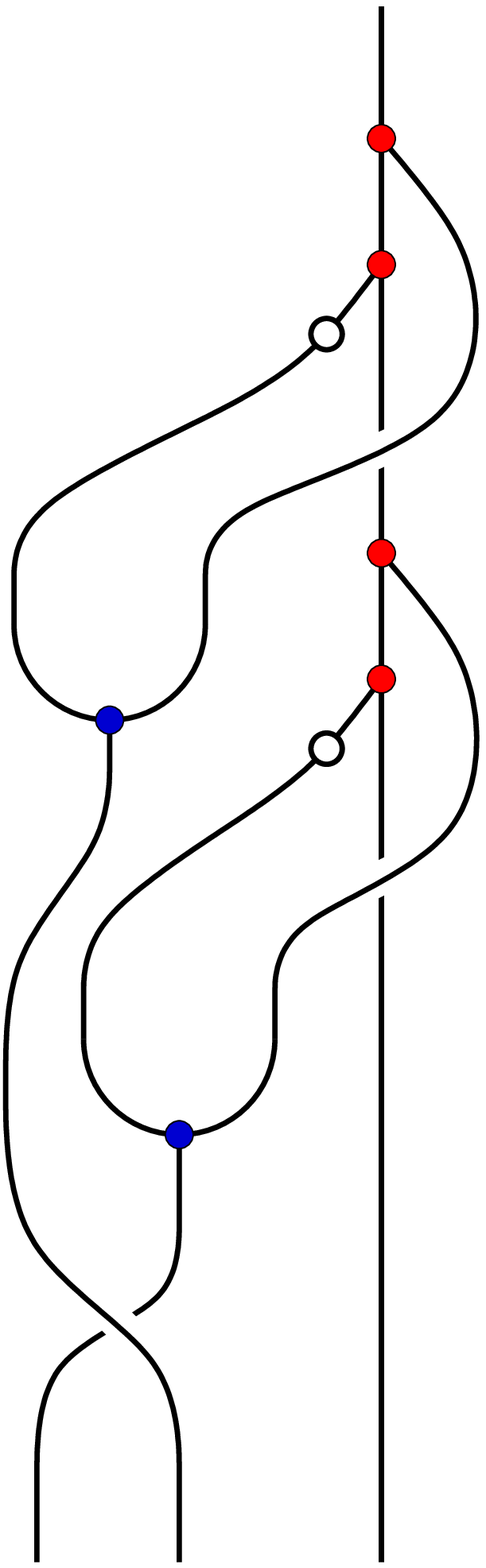}} 
  \put(100,98)  {$ = $}
  \put(131,0) { \Includepichopfsm{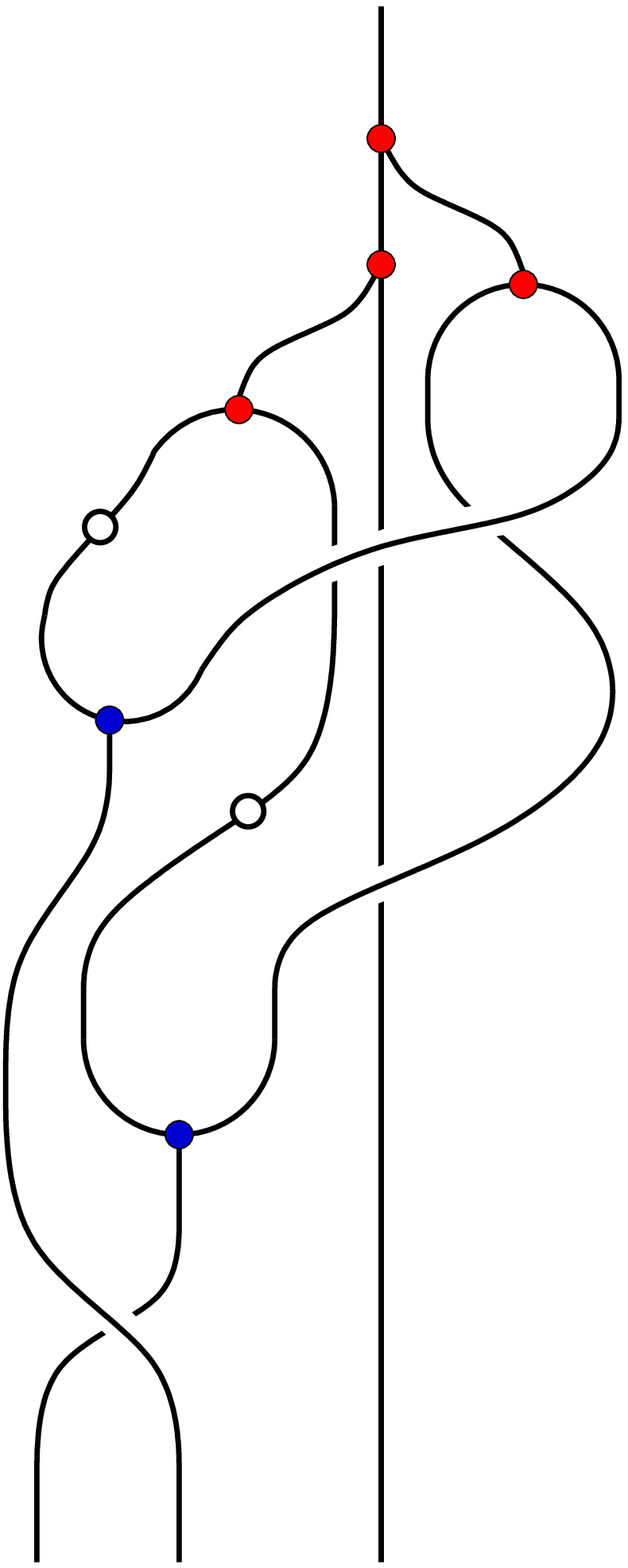}}
  \put(248,98)  {$ = $}
  \put(284,0) { \Includepichopfsm{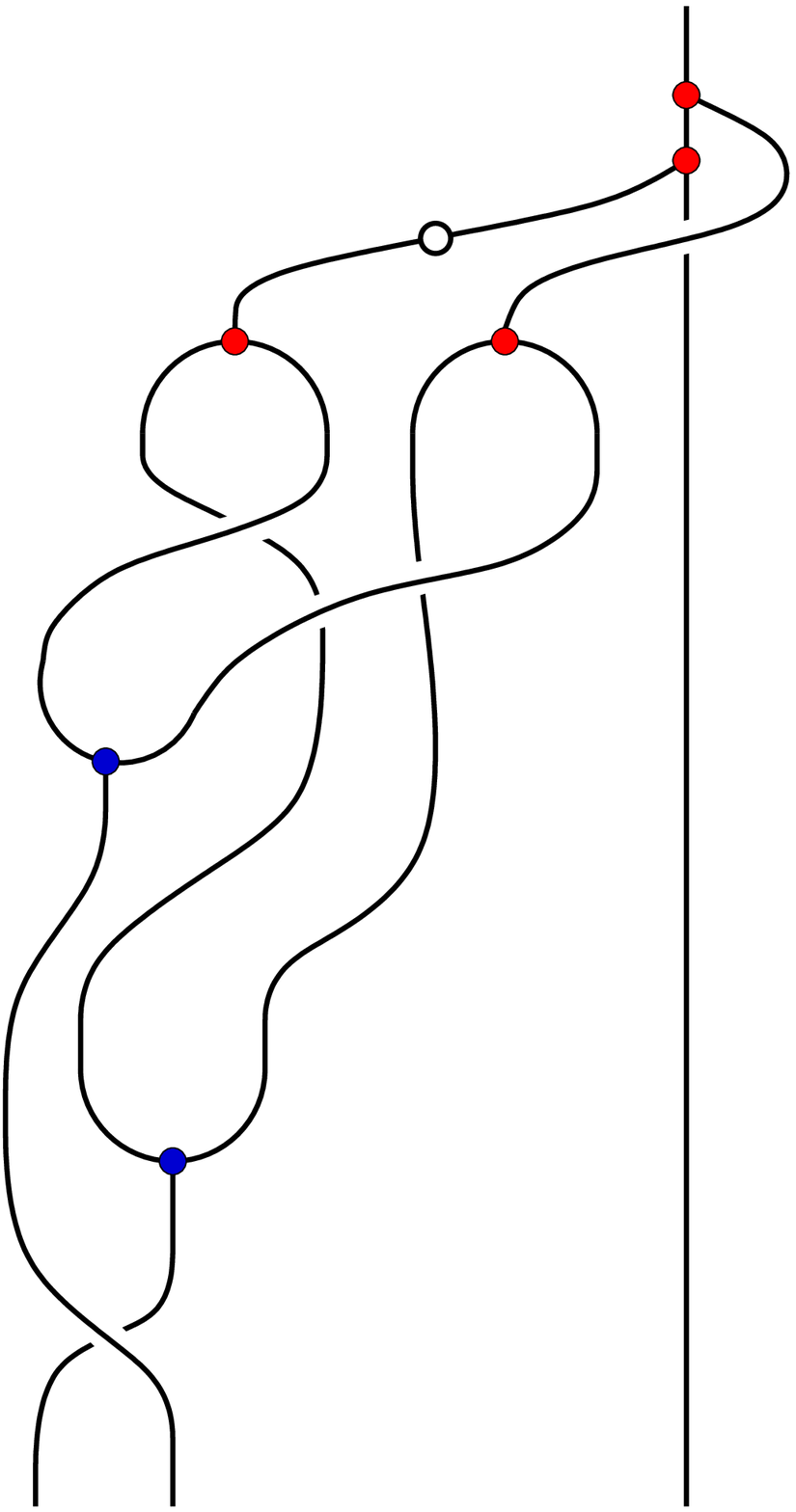}}
  }
  \put(285,25){
  \put(100,98)   {$ = $}
  \put(126,0) { \Includepichopfsm{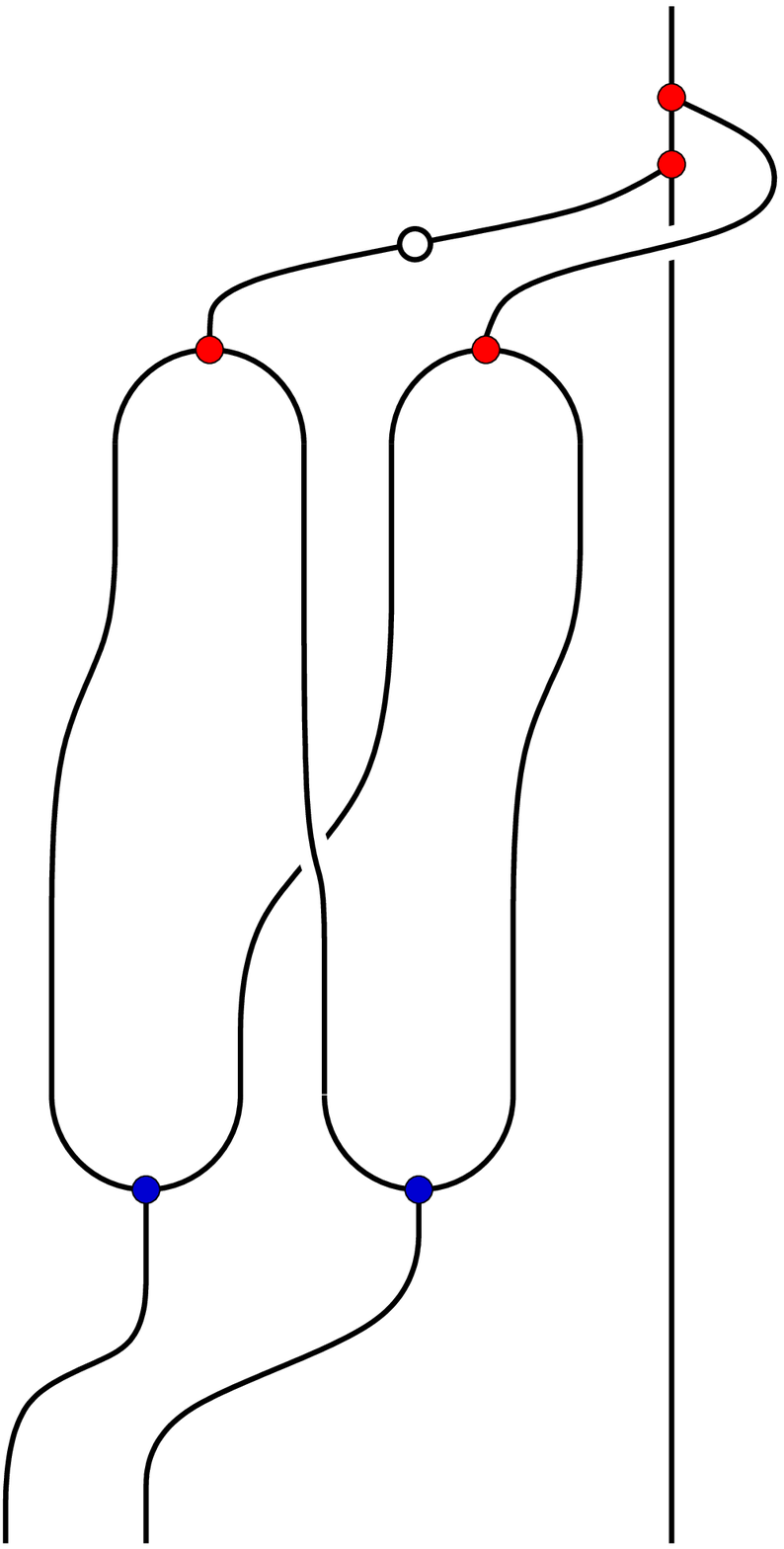}}
  \put(259,98)  {$ = $}
  \put(278,0) { \Includepichopfsm{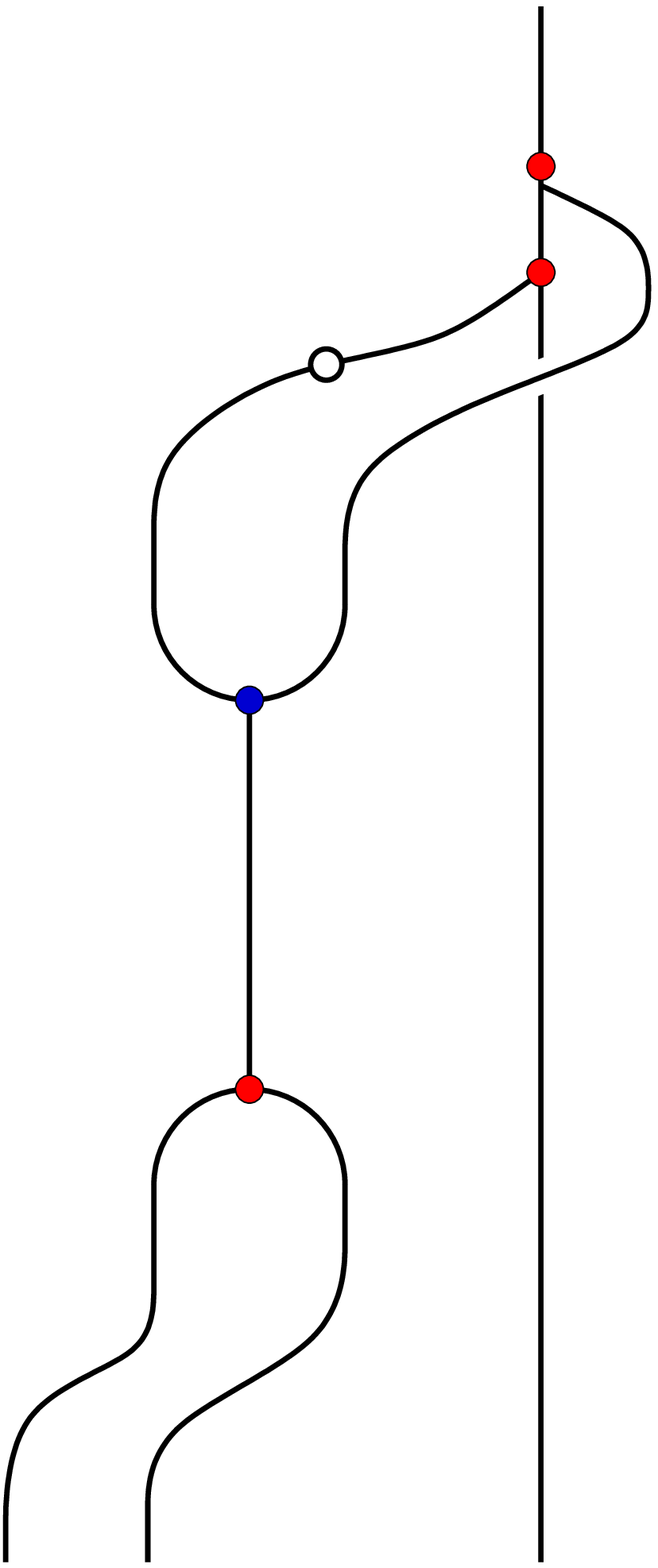}}
  } \eP
  \ee

It is interesting to note that the self-braiding $c_{H,H}$ of a Hopf algebra is 
already determined by the structural morphisms of $H$. Indeed \cite{schau8},
  \Eqpic{cHH} {120} {41} {\setulen70
  \put(0,0)   {\Includepichopfsm{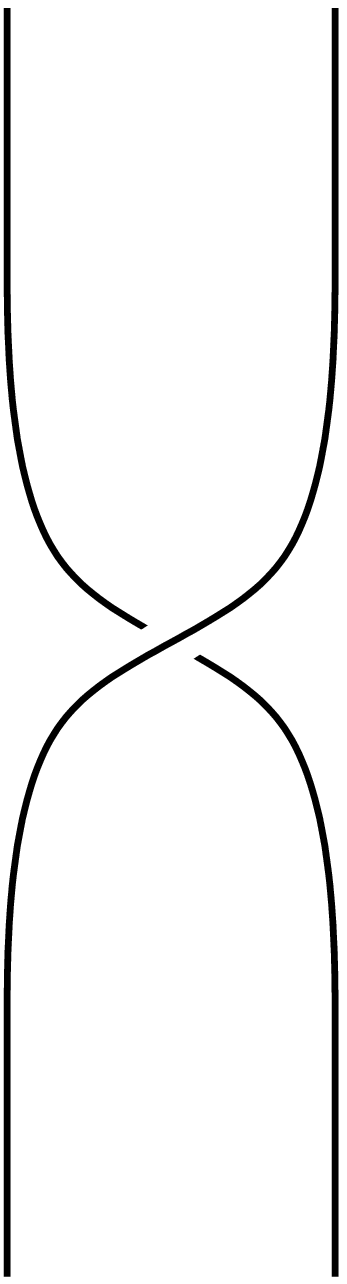}}
  \put(69,65)   {$ = $}
  \put(108,0)  {\Includepichopfsm{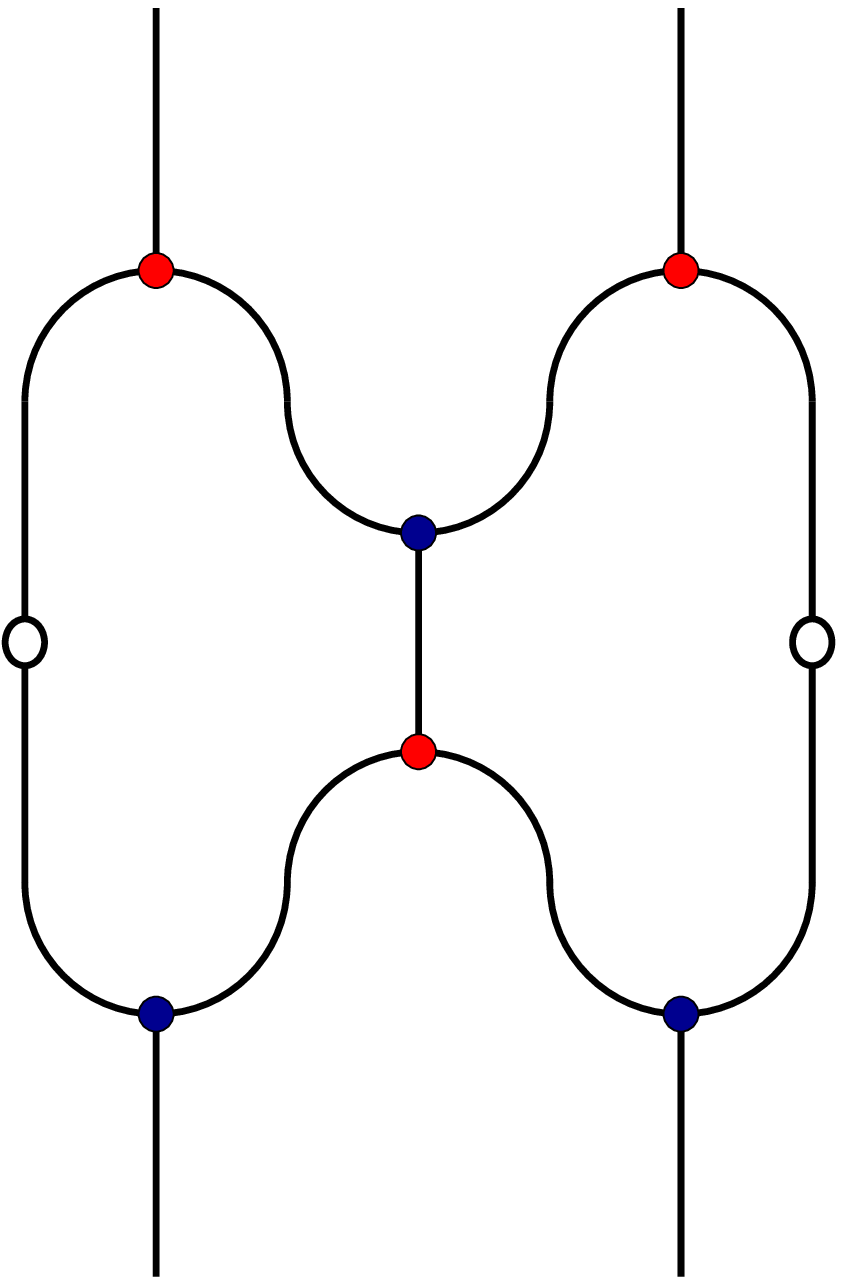}}
  }
In particular if $H$ is commutative or cocommutative, then the self-braiding is 
symmetric, $c_{H,H}^2 \eq \id_{H\otimes H}$; accordingly, in this case in 
pictures like those above the distinction between over-
and underbraiding is immaterial, even if the braiding on \C\ is not symmetric.

Below we will also need the notion of a \emph{Hopf pairing} of a Hopf algebra
$H$; this is  a morphism $\omega\colon\, H \oti H \To \one$ satisfying
  \Eqpic{hopf_pair} {420} {36} {
  \put(0,0)   {\Includepichtft{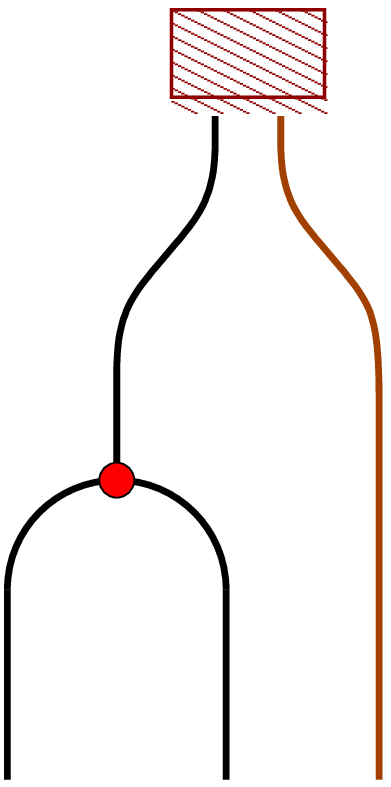}}
  \put(63,38)   {$ = $}
  \put(94,0)  {\Includepichtft{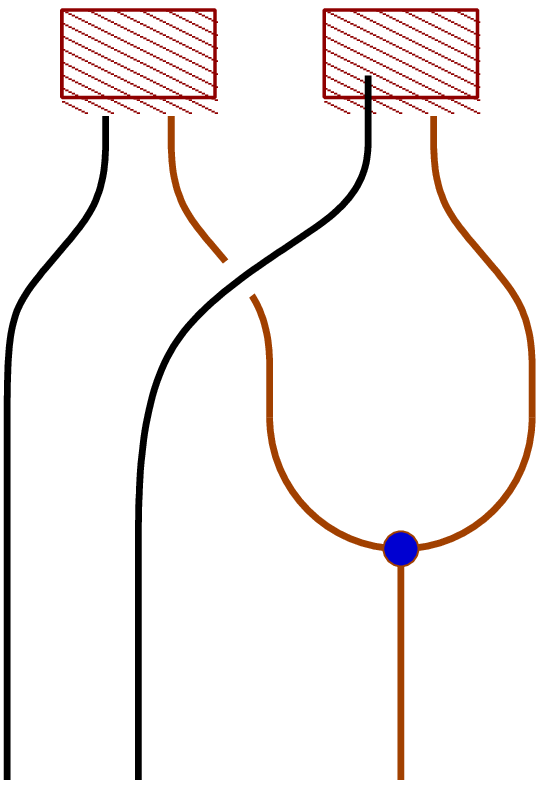}}
  \put(192,38)  {and}
   \put(250,0){
  \put(0,0)   {\Includepichtft{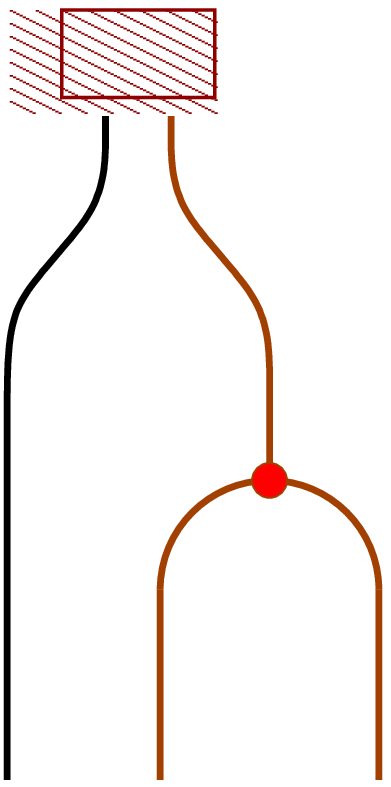}}
  \put(63,38)   {$ = $}
  \put(94,0)  {\Includepichtft{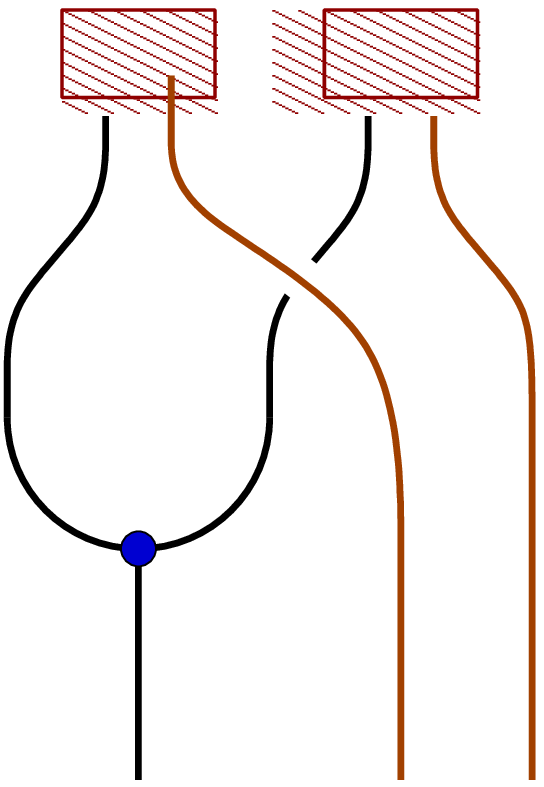}}
  } }
as well as
  \Eqpic{hopf_pair-II} {305} {21} {
  \put(0,0)   {\Includepichtft{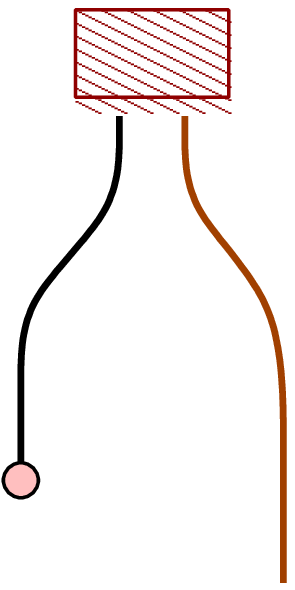}}
  \put(55,28)   {$ = $}
  \put(86,0)  {\Includepichtft{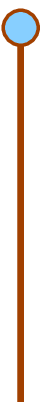}}
  \put(143,28)  {and}
   \put(211,0){
  \put(0,0)   {\Includepichtft{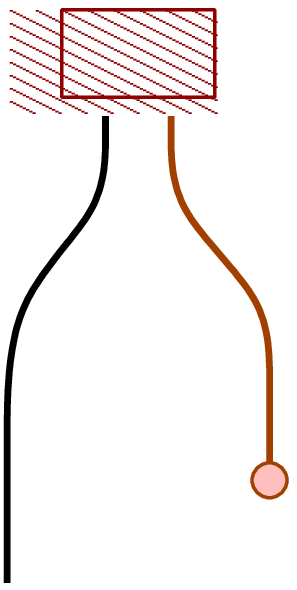}}
  \put(55,28)   {$ = $}
  \put(86,0)  {\Includepichtft{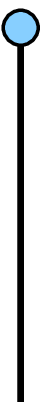}}
  } }
As is easily checked, a non-degenerate Hopf pairing gives an isomorphism
$H\To H^\vee$ of Hopf algebras.


\subsection{Dinatural transformations and coends}

We now summarize some pertinent information about coends.
For \C\ and \D\ categories and $F$ a functor $F\colon\, \CopC\To\D$, a \emph{dinatural 
transformation} $F\,{\Rightarrow}\,B$ from  $F$ to an object $B\iN\D$ is a family of 
morphisms $\varphi \eq \{ \varphi_X\colon F(X,X)\To B \}_{\!X\in\C}^{}$ such that the 
diagram
  \bee4010{
  \xymatrix @R+13pt{
  F(Y,X)\ \ar^{F(\id_Y,f)}[rr]\ar_{F(f,\id_X)}[d]&&\ F(Y,Y)\ar^{\varphi_Y^{}}[d] \\
  F(X,X)\ \ar^{\varphi_X^{}}[rr] && \, B
  } }
commutes for all $X,Y\iN\C$ and all $f\colon X\To Y$. Pictorially:
  \Eqpic{dinat_trafo} {95} {32} {
  \put(0,2)   {\Includepichtft{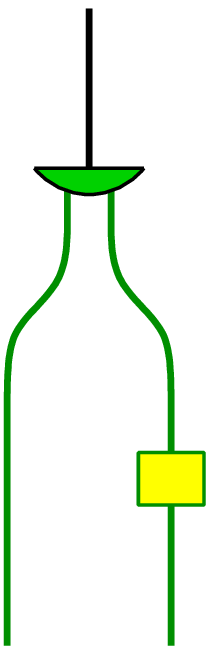}}
  \put(-5,-8)   {$\sse Y^{\!\vee} $}
  \put(6.5,74)  {$\sse B $}
  \put(15,-8)   {$\sse X $}
  \put(17,38)   {$\sse Y $}
  \put(17.5,51) {$\sse \varphi_Y^{} $}
  \put(24,18.5) {$\sse f $}
  \put(52,25)   {$ = $}
  \put(84,2) { {\Includepichtft{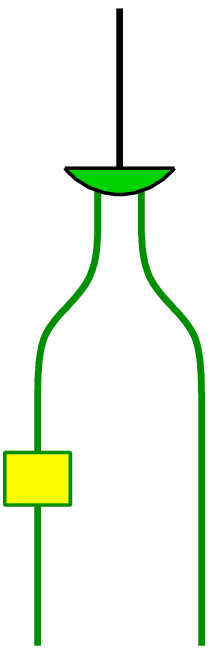}}
  \put(-2,-8)   {$\sse Y^{\!\vee} $}
  \put(9.5,74)  {$\sse B $}
  \put(18.5,51) {$\sse \varphi_X^{} $}
  \put(18,-8)   {$\sse X $}
  \put(-6,35)   {$\sse X^{\!\vee} $}
  \put(-9,16.5) {$\sse f^{\!\vee} $}
  } }
for all $X,Y\iN\C$ and all $f\colon X\To Y$.

A \emph{Coend} $(A,\iota)$ for a functor $F\colon~ \CopC\To\D$ is an initial object in 
the category of dinatural transformations from $F$ to a constant. In other words,
it is a dinatural transformation $(A,\iota)$ with the universal property that 
any dinatural transformation $(B,\varphi)\colon\, F\,{\Rightarrow}\,B$ 
uniquely factorizes, i.e.\ there is a unique morphism $A\To B$ such that the
two triangles in
  \bee6232{
  \xymatrix @R+13pt{
  F(Y,X)\ \ar^{F(\id_Y,f)}[rr]\ar_{F(f,\id_X)}[d] 
  && F(Y,Y)\ar^{\iota_Y^{}}[d]  \ar @/^2pc/^{\varphi_Y^{}}[ddr]& \\
  F(X,X)\ar @/_2pc/_{\varphi_X^{}}[drrr] \ar^{\iota_X^{}}[rr] && \, A \ar@{-->}[dr] & \\
  &&& \, B
  } }
commute for all $X,Y\iN\C$ and all $f\colon X\To Y$.

If the coend exists, it is unique up to unique isomorphism. One denotes it by
  \be
    A = \Coend FX \,,
  \ee
so that in particular $\iota_X\colon\, F(X,X) \,\To \Coend FX$. 
Assuming that arbitrary coproducts exist, an equivalent description of $\Coend FX$ 
(see e.g.\ section V.1 of \cite{MAy4}) is as the coequalizer of the morphisms
  \be
  \xymatrix{
  {\dsty\coprod_{f\colon Y\to Z}} F(Y,Z)\, \ar@<8pt>[r]^s \ar@<2pt>[r]_t
  & {\dsty\coprod_{X\in\,\C} F(X,X)}
  } \labl{coendalizer}
whose restrictions to the `$f$th summand' are $s_f \eq F(f,\id)$ and $t_f \eq F(\id,f)$,
respectively.
An immediate consequence of the universal property of the coend is that to give a 
morphism with domain $\Coend FX$ and codomain $Y$ is equivalent to give a family $\{f_X\}$ 
of morphisms from $F(X,X)$ to $Y$ such that $(Y,f)$ is a dinatural transformation.

Provided that the coends exist, the assignment of the coend of $F$ to a functor $F$ 
is functorial, i.e.\ extends to a functor
  \be
  \coen X \colon\quad \FUN(\CopC,\D) \to \D \,.
  \ee
This can be seen by considering, for a given a natural transformation $\alpha\colon F\To F'$
between functors $F,F'\colon \CopC\To\D$, the diagram
  \bee8757{
  \xymatrix @R+10pt{
  & F(Y,X) \ar[dl]_{F(f^\vee\!,\idsm_X)} \ar[d]_{\alpha_{Y,X}^{}\!} \ar[dr]^{F(\idsm_Y\!,f)}
  \\
  F(X,X) \ar[d]_{\alpha_{X,X}^{}\!} & F'(Y,X) \ar[dl]^{F'(f^\vee\!,\idsm_X)}
  \ar[dr]_{F'(\idsm_Y\!,f)} & F(Y,Y) \ar[d]^{\alpha_{Y,Y}^{}\!}
  \\
  F'(X,X) \ar[dr]_{i'_X} && F'(Y,Y) \ar[dl]^{i'_Y}
  \\
  & \Coend{F'}X } }
in which the upper left and upper right squares commute by the defining property
of the natural transformation $\alpha$, while the lower square commutes by
the dinaturality of the coend of $F'$. Hence also the outer square commutes,
and thus it defines a dinatural transformation to the object $\Coend{F'}X$.
By the universal property of $\Coend FX$ there is then a unique morphism 
$\coen X\alpha\colon \Coend FX \to \Coend{F'}X$ in \D.


\subsection{Coends in braided finite tensor categories}

The following result has been established in \cite{lyub6,kerl5}: For \C\ a braided 
finite tensor category, the coend
  \be
  \H = \coend X 
  \ee
of the functor
$F\colon\, \COPC \,{\ni}\, (X,Y) \longmapsto X^\vee{\otimes}\,Y \, \in\C$
exists and has a natural structure of a Hopf algebra in \C.

The proof of this result can e.g.\ be found in \cite{vire4}. Here we only
present the structure morphisms, which are described (using the universal property
of the coend) in terms of families of morphisms from $X^\vee\oti X$ to \H:
The algebra structure is given by
  \be
  m_\H \circ (\iota_X\oti \iota_Y) := \iota_{Y\otimes X} \circ (\gamma_{X,Y}\oti\id_{Y\otimes X})
  \circ (\id_{X^\vee} \oti c_{X,Y^\vee\otimes Y})  \qquand
  \eta_\H := \iota_\one
  \ee
($\gamma_{X,Y}$ is the canonical identification of $X^\vee{\otimes}\,Y^\vee$ with
$(Y{\otimes}X)^\vee$; in the definition of $\eta_\H$ it is used that 
$\one^\vee{\otimes}\,\one\cong\one$), the coalgebra structure by
  \be
  \Delta_\H \circ \iota_X := (\iota_X\oti \iota_X) \circ (\id_{X^\vee} \oti b_X \oti \id_X)  \qquand
  \eps_\H \circ \iota_X := d_X 
  \ee
(note that the braiding does not enter the coproduct and counit), and the antipode is
  \be
  \apo_\H \circ \iota_X := (d_X \oti \iota_{X^\vee}) \circ (\id_{X^\vee} \oti c_{X^{\vee\!\vee}_{}\!,X}
  \oti \id_{X^\vee}) \circ (b_{X^\vee} \oti c_{X^\vee\!,X}) \,.
  \ee
In pictures, the structure morphisms look as follows:
  \Eqpic{p7} {395}{102} { 
   \put(13,103) {
    \put(0,0) {
  \put(0,0)   {\Includepichtft{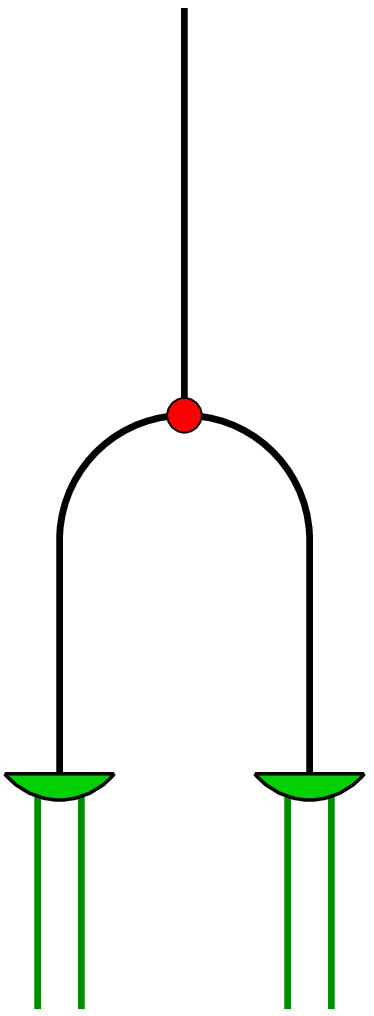} 
  \put(-6,-9)   {$\ssg X^{\!\vee} $}
  \put(7,-9)    {$\ssg X $}
  \put(15.6,115){$\sse \H $}
  \put(22,69)   {$\sse m_\H $}
  \put(23,-9)   {$\ssg Y^{\!\vee} $}
  \put(36,-9)   {$\ssg Y $} 
  \put(-9.6,23) {$\scs \iota_{\!X}^{} $} 
  \put(41,23)   {$\scs \iota_{\!Y}^{} $} 
  \put(60,50)   {$ = $}
   }
  \put(82,0) { {\Includepichtft{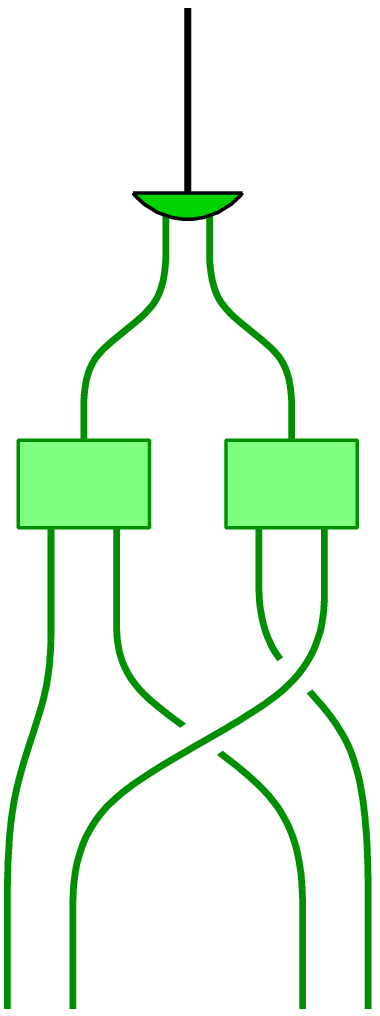}} 
  \put(-10.8,66)  {$\sse \gamma_{X,Y} $}
  \put(34,65.2)   {$\sse \id_{Y|X} $}
  \put(-7,-8)   {$\ssg X^{\!\vee} $}
  \put(6,-8)    {$\ssg X $}
  \put(16,115)  {$\sse \H $}
  \put(25,-9)   {$\ssg Y^{\!\vee} $}
  \put(38,-8)   {$\ssg Y $} 
  \put(-11,79)  {$\sss (Y{\otimes}X)^{\!\vee}_{} $}
  \put(25.5,79)   {$\sss Y{\otimes}X $}
   } }
    \put(210,8) {
  \put(0,0) { {\Includepichtft{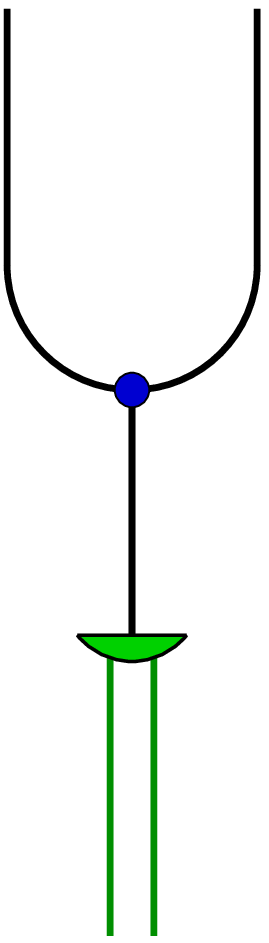}} 
  \put(-3.4,107){$\sse \H $}
  \put(23.4,107){$\sse \H $}
  \put(2,-9)    {$\ssg X^{\!\vee} $}
  \put(15,-9)   {$\ssg X $}
  \put(15.9,53.5) {$\sse \Delta_\H $}
  \put(55,50)   {$ = $}
  \put(85,0) { {\Includepichtft{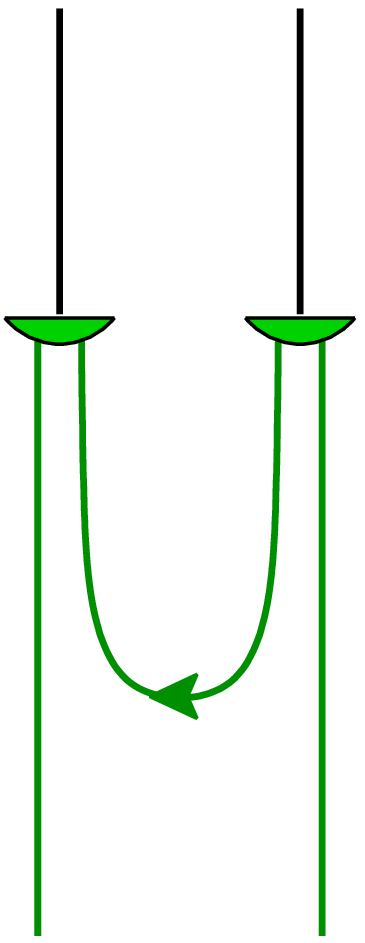}}
  \put(1.5,107) {$\sse \H $}
  \put(28.2,107){$\sse \H $}
  \put(-2,-9)   {$\ssg X^{\!\vee} $}
  \put(30.5,-9) {$\ssg X $}
   } } }
   }
    \put(13,2) {
  \put(0,17) { {\Includepichtft{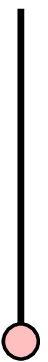}} 
  \put(-2.8,42) {$\sse \H $}
  \put(5.7,1)   {$\sse \eta_\H^{} $}
  \put(25,23)   {$ = $}
  \put(52,-18) {\Includepichtft{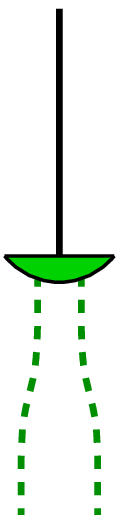}}
  \put(54,42)   {$\sse \H $}
   }
  \put(114,4) { {\Includepichtft{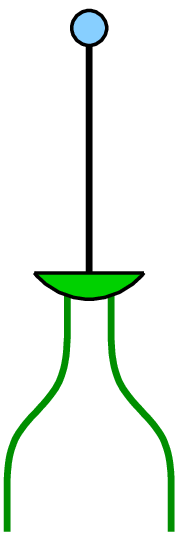}} 
  \put(13.7,54.8) {$\sse \eps_\H^{} $}
  \put(-5.2,-9) {$\ssg X^{\!\vee} $}
  \put(14,-9)   {$\ssg X $}
  \put(38,30)   {$ = $}
  \put(67,10)   {\Includepichtft{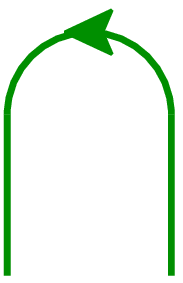}}
  \put(62,1)    {$\ssg X^{\!\vee} $}
  \put(81,1)    {$\ssg X $}
   }
  \put(260,-13) { {\Includepichtft{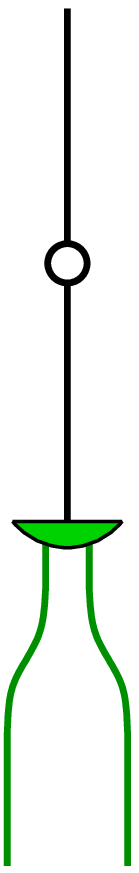}} 
  \put(11.2,64.5) {$\sse \apo_\H $}
  \put(2.8,98)  {$\sse \H $}
  \put(34,44)   {$ = $}
  \put(63,0)   {\Includepichtft{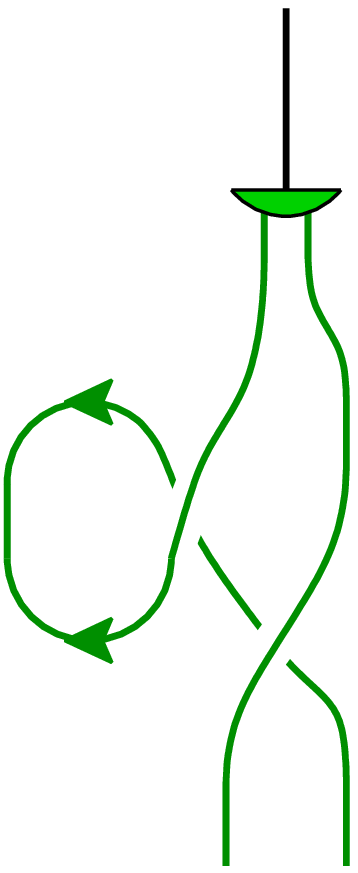}}
  \put(89.7,98) {$\sse \H $}
  \put(76.5,62) {$\sss X^{\!\vee\!\vee} $}
  \put(99.9,62) {$\sss X^{\!\vee} $}
   } } }
(In the picture for $m_\H$, $\id_{X|Y}$ is the identification of $\id_X\oti \id_Y$ 
with $\id_{X\otimes Y}$.)

\medskip

The situations discussed in sections \ref{sec.corr.cb} and \ref{sec.fac.rib} -- modular 
tensor categories and \findim\ Hopf algebras -- provide concrete examples for the
Hopf algebra \H. First, if \C\ is semisimple with finite set $\I$ of isomorphism 
classes of simple objects $\U_i$, then \H\ decomposes as an object as
\cite[Sect.\,3.2]{vire4}
  \be
  \H \,\cong\, \bigoplus_{i\in\I} \U_i^\vee {\otimes}\, \U_i^{} \,,
  \ee
and the structural morphisms $\iota_X$ are just combinations of the embedding
morphisms of the simple subobjects of $X$ and $X^\vee$ in \H.

Second, if \C\ is equivalent to the \rep\ category $H\Mod$ of a \findim\ ribbon Hopf 
algebra $H$, then \H\ is given by the dual space $\Hs\eq \Homk(H,\ko)$ endowed with
the left coadjoint $H$-action \erf{def_lads}, and with structural morphisms 
(see \cite[Lemma\,3]{kerl5} \cite[Sect.\,4.5]{vire4})
  \be
  \iota_X\colon~~ X^\vee{\otimes}X \ni\,
  \tilde x \oti x \,\longmapsto\, \big(\,h\,{\mapsto}\,\langle\tilde x,h.x\rangle\,) \,,
  \ee 
i.e.
  \Eqpic{p8} {230}{37} { \put(65,5) {
  \put(0,0)   {\Includepichtft{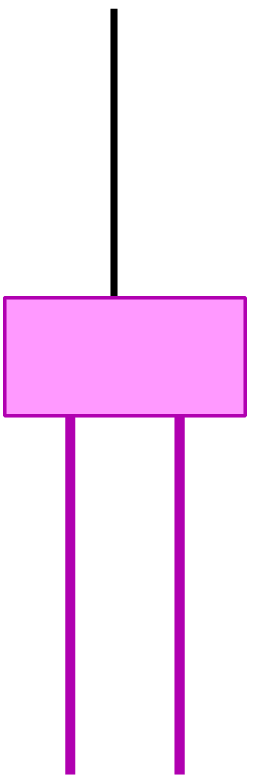}} 
  \put(80,0)  {\Includepichtft{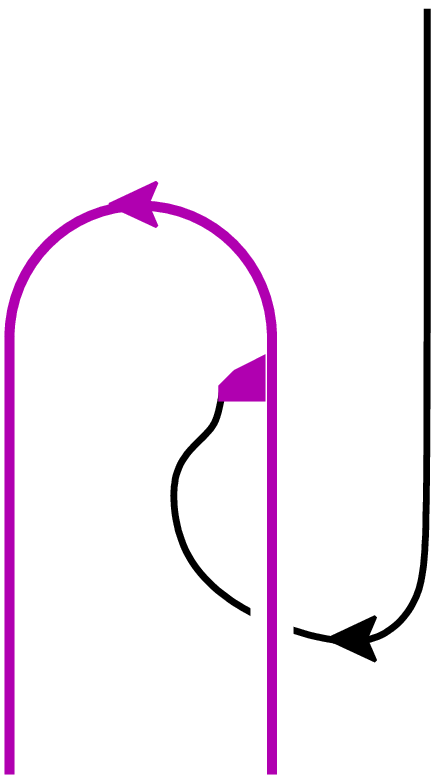}} 
  \put(0,-9)    {$\ssg X^{\!\vee} $}
  \put(8.8,45.5){$\scriptstyle i_{\!X}^{} $}
  \put(8.1,89)  {$\sse \Ha $}
  \put(16,-9)   {$\ssg X $}
  \put(48,38)   {$ = $}
  \put(111,43)  {$\sss \rho_{\!X}^{} $}
  \put(75,-9)   {$\ssg X^{\!\vee} $}
  \put(104.6,-9){$\ssg X $}
  \put(122.5,89){$\sse \Ha $}
  } }


\subsection{Three-manifold invariants from the coend \H}\label{subsec.invH}

For any finite braided tensor category \C, the coend \H\ is not just a Hopf algebra
in \C, but comes with further structure, in particular with integrals and with a
Hopf pairing. This way \H\ gives rise both to three-manifold invariants and to
representations of mapping class groups \cite{lyub6,lyub8,kerl5,KEly,vire4}.

\medskip

As can be seen by a similar argument as for \findim\ Hopf algebras in \Vectk\ 
\cite{lyub8}, \H\ has left and right integrals. If the left and right integrals coincide 
(when properly normalized), then they can be used as a Kirby element and thus provide
an invariant of three-manifolds \cite{vire4}. This invariant has been first studied in
\cite{lyub6}. If $\,\C\eq H\Mod$ for a \findim\ rib\-bon Hopf algebra $H$, the invariant
reduces to the one constructed in \cite{kaRad3,henni}.

If \,\C\ is a (semisimple) modular tensor category, then the integral of \H\ is 
two-sided and can be given explicitly in terms of the duality morphisms 
$b_i\iN\Hom(\U_i^{}\oti \U_i^\vee)$ \cite[Sect.\,2.5]{kerl5}:
  \be
  \Lambda_\H = \bigoplus_{i\in I}\dim(\U_i) \, b_{\U_i} \,.
  \ee
The three-manifold invariant obtained from this integral is, up to normalization 
\cite{chkS}, the same as the one  constructed in \cite{retu2}.


\subsection{A Hopf pairing for \H\ and modular tensor categories}\label{subsec.mtcH}

The second structure on \H\ that is of interest to us is a symmetric Hopf pairing 
$\omega_\H^{}$ of \H. It is given by \cite{lyub6}
  \Eqpic{def.hopa} {120}{31} { \put(0,-3){
  \put(0,4) { {\Includepichtft{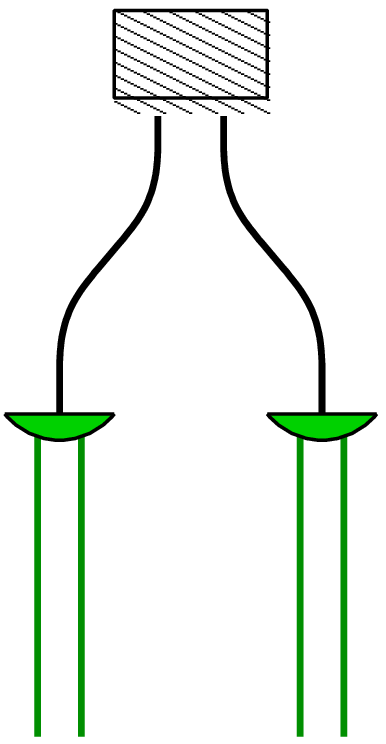}}
  \put(-6,-9)   {$\ssg X^{\!\vee} $}
  \put(7,-9)    {$\ssg X $}
  \put(24,-9)   {$\ssg Y^{\!\vee} $}
  \put(37,-9)   {$\ssg Y $}
  \put(14.7,74.9) {\boldmath$\sse \omega_\H^{}$}
  \put(61,38)   {$ := $}
  \put(95,0)  {\Includepichtft{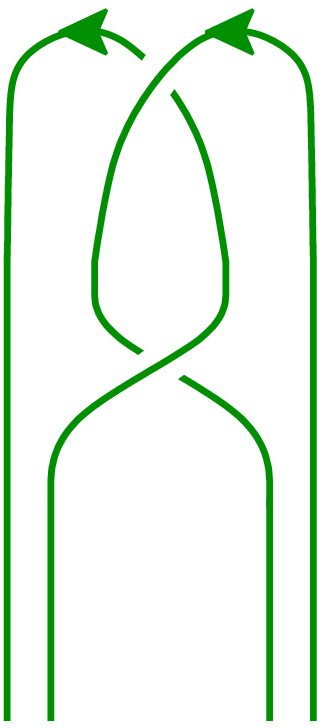}}
    \put(91,0){
  \put(-6,-9)   {$\ssg X^{\!\vee} $}
  \put(7,-9)    {$\ssg X $}
  \put(24,-9)   {$\ssg Y^{\!\vee} $}
  \put(37,-9)   {$\ssg Y $}
    } } } }
i.e.\ is induced by the dinatural family of morphisms $(d_X \oti d_Y) 
\cir [\id_{X^\vee} \oti (c_{Y^\vee,X}^{}\cir c_{X,Y^\vee}^{}) \oti\id_Y]$.

Now notice the similarity of the morphism on the right hand side of \erf{def.hopa}
with the one in the expression \erf{def.s} for the matrix $\soo$. Since
invertibility of $\soo$ is the crucial ingredient of the definition of a (semisimple)
\mtc, the following definition is a natural generalization to the non-semisimple 
case: A \emph{modular finite tensor category} is a braided finite tensor category 
for which the Hopf pairing $\omega_\H$ given by \erf{def.hopa} is non-degenerate.

It can be shown \cite[Thm.\,6.11]{lyub8} that if \C\ is modular in this sense,
then the left and right integrals of \H\ coincide. An example for a (non-semisimple)
\mtc\ is provided by the category $H\Mod$ of left modules over any (non-semisimple)
finite-dimensional factorizable ribbon Hopf algebra $H$ {\rm \cite{lyma,lyub6}}.

\medskip

The terminology suggests that there is a relation with the modular group \slz. In
the case of semisimple \mtcs, \slz\ arises as the mapping class group of the torus, 
which as noted in section \ref{sec.coco} naturally acts on the space of zero-point
conformal blocks on the torus. But \slz\ appears in the non-semisimple case as well:
there are endomorphisms $S_\H$ and $T_\H$ of \H\ which satisfy the relations for 
the generators of (a twisted group algebra of) \slz\ \cite{lyub8}. These endomorphisms
will be given below.

Denote by $Z(\C)$ the \emph{center} of \C, i.e.\ the algebra of natural 
endotransformations of the identity functor of \C. An element of $Z(\C)$ is a family
$(\phi_X)_{X\in\C}$ of endomorphisms $\phi_X\iN\End(X)$, and one can check that the
composition $(\iota_X\cir(\id_{X^\vee_{}}\oti\phi_X))_{X\in\C}$ gives a dinatural
family. By the universal property of the coend there is thus a unique 
endomorphism $\overline\phi_\H$ of \H\ such that the diagram
  \bee3909{ \label{ZC.EndH} 
  \xymatrix @R+4pt{
  X^\vee\oti X\ \ar^{\id\otimes\phi_X^{}}[rr]\ar_{\iota_X}[d]
  && \, X^\vee\oti X \ar^{\iota_X}[d]\\
  \H\, \ar@{-->}_{\overline\phi_\H^{}}[rr]&&\, \H
  } }
commutes for all $X\iN\C$; this furnishes an injective linear map
$Z(\C) \To \End(\H)$. By concatenation with the counit $\eps_\H$ this gives a map
  \be
  Z(\C) \longrightarrow \mathrm{End}(\H) \stackrel{{(\eps_\H^{})}_*}
  {-\!\!\!-\!\!\!\longrightarrow} \Hom(\H,\one) \,.
  \labl{ZC.HomH1}
Now the vector space $\Hom(\H,\one)$ has a natural structure of a \ko-algebra 
(since \H\ is in particular a coalgebra and $\one$ is an algebra); it can be
shown \cite[Lemma\,4]{kerl5} that \erf{ZC.HomH1} is an isomorphism of \ko-algebras.

If the category \C\ is a ribbon category (e.g.\ if \C\ is sovereign, i.e.\ if
the left and right dualities are equal as functors), then the family 
$(\theta_X)_{X\in\C}$ of twist isomorphisms constitutes an element $\nu\iN Z(\C)$,
the ribbon element. Denote by
  \be
  T_\H := \overline\nu_\H^{} \,\,\in\,\End(\H) 
  \labl{def.Th}
the endomorphism of \H\ obtained by applying the map defined in \erf{ZC.EndH} to
the ribbon element. Pictorially,
  \Eqpic{p9} {77}{22} {
  \put(0,0)   {\Includepichtft{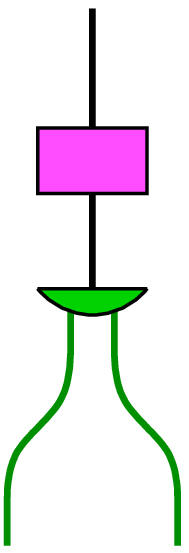}
  \put(-11.8,40.7) {\small$\overline\nu_{\!\H}^{}$}
  \put(-6,-9)   {$\sse X^{\!\vee} $}
  \put(14.9,-9) {$\sse X $}
  \put(6.8,62.7){$\sse \H $}
  }
  \put(46,28) {$=$}
  \put(81,0)  {\Includepichtft{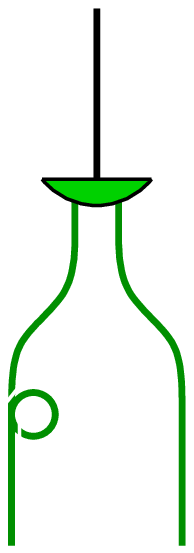}
  \put(-10.9,14) {$\sse \nu_{\!X^{\!\vee}_{}}^{}$}
  \put(-3,-9)   {$\sse X^{\!\vee} $}
  \put(17.9,-9) {$\sse X $}
  \put(8.4,62.7){$\sse \H $}
  } }

Next consider the family of morphisms on the right hand side of
  \Eqpic{def.Sig} {135}{35} {
  \put(0,0)   {\Includepichtft{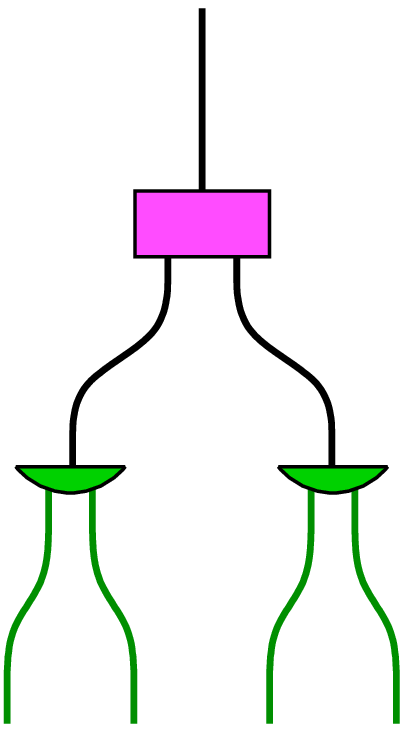}
  \put(3.5,51.6) {$\varSigma$}
  \put(-6,-9)   {$\sse X^{\!\vee} $}
  \put(9.9,-9)  {$\sse X $}
  \put(25,-9)   {$\sse Y^{\!\vee} $}
  \put(40.2,-9) {$\sse Y $}
  \put(18.8,83) {$\sse \H $}
  }
  \put(69,35) {$:=$}
  \put(108,0) {\Includepichtft{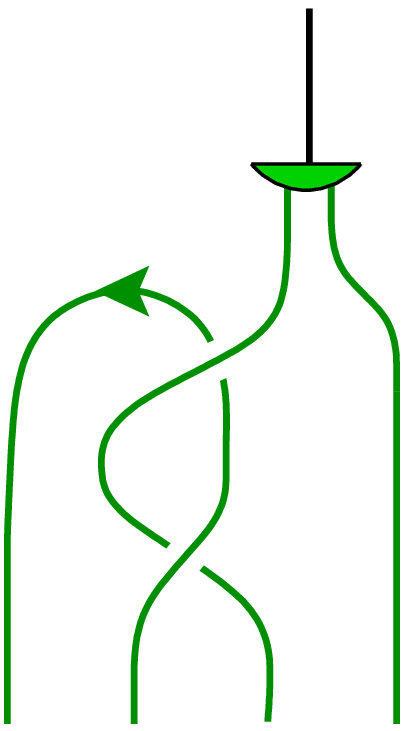}
  \put(-6,-9)   {$\sse X^{\!\vee} $}
  \put(9.9,-9)  {$\sse X $}
  \put(25,-9)   {$\sse Y^{\!\vee} $}
  \put(40.2,-9) {$\sse Y $}
  \put(29.8,83) {$\sse \H $}
  } }
This family is dinatural both in $X$ and in $Y$, and hence \erf{def.Sig} defines
a  morphism $\varSigma\colon \H\oti\H\To\H$. By composition with a left or right 
integral $\Lambda$ of \H\ one arrives at an endomorphism 
  \be
  S_\H := \varSigma \circ (\id_\H \oti \Lambda) \,\,\iN \,\End(\H) \,.
  \labl{def.Sh}

It was established in \cite[Sect.\,6]{lyub8} that for any (not necessarily
semisimple) \mtc\ \C, the two-sided integral of \H\ can be suitably normalized 
in such a way that the endomorphisms \erf{def.Th} and \erf{def.Sh} furnish
a morphism
  \be
  \ko_\xi\slz \longrightarrow \End(\H)
  \ee
of \ko-algebras, where for $\xi\iN \ko^\times$, $\ko_\xi\slz$ denotes the twisted 
group algebra of \slz\ with relations $S^4\eq 1$ and $(ST)^3\eq \xi\, S^2$.

Since for every object $U$ of \C\ the morphism space $\Hom(U,\H)$ is, by 
push-forward, a left module over the algebra $\End(\H)$, one obtains this way
projective representations of \slz\ on all vector spaces $\Hom(U,\H)$.

If \C\ is semisimple, the vector space $\Hom(\one,\H)$ coincides with the space of 
conformal blocks of the torus, $\Hom(\one,\H) \,{\cong}\, \tftc(\torus)$, and
the \slz-\rep\ obtained this way is precisely the \rep\ on the characters 
\erf{def_chi} of a rational CFT as described in section \ref{sec.coco}.
Also note that one can think of $\Hom(\one,\H)$, which is dual to 
the space $\Hom(\H,\one)$ on the right hand side of \erf{ZC.HomH1}, as the appropriate 
substitute for the space of class functions. $\Hom(\one,\H)$ would therefore be a 
natural starting point for constructing a vector space assigned
to the torus $\torus$ by a topological field theory based on \C.

\medskip

Let us finally comment on a relationship with the pseudo-characters that we
encountered in section \ref{verl.L1p}. Consider the map 
$\Chi\colon \mathrm{Obj}(\C) \To \Hom(\one,\H)$ given by
  \Eqpic{def_Chi} {65} {19} {
  \put(-40,28) {$ U \,\longmapsto\, \Chi_U \,:=\, \iota_U \cir \tilde b_U ~= $}
  \put(100,-6) { {\Includepichtft{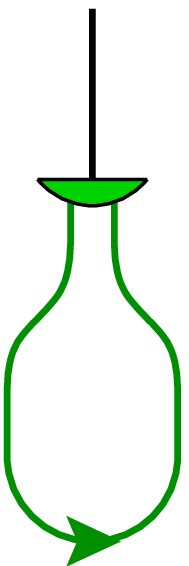}}
  \put(5.8,63.3) {$\sse \H $}
  \put(21.5,13){$\sse U $}
  } }
The morphisms $\Chi_U$ can be regarded as generalizations of characters:
The map $\Chi$ factorizes to a morphism of rings $K_0(\C)\To \Hom(1,\H)$,
i.e.\ it is additive under exact sequences and one has
  \be
  \Chi_X \cdot \Chi_Y := m_\H \cir (\Chi_Y\oti\Chi_X) = \Chi_{X\otimes Y} \,.
  \ee
The latter is seen as follows:
  \Eqpic{def_Chi2} {380} {59} {
  \put(-5,59) {$ m_\H \cir (\Chi_Y\oti\Chi_X) ~= $}
  \put(112,0) { {\Includepichtft{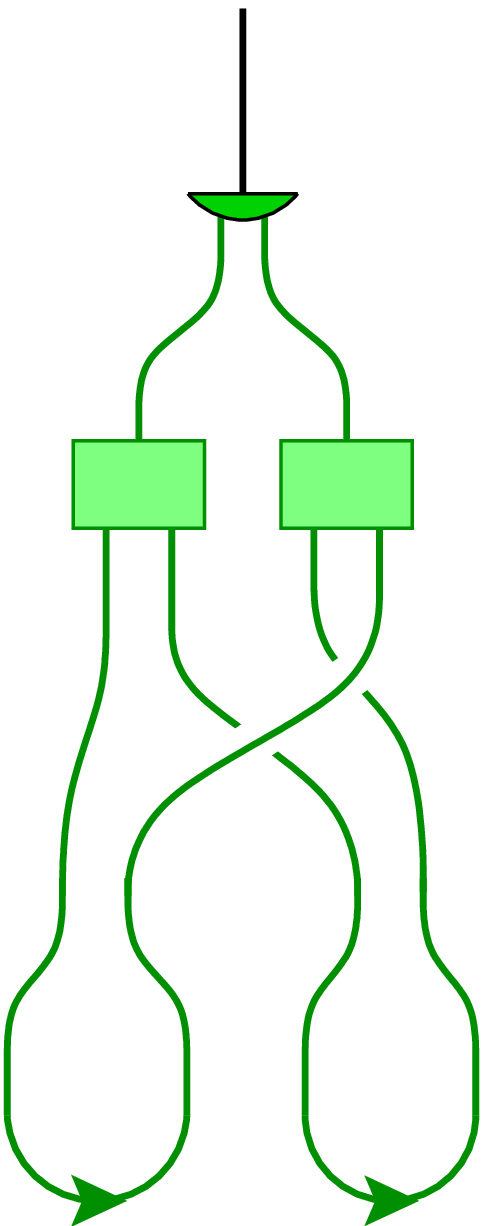}}
  \put(21.1,13){$\sse Y $}
  \put(22.8,136){$\sse \H $}
  \put(32.8,102){$\sse X{\otimes}Y $}
  \put(52.9,13){$\sse X $}
  \put(-4.9,89)  {$\sse \gamma_{X,Y} $}
  \put(40,88)  {$\sse \id_{Y|X} $}
  }
  \put(193,59) {$ = $}
  \put(223,0) { {\Includepichtft{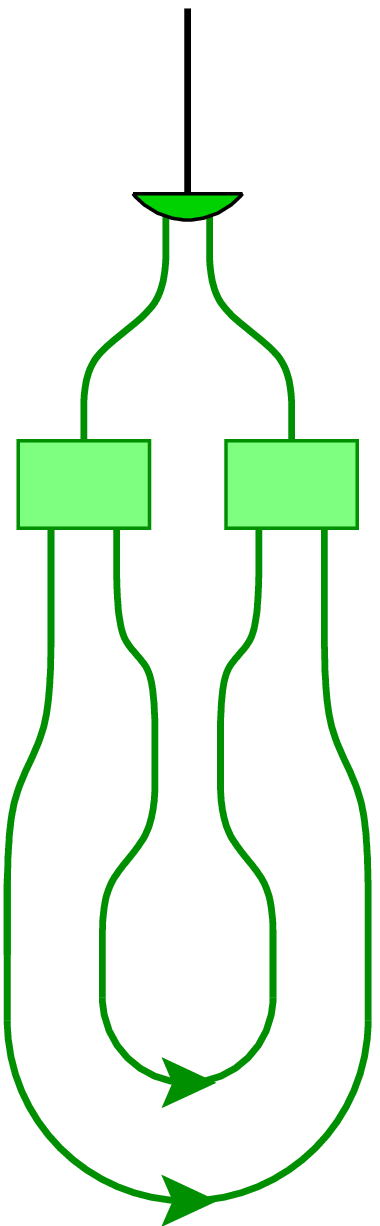}}
  \put(30.8,26){$\sse X $}
  \put(16.8,136){$\sse \H $}
  \put(42.2,26){$\sse Y $}
  }
  \put(293,59) {$ = $}
  \put(321,22) { {\Includepichtft{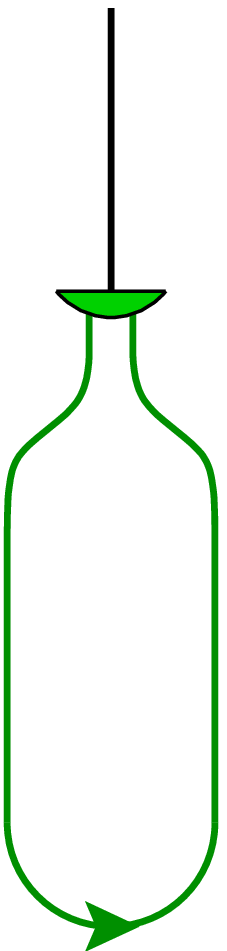}}
  \put(7.8,107){$\sse \H $}
  \put(24.9,27){$\sse X{\otimes}Y $}
  } }
Furthermore, if the category \C\ is semisimple, then
$\Hom(\one,\H) \,{\cong}\, \bigoplus_{i\in\I} \Hom(\one,\U^\vee_i\oti \U_i^{})$,
so that $\{\Chi_{\U_i}^{}\}_{i\in \I}^{}$ constitutes a basis of the vector 
space $\Hom(\one,\H)$. If \C\ is not semisimple, these elements are still 
linearly independent, but they do not form a basis any more. A (non-canonical)
complement of this linearly independent set is in certain cases provided by an 
analogue of pseudo-characters.

  \newpage

\appendix \section{Appendix}

\subsection{Semisimple modular tensor categories}\label{app.mtc}

The category of chiral data of a rational \cft\ is a semisimple \mtc. It arises 
as the \rep\ category of a conformal vertex algebra that is rational in the sense 
that it obeys the $C_2$-cofiniteness condition and certain conditions on its 
homogeneous subspaces \cite{huan21}. But there are also various other algebraic 
structures which give rise to \mtcs, for instance (for some more details and 
references see e.g.\ section 3.3 of \cite{fuRs13}) nets of von Neumann algebras 
on the real line which have finite $\mu$-index, are strongly additive and 
possess the split property \cite{kalm}, as well as connected ribbon 
factorizable weak Hopf algebras over \complex\ with a Haar integral \cite{nitv}.

\medskip

A (semisimple) \emph{modular tensor category} is, by definition, a semisimple abelian 
\complex-linear monoidal category with simple tensor unit $\one$ and with the
set $\I$ of isomorphism classes of simple objects being finite, which has a ribbon 
structure for which the braiding is maximally non-degenerate. 

\medskip

We will spell out the last two ingredients of this definition. We denote the tensor 
product bifunctor by $\otimes\,{:}~\,\C{\times}\C\To\C$~ and the monoidal unit by 
$\one$. A \emph{ribbon} category \cite[Ch.\,XIV.3]{KAss} 
$\C \eq (\C,\otimes,\one,c,b,d,\theta)$ is a monoidal category endowed with four 
specific families of morphisms: a duality (evaluation and coevaluation), a braiding, 
and a twist.  A (right) {\em duality\/} assigns to every $U\iN\C$ another object 
$U\Vee$, called the (right-)\,dual of $U$, and morphisms $b_U \iN\Hom(\one,U\Oti U\Vee)$ 
and $d_U \iN \Hom(U\Vee\Oti U,\one)$, called coevaluation and evaluation, respectively.
A {\em braiding\/} is a family of isomorphisms $c_{U,V}\iN\Hom(U\Oti V,V\Oti U)$,
one for each pair $U,V\iN\C$, and a {\em twist\/} is a family of
isomorphisms $\theta_U\iN\End(U)$, one for each $U\iN\C$.

Every monoidal category is equivalent to one that is \emph{strict} monoidal, i.e.\ for 
which one has equalities $U\oti(V\oti W) \eq (U\oti V) \oti W$ and $U\oti\one \eq U 
\eq \one\oti U$ for $U,V,W\iN\,\C$ rather than just natural isomorphisms of objects. 
We replace any non-strict monoidal category by an equivalent strict one; this allows 
for a simple graphical notation for morphisms, in which the tensor product of 
morphisms is just juxtaposition, see section \ref{sec.frob}. In this notation the 
braiding and twist and their inverses and the duality are depicted as follows:
  \be
  \begin{array}{llll}
c_{U,V}^{} =~
\begin{picture}(64,28)(0,18) \apppicture{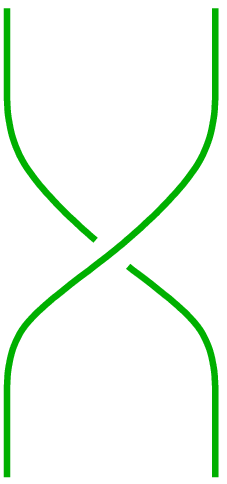}  {7}
\put(4.2,-8.8)   {$\sse U$}
\put(4.8,43.3)   {$\sse V$}
\put(21.2,-8.8)  {$\sse V$}
\put(22.9,43.3)  {$\sse U$} \eP
&
c_{U,V}^{-1} =~
\begin{picture}(64,20)(0,18) \apppicture{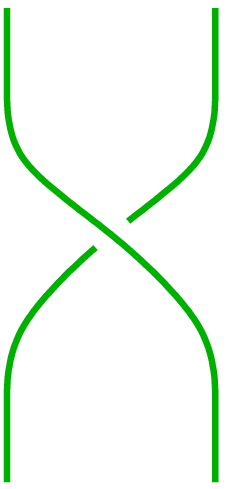}  {7}
\put(4.2,-8.8)   {$\sse V$}
\put(4.8,43.3)   {$\sse U$}
\put(21.2,-8.8)  {$\sse U$}
\put(22.9,43.3)  {$\sse V$} \eP
&
\theta_{U}^{} =~
\begin{picture}(54,20)(0,18) \apppicture{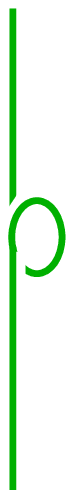}  {7}
\put(5.5,-8.8)   {$\sse U$}
\put(6.4,44.3)   {$\sse U$} \eP
&
\theta_{U}^{-1} =~
\begin{picture}(26,20)(0,18) \apppicture{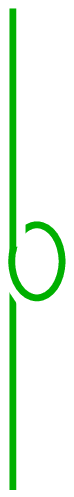}  {7}
\put(5.5,-8.8)   {$\sse U$}
\put(6.4,44.3)   {$\sse U$} \eP
\\
~~b_{U}^{} =~
\begin{picture}(36,60)(0,18) \apppicture{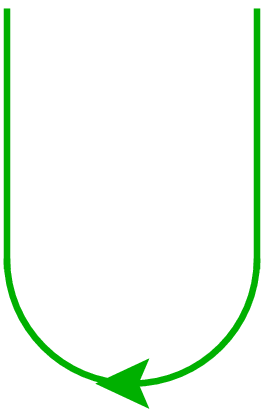}  {7}
\put(4.8,37.3)   {$\sse U$}
\put(24.9,37.3)  {$\sse U^\vee$} \eP
&
~~d_{U}^{} =~
\begin{picture}(36,30)(0,11) \apppicture{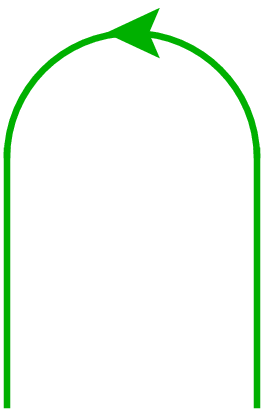}  {7}
\put(3.2,-8.8)   {$\sse U^\vee$}
\put(25.2,-8.8)  {$\sse U$} \eP
\\[.5em]~ \eear
  \ee
(Since $\one$ is a strict tensor unit, we have adopted the convention that it is 
invisible in such pictures.) With these conventions, the relations to be satisfied 
by the duality, braiding and twist are
  \be
  \hspace*{-1em}\begin{array}{llll} 
\begin{picture}(115,66) \apppicture{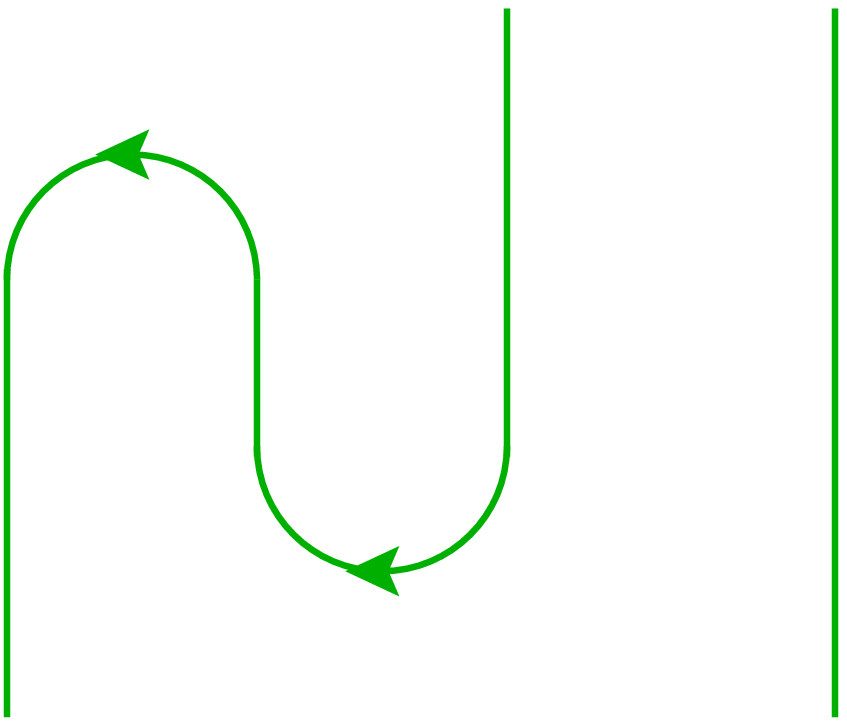}  {14}
\put(10.2,-9.4)  {$\sse U^\vee$}
\put(51.8,64.2)  {$\sse U^\vee$}
\put(65.5,25)    {\small$=$}
\put(79.2,-9.4)  {$\sse U^\vee$}
\put(79.8,64.2)  {$\sse U^\vee$} \eP
&
~~~~\begin{picture}(97,49) \apppicture{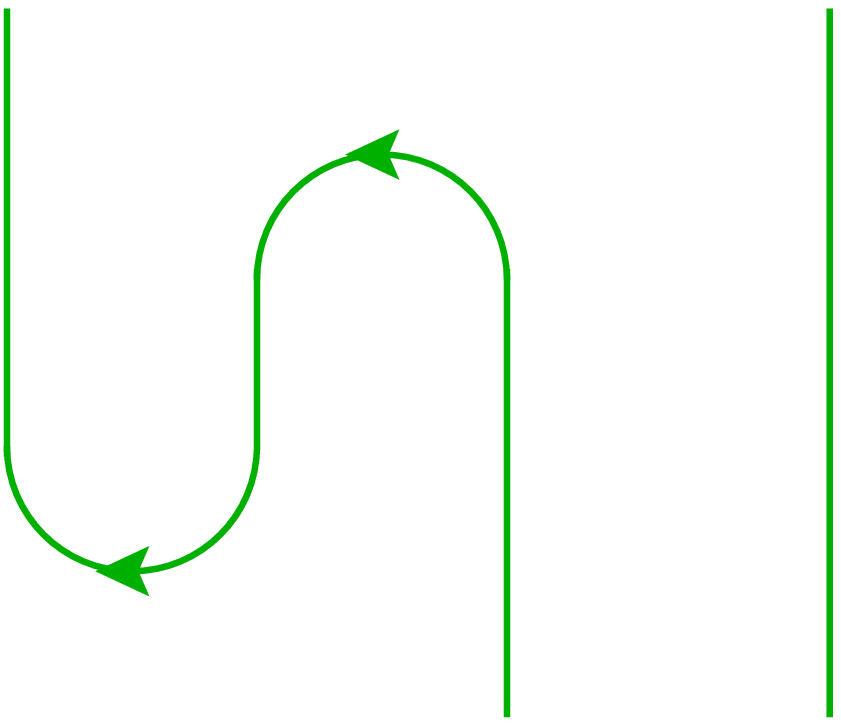}  {14}
\put(12.2,64.2)  {$\sse U$}
\put(52.8,-8.8)  {$\sse U$}
\put(65.5,25)    {\small$=$}
\put(80.4,-8.8)  {$\sse U$}
\put(81.1,64.2)  {$\sse U$} \eP
&
\begin{picture}(83,36)(0,-3) \apppicture{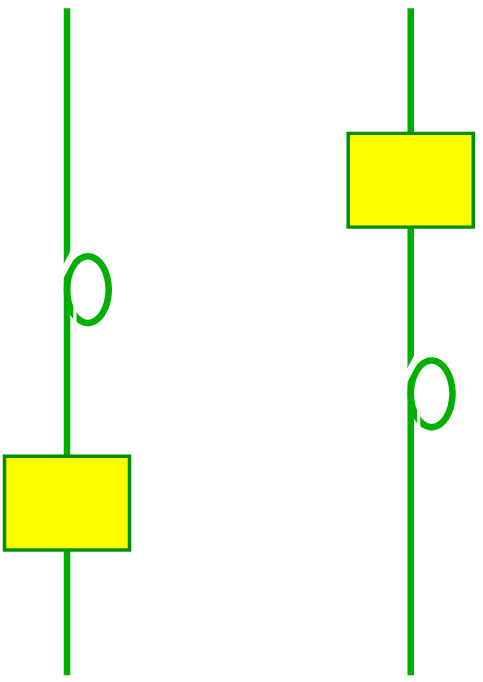}  {10}
\put(12.4,-8.5)  {$\sse U$}
\put(13.2,59.9)  {$\sse V$}
\put(13.5,13.6)  {\tiny$f$}
\put(26,26)      {\small$=$}
\put(41.4,-8.5)  {$\sse U$}
\put(42.2,59.9)  {$\sse V$}
\put(42.5,40.6)  {\tiny$f$} \eP
&
\begin{picture}(60,29)(0,-6) \apppicture{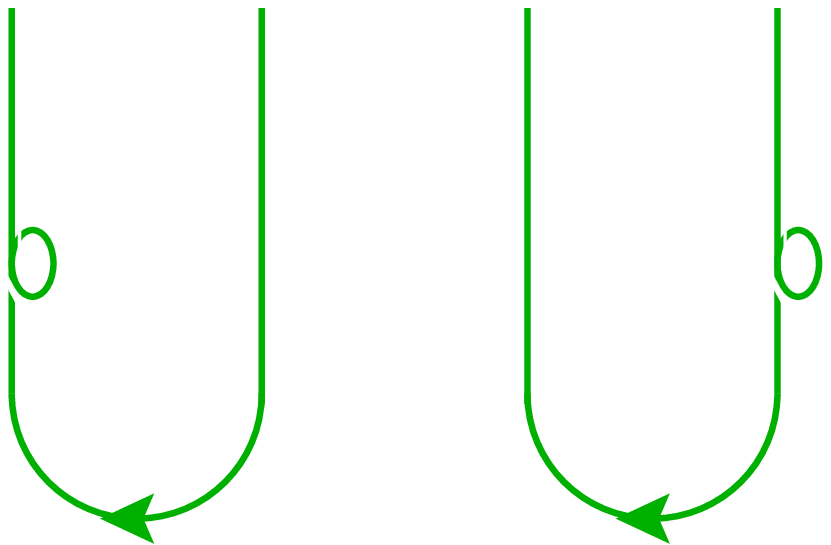} {-2}
\put(-2.8,49.3)  {$\sse U$}
\put(18.5,49.3)  {$\sse U$}
\put(30.5,18)    {\small$=$}
\put(40.8,49.3)  {$\sse U$}
\put(61.8,49.3)  {$\sse U$} \eP
\\
\begin{picture}(97,105)(0,-8) \apppicture{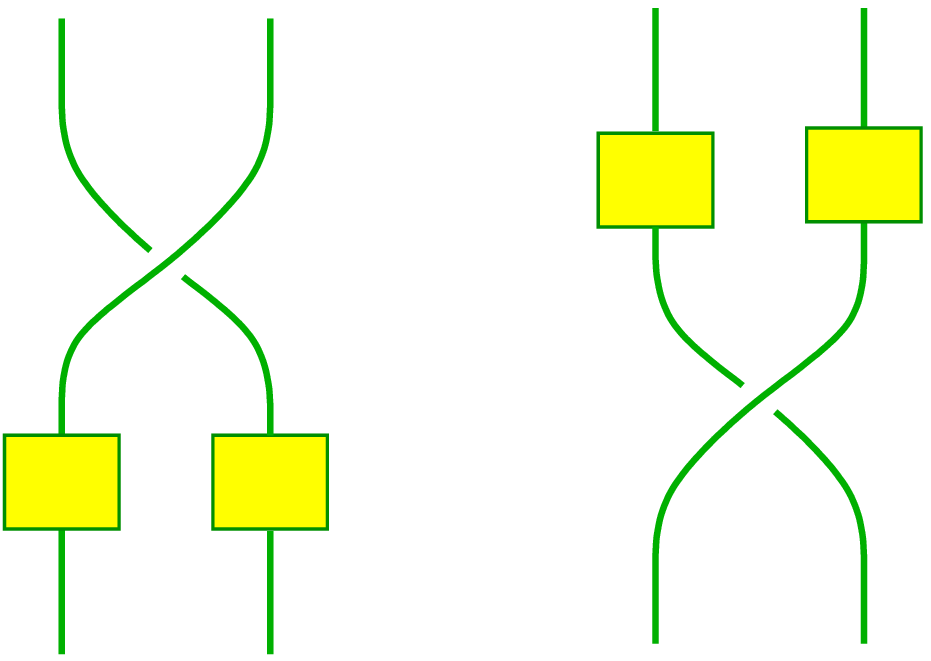}  {10}
\put(11.8,-8.5)  {$\sse U$}
\put(11.8,58.1)  {$\sse W$}
\put(12.5,13.5)  {\tiny$f$}
\put(31,14)      {\tiny$g$}
\put(28.8,-8.5)  {$\sse V$}
\put(29.2,58.1)  {$\sse X$}
\put(45,25)      {\small$=$}
\put(61.3,-8.5)  {$\sse U$}
\put(61.3,58.1)  {$\sse W$}
\put(63,39.7)    {\tiny$g$}
\put(80,39.3)    {\tiny$f$}
\put(78.8,-8.5)  {$\sse V$}
\put(79.2,58.1)  {$\sse X$} \eP
&
\begin{picture}(144,62) \apppicture{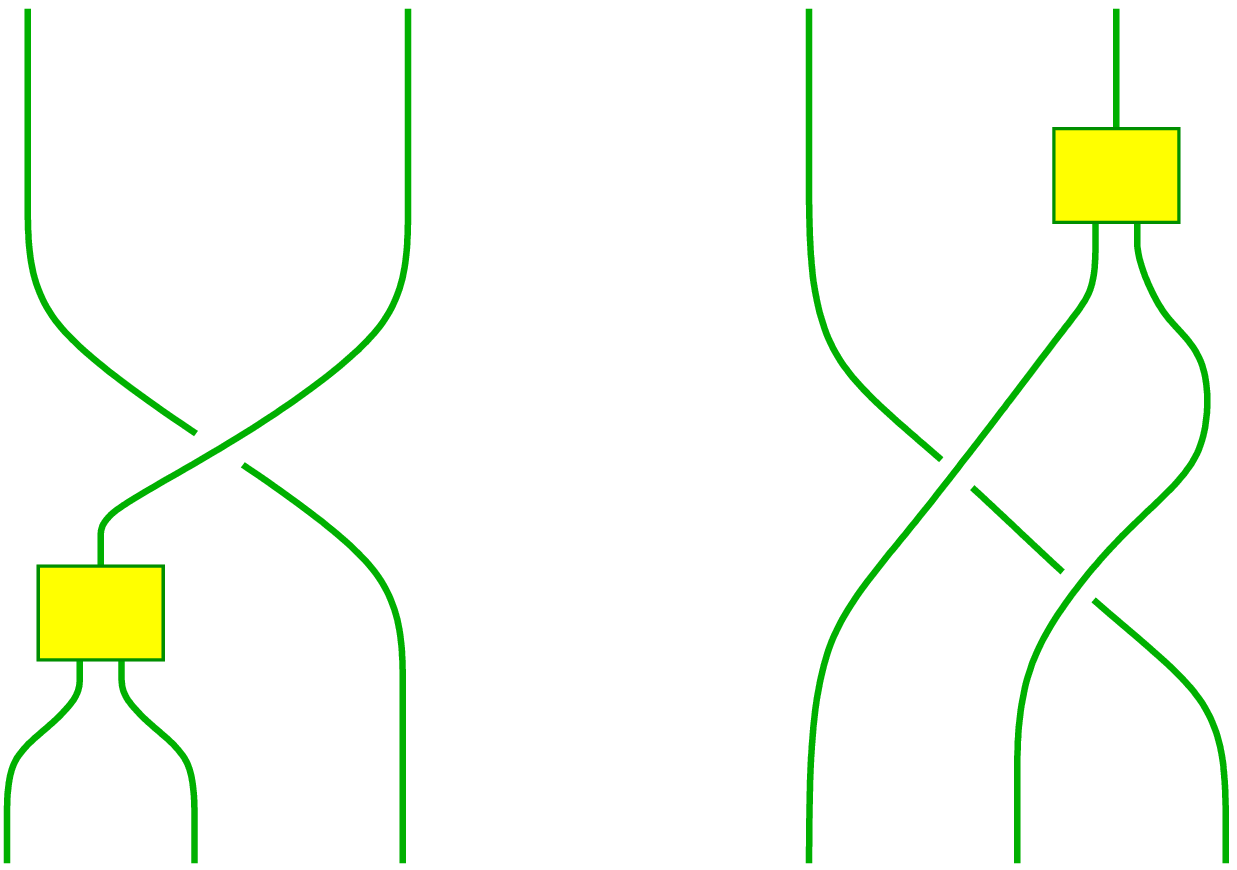}  {10}
\put(7.4,-8.5)   {$\sse U$}
\put(8.4,76.4)   {$\sse W$}
\put(16.3,19.9)  {\tiny$f$}
\put(22.8,-8.5)  {$\sse V$}
\put(34.4,76.4)  {$\sse U\Oti V$}
\put(39.4,-8.5)  {$\sse W$}
\put(58.5,31)    {\small$=$}
\put(74.1,-8.5)  {$\sse U$}
\put(74.4,76.4)  {$\sse W$}
\put(91.5,-8.5)  {$\sse V$}
\put(93.5,76.4)  {$\sse U\Oti V$}
\put(101.3,56.9) {\tiny$f$}
\put(107.7,-8.5) {$\sse W$} \eP
&
\begin{picture}(89,94)(0,8) \apppicture{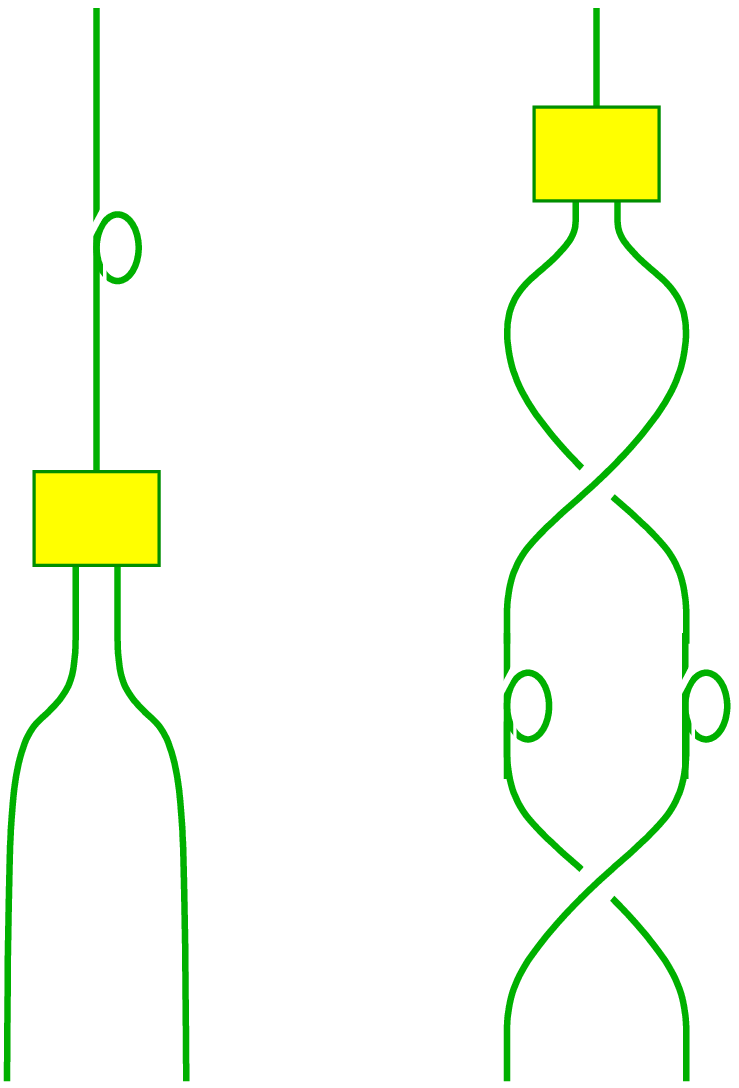}  {18}
\put(15.0,-8.5)  {$\sse U$}
\put(16.2,94.9)  {$\sse U\Oti V$}
\put(23.6,46.1)  {\tiny$f$}
\put(29.7,-8.5)  {$\sse V$}
\put(41,42)      {\small$=$}
\put(57.1,-8.5)  {$\sse U$}
\put(58.2,94.9)  {$\sse U\Oti V$}
\put(65.6,76.6)  {\tiny$f$}
\put(71.9,-8.5)  {$\sse V$} \eP
\\[-.1em]~ \eear
  \labl{ax.rib}
One may think of the lines in these pictures as ribbons, with the twist $\theta$
describing a $2\pi$-rotation of a ribbon around its core. The equalities \erf{ax.rib}
are then precisely the relations satisfied by ribbons embedded in the three-sphere.

\medskip

Finally, that the braiding is maximally non-symmetric means that the 
$|\I|\Times|\I|$-matrix $\soo$ with entries
  \Eqpic{def.s} {120} {16} {  \put(0,-5){
  \put(0,29) {$ \sOO_{i,j} :=~$}
  \put(43,0) { \scalebox{.42}{\includegraphics{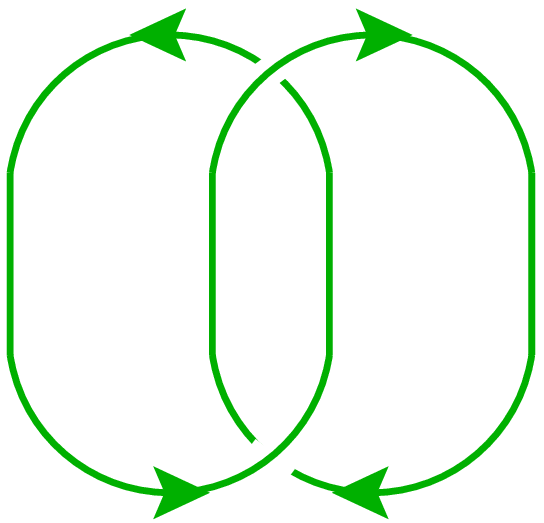}}
  \put(-52,31){\Turlabg{\U_{\! i}^{}}}
  \put(2,25){\turlabg{\U_{\! j}^{}}}
  } } }
(in $\End(\one)\eq \complex\,\id_\one$, which we identify with $\complex$) is 
non-degenerate.

  \newpage

\subsection{Invertibility of the Frobenius map}\label{app.fmap}

Here we give a graphical proof of the following statement:
\\
For any Hopf algebra $H$ in an additive ribbon category \C\ that has an invertible 
antipode and a left integral $\Lambda \iN \Hom(\one,H)$ and right cointegral $\lambda 
\iN \Hom(H,\one)$ such that $\lambda \cir \Lambda$ is invertible, the morphisms 
\fmap\ (the Frobenius map) and \fmapi\ defined in \erf{def_fmap} are mutually inverse.

\medskip

Without loss of generality, we may normalize the integrals such that $\lambda \cir
\Lambda \eq \id_\one$. Then we have the following chain of equalities:
  \Eqpic{proof_fmap} {400} {185} {
  \put(-85,200){
  \put(72,0)   {\Includepichopf{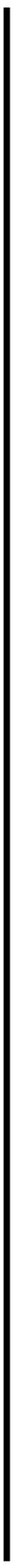}}
  \put(100,91)  {$=$}
  \put(132,0)  {\Includepichopf{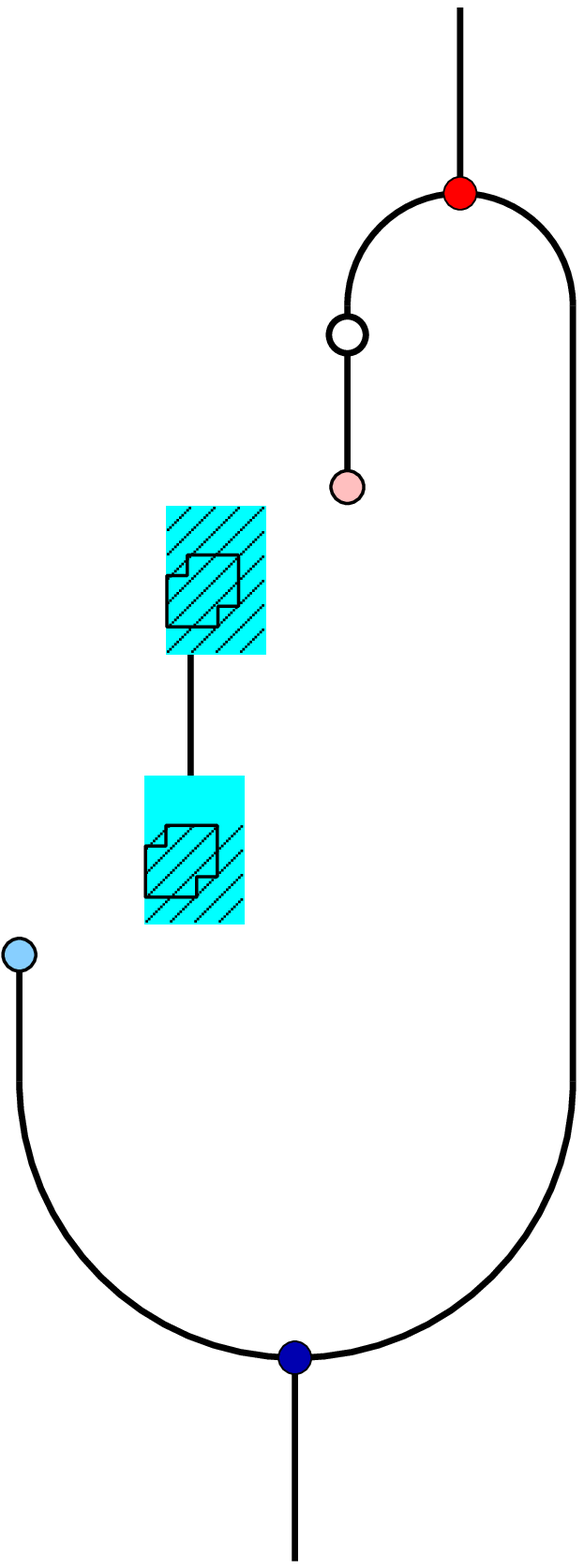}}
  \put(224,91) {$=$}
  \put(256,0)  {\Includepichopf{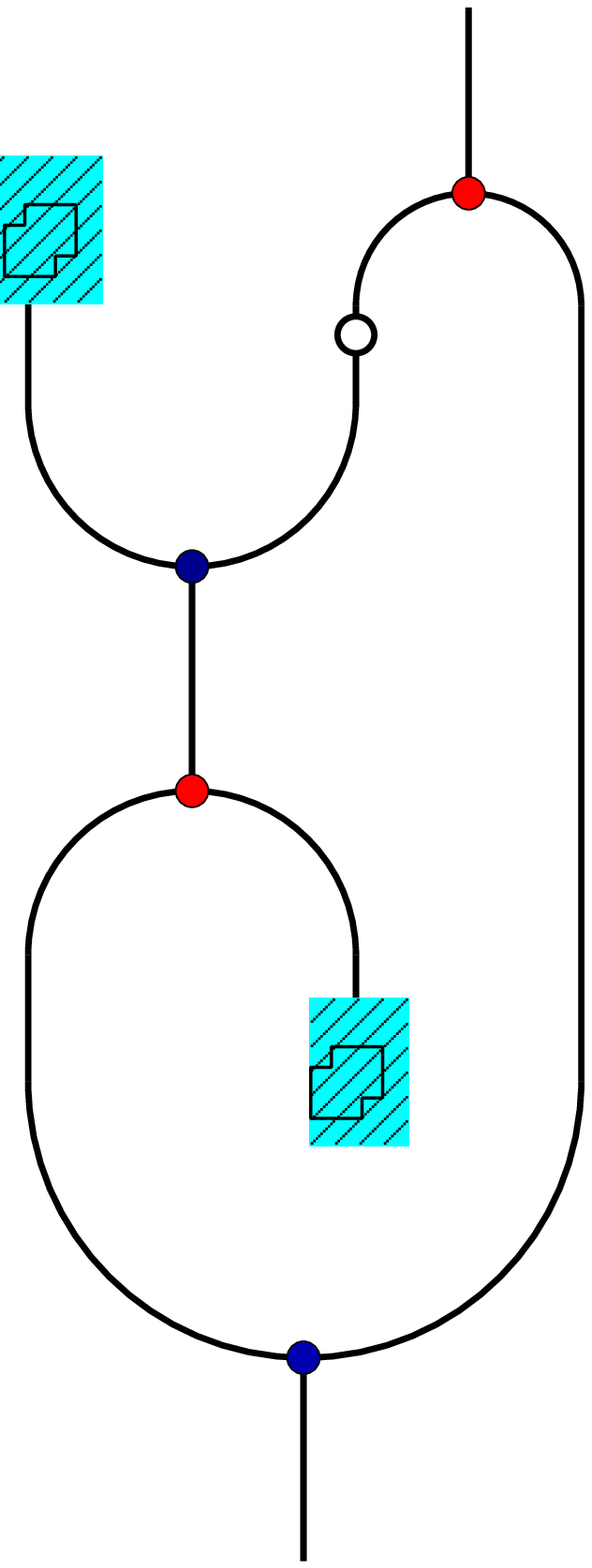}}
  \put(348,91) {$=$}
  \put(380,0)  {\Includepichopf{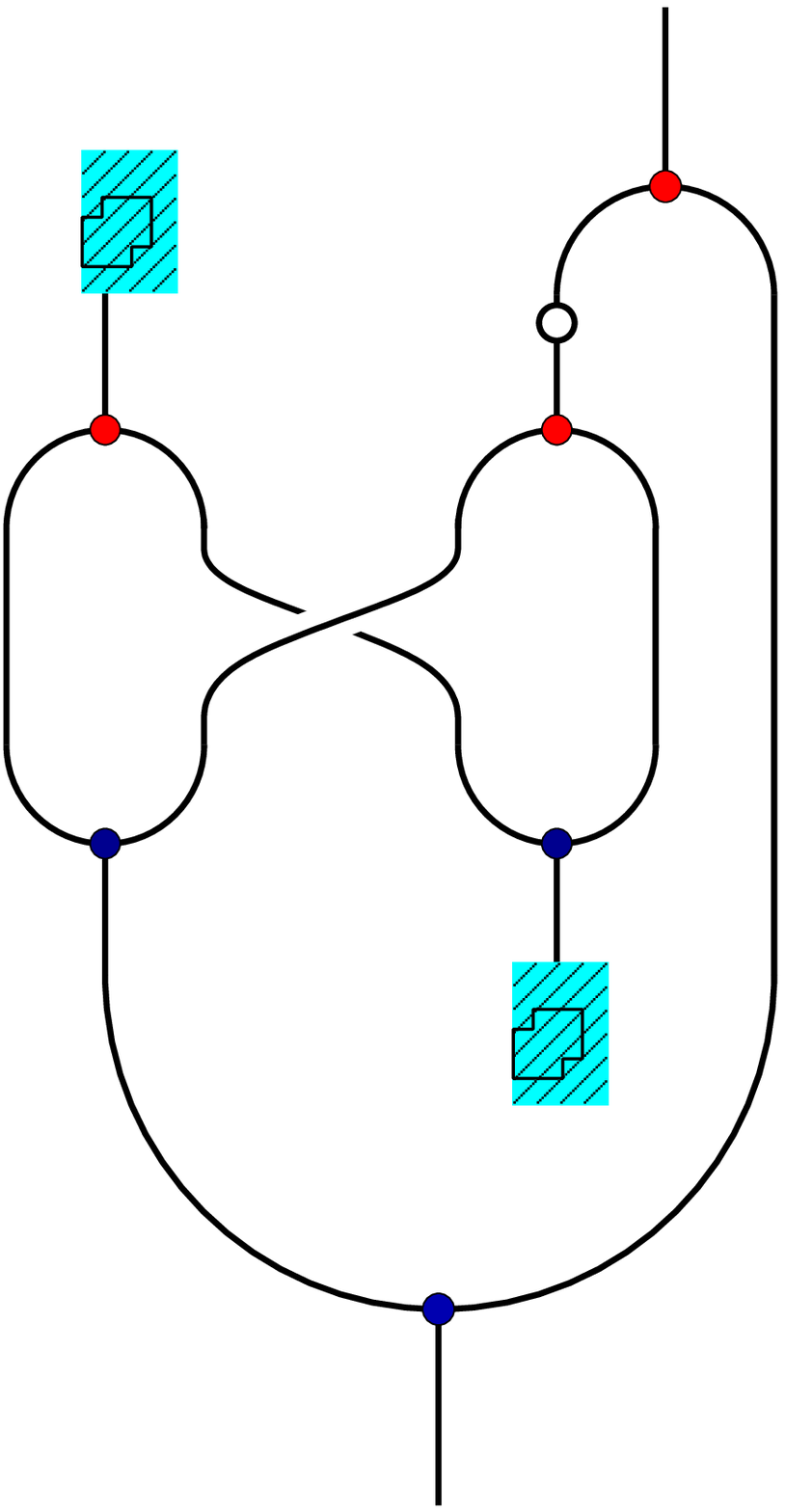}}
 }
  \put(-88,-3){
  \put(103,91) {$=$}
  \put(130,0)  {\Includepichopf{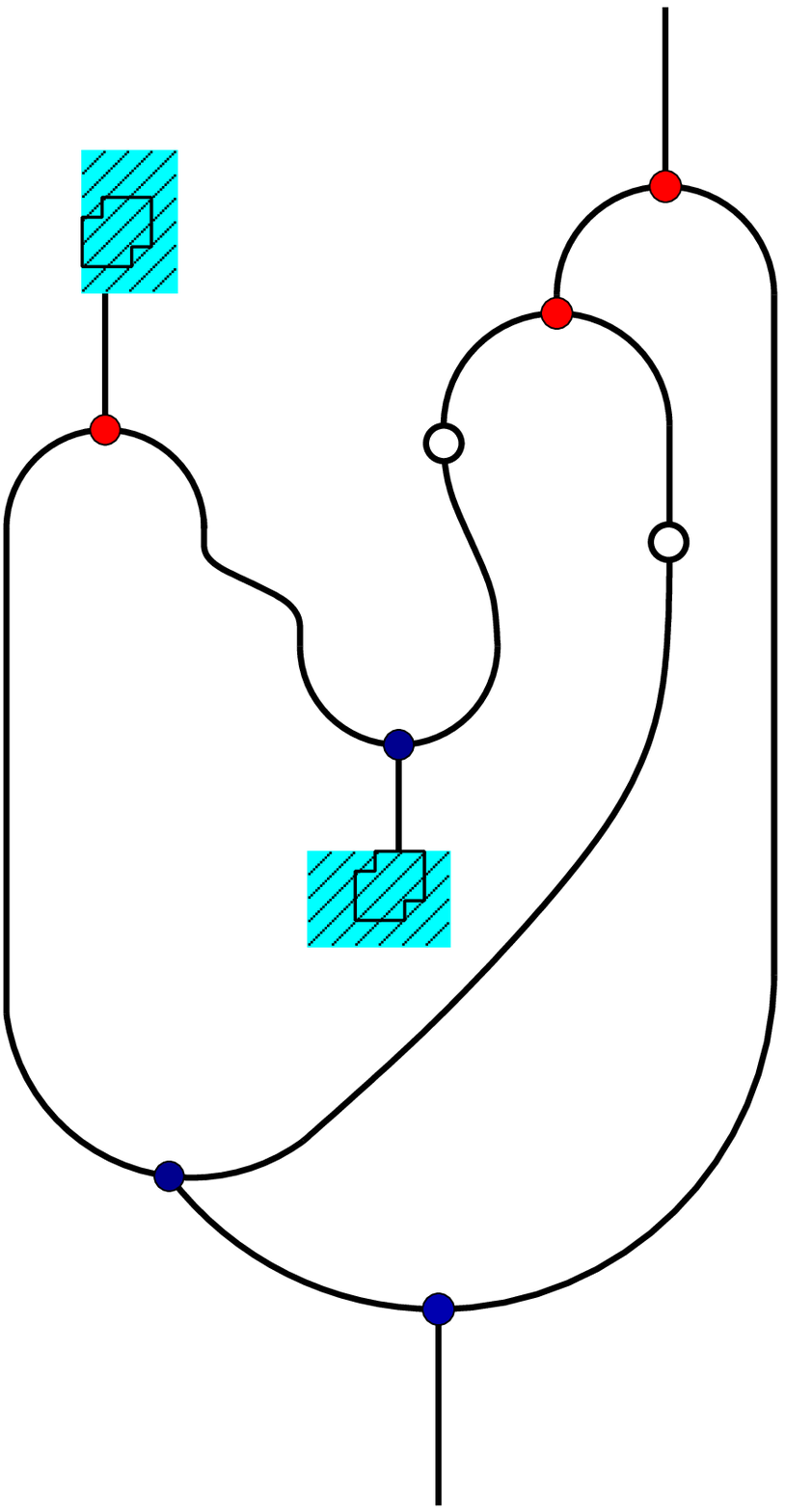}}
  \put(243,91) {$=$}
  \put(270,0)  {\Includepichopf{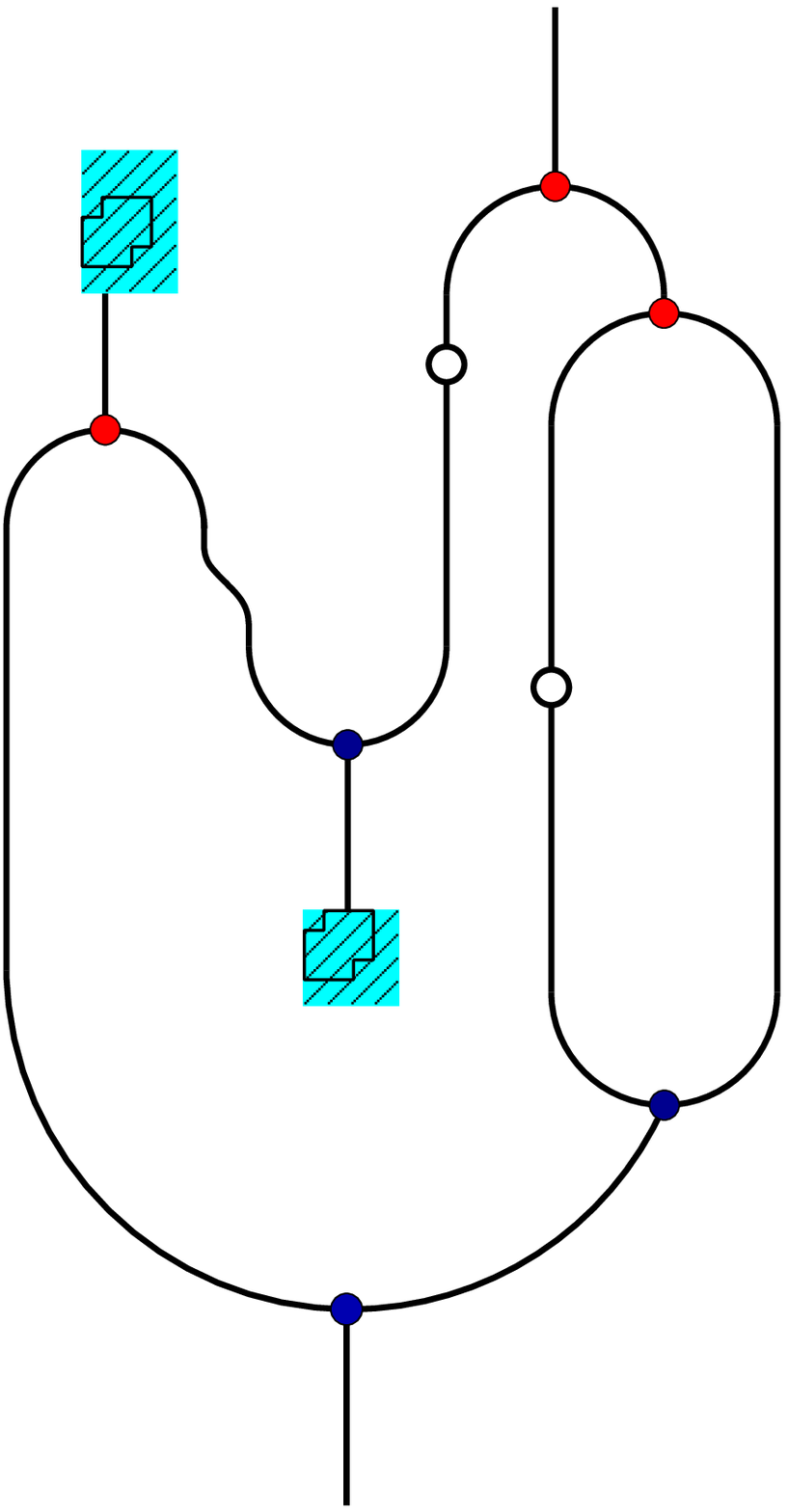}}
  \put(383,91) {$=$}
  \put(412,0)  {\Includepichopf{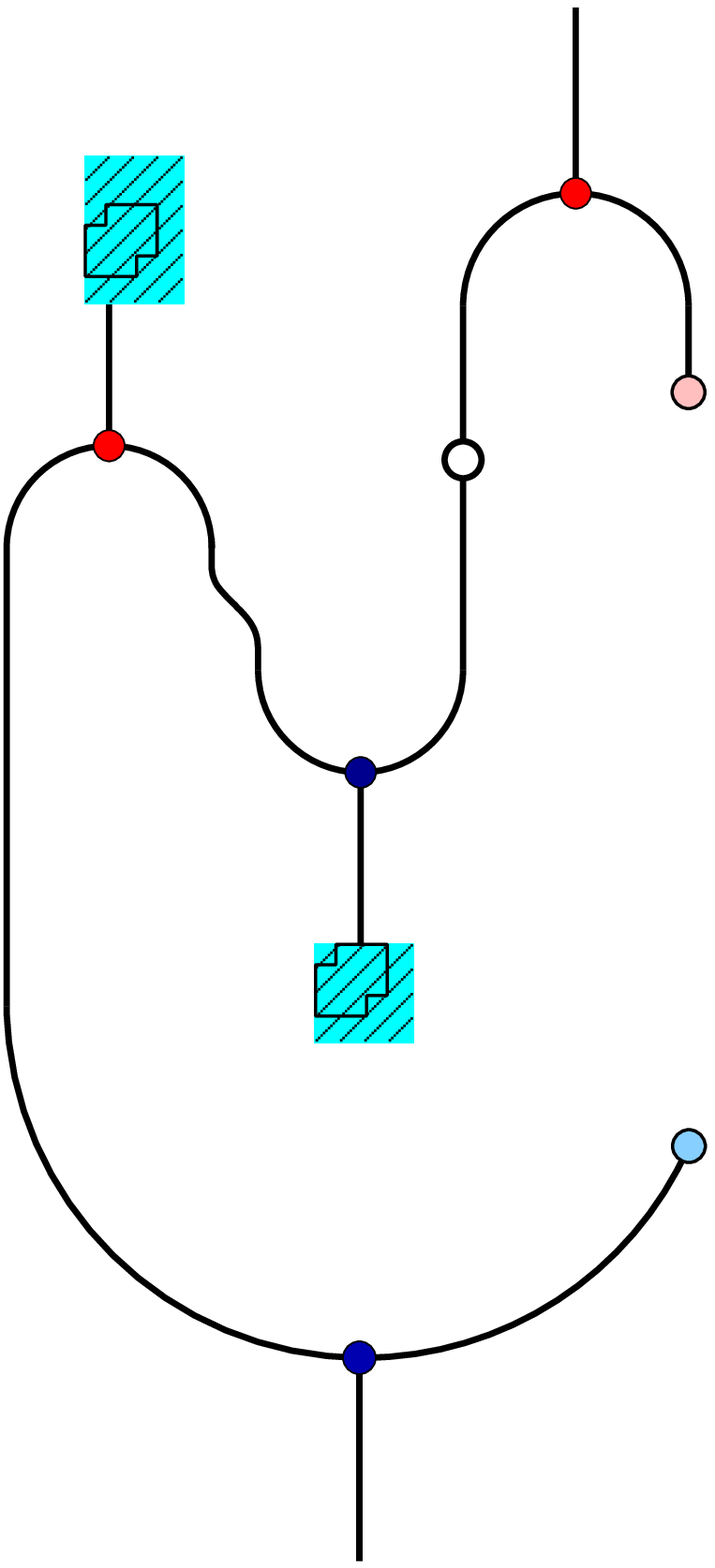}}
 }
  }
The first equality combines $\lambda \cir \Lambda \eq \id_\one$ with simple properties 
of unit, counit and antipode, the second implements the defining properties of $\lambda$ 
and $\Lambda$, the third is the compatibility of product and coproduct in a bialgebra, 
the fourth is the antialgebra property of the antipode together with a slight 
deformation of the graph, the fifth uses associativity and coassociativity, and the 
last follows by the defining property of the antipode. Using finally the defining 
properties of the unit and counit and composing from the left with $\apo^{-1}$ and 
from the right with $\apo$, one arrives at 
  \be
  \fmapi \circ \fmap = \idH \,.
  \ee 
To show that \fmapi\ is also a right-inverse, we proceed in two steps. First we 
show that
  \Eqpic{proof_fmap2} {380} {190} {
  \put(-80,203){
  \put(68,0)   {\Includepichopf{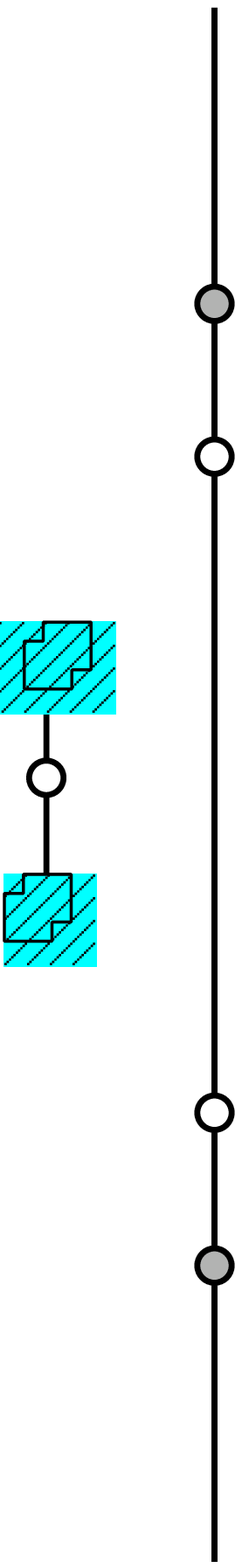}}
  \put(115,96) {$=$}
  \put(145,0)  {\Includepichopf{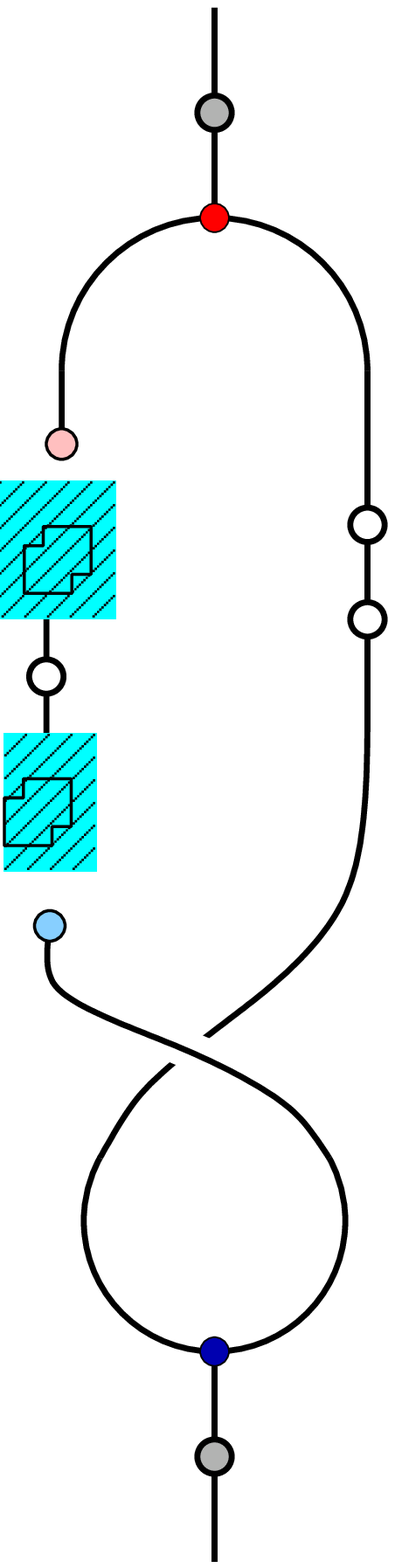}}
  \put(216,96) {$=$}
  \put(246,0)  {\Includepichopf{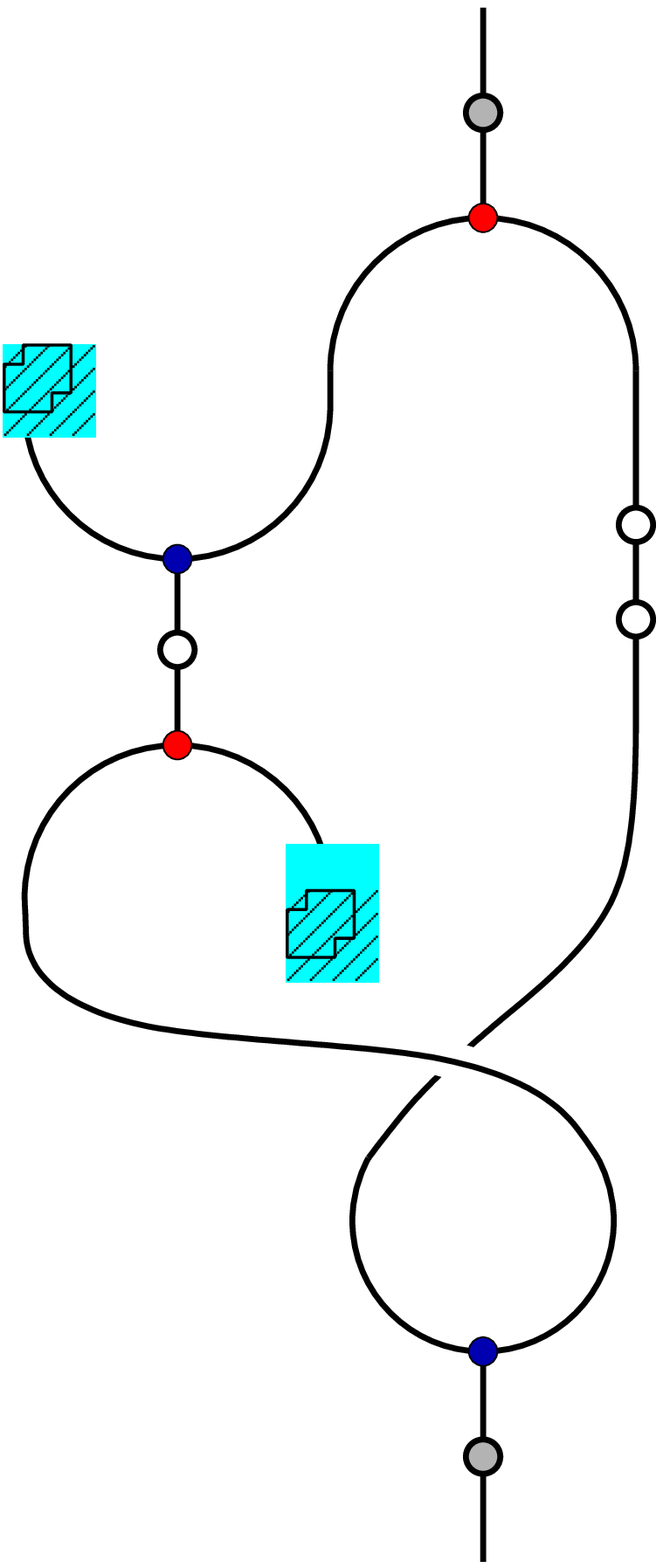}}
  \put(351,96) {$=$}
  \put(380,0)  {\Includepichopf{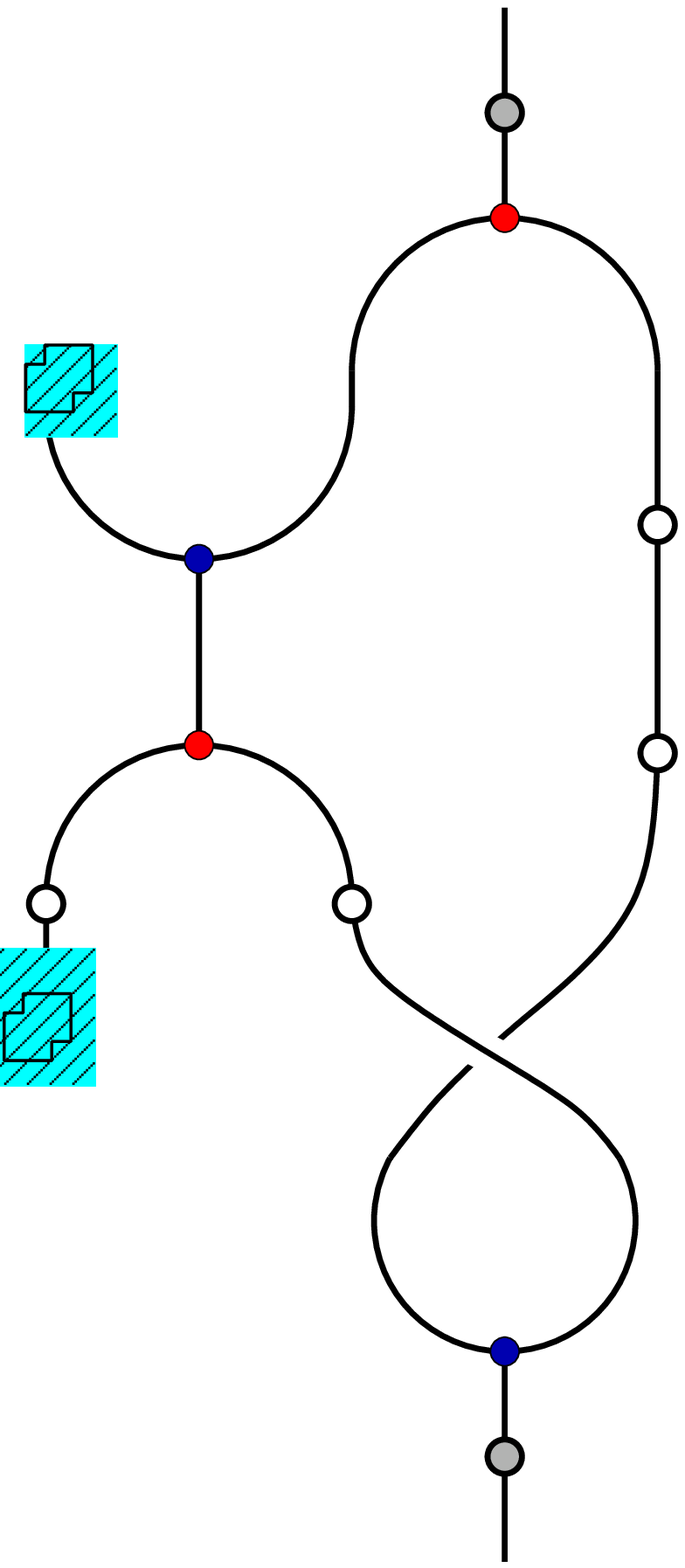}}
 }
  \put(52,-6){
  \put(-27,96) {$=$}
  \put(0,0)    {\Includepichopf{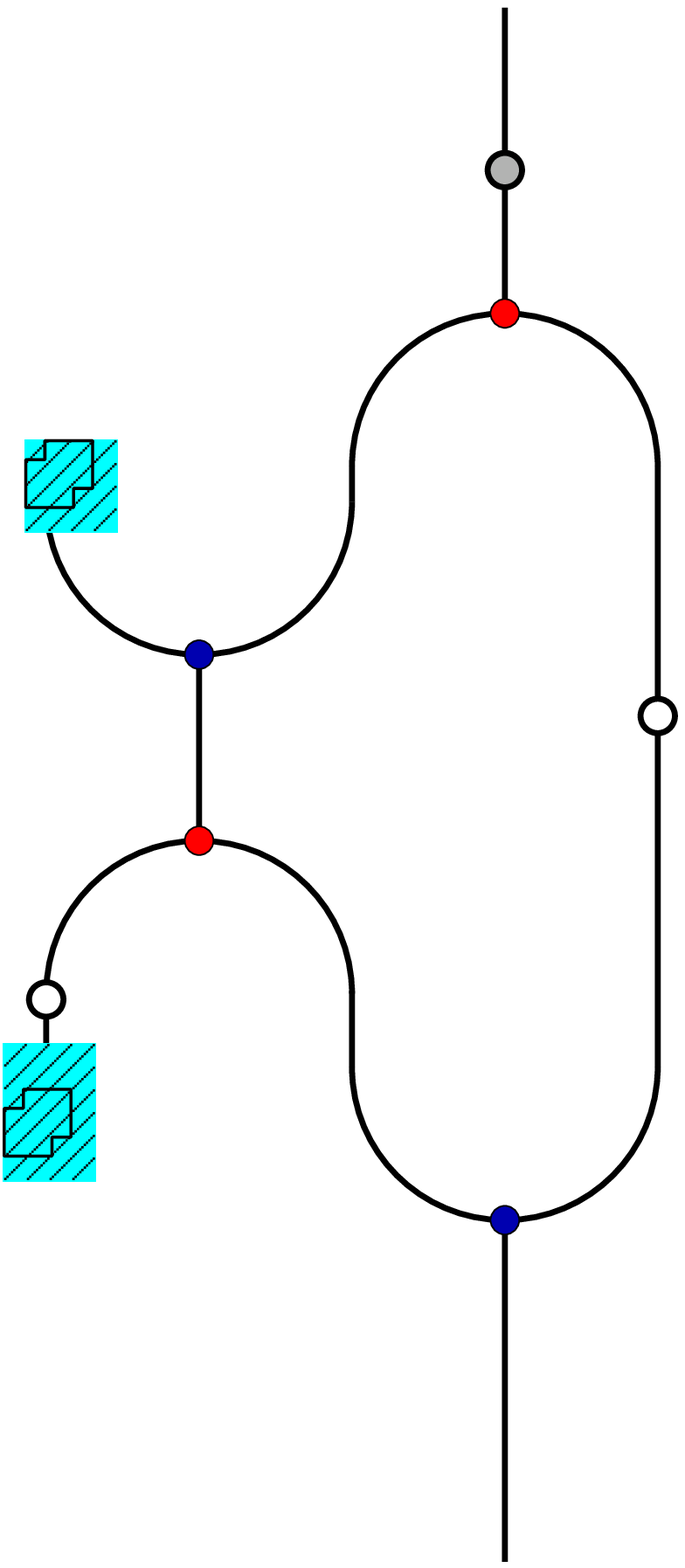}}
  \put(109,96) {$=$}
  \put(151,0)  {\Includepichopf{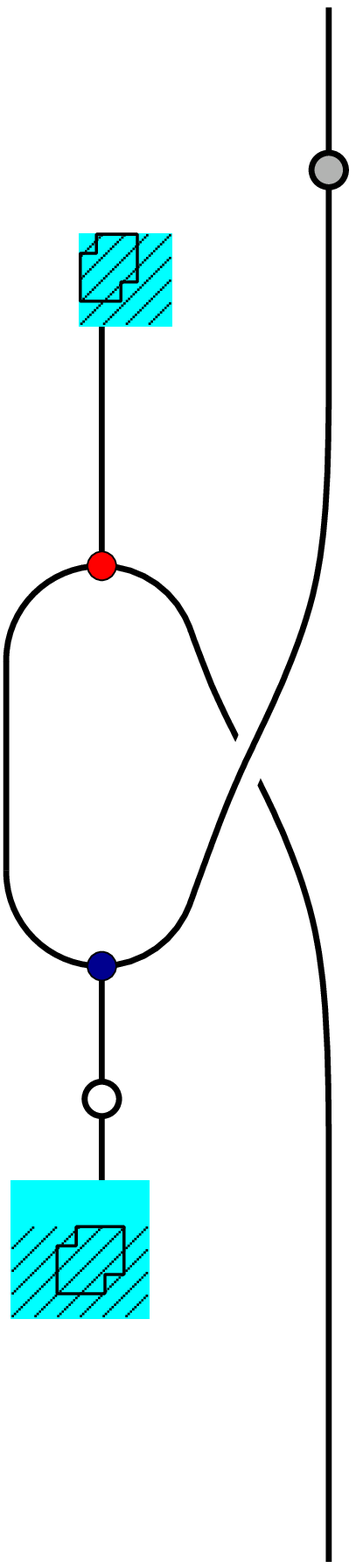}}
  \put(218,96) {$=$}
  \put(253,0)  {\Includepichopf{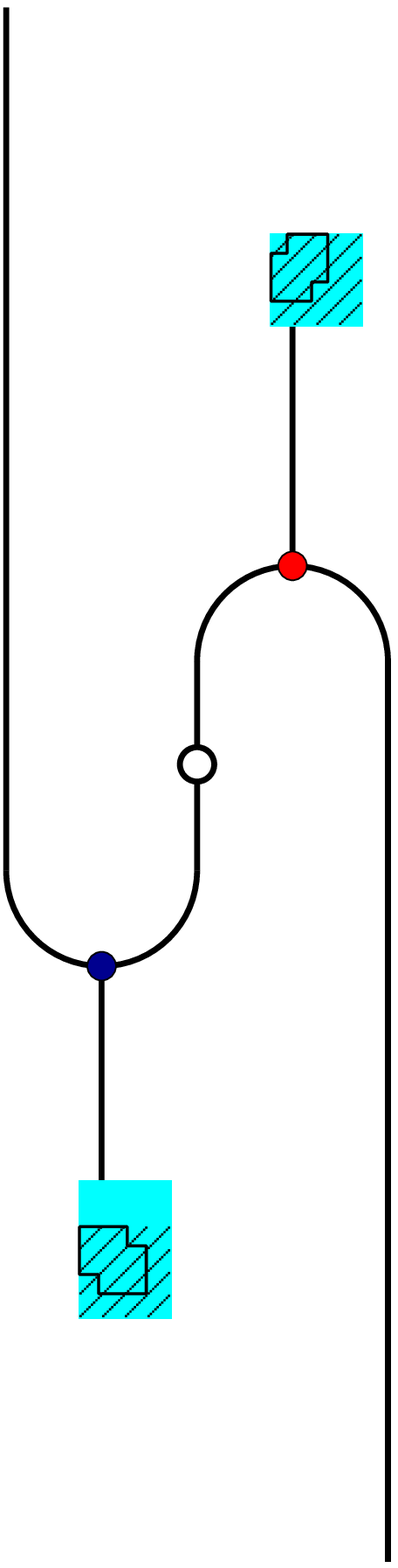}}
 } }
Here the first equality uses the defining properties of the unit and counit and
of $\lambda$ and $\Lambda$, the fifth follows by a combination of the compatibility 
of product and coproduct and of properties of the antipode, while all other 
equalities are just properties of the antipode. 
In the second step we compose the result just obtained with the 
(co)integrals and then use again the defining property of $\Lambda$, yielding
  \Eqpic{proof_fmap3} {355} {59} {
  \put(0,0)     {\Includepichopf{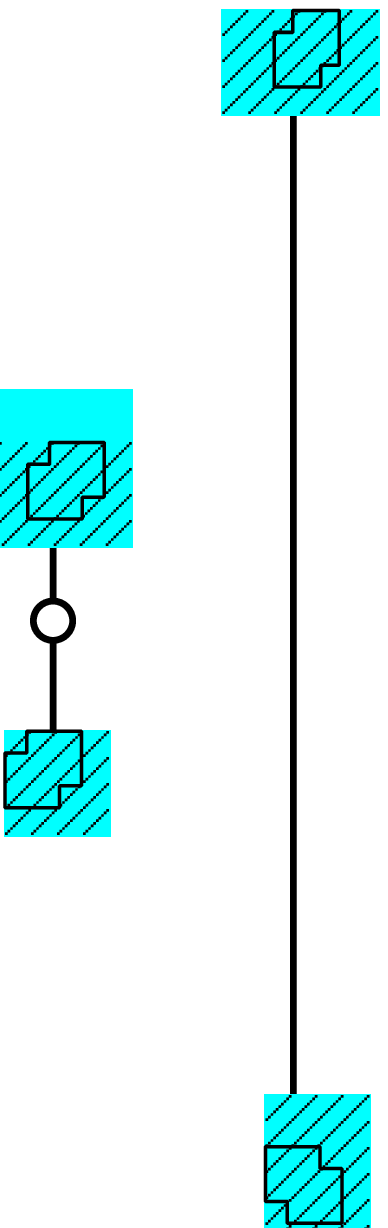}}
  \put(62,64)   {$ = $}
  \put(97,0)    {\Includepichopf{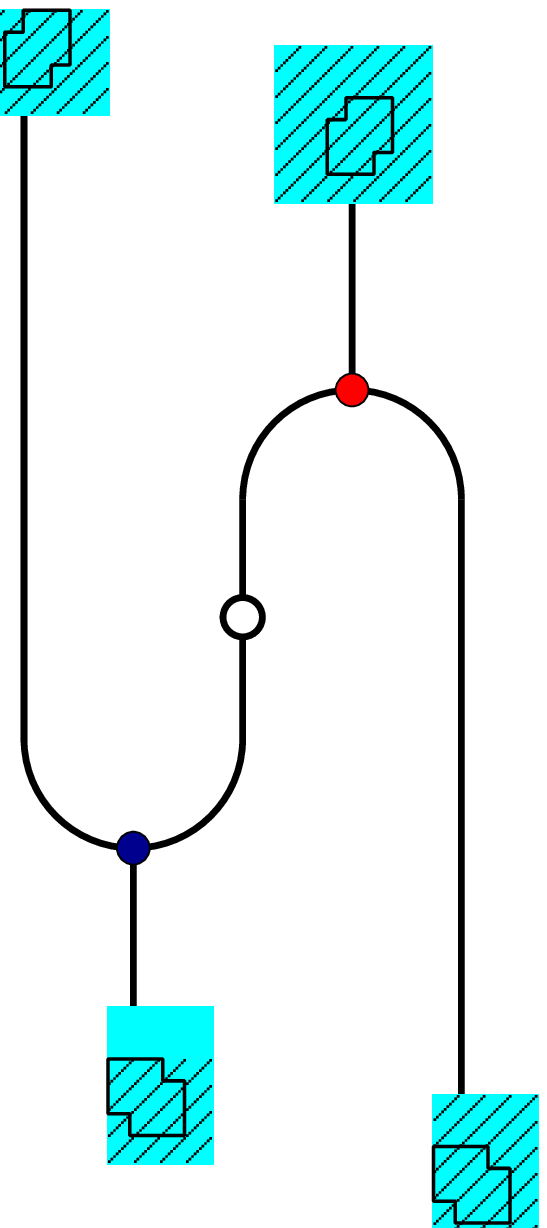}}
  \put(177,64)  {$ = $}
  \put(213,3)   {\Includepichopf{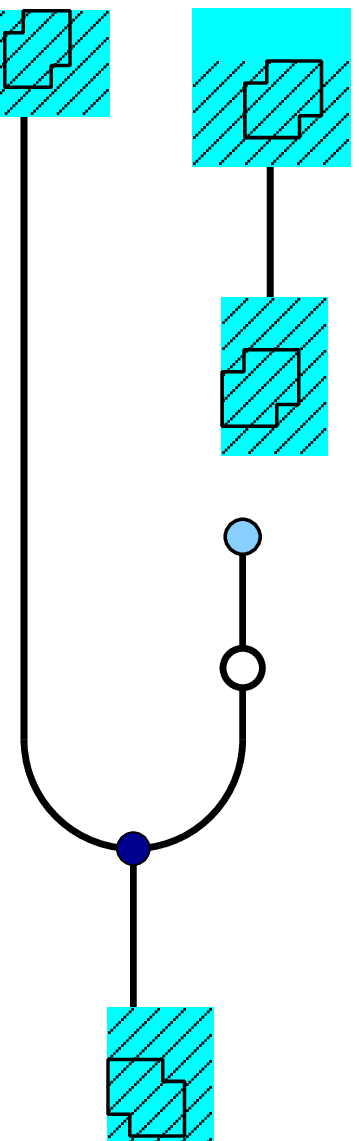}}
  \put(274,64)  {$ = $}
  \put(307,47)  {\Includepichopf{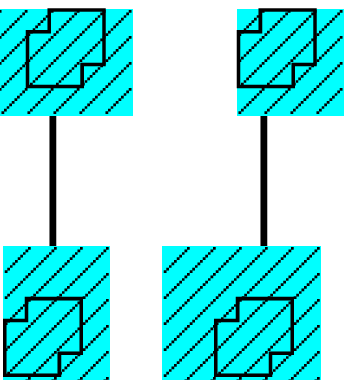}}
 }
and thus
  \be
  \lambda \circ \apo \circ \Lambda = \id_\one \,.
  \ee
Re-inserting this identity into \erf{proof_fmap2} finally proves that
  \be
  \fmap \circ \fmapi = \idH \,.
   \ee

\medskip

Similarly as for \findim\ Hopf algebras over a field or, more generally, over
a commutative ring (see e.g.\ \cite{pare7,kaSt2}), as a corollary of the
result above one shows that any Hopf algebra $(H,m,\eta,\Delta,\eps,\apo)$ 
of the form assumed above is naturally also a Frobenius algebra
$(H,m,\eta,\Deltaf,\epsf)$, with the same algebra structure and with 
Frobenius counit $\epsf \eq \lambda$ and Frobenius coproduct
  \Eqpic{Deltaf} {250} {47} {
  \put(-52,46)  {$ \Deltaf ~= $}
  \put(0,0)   {\Includepichopf{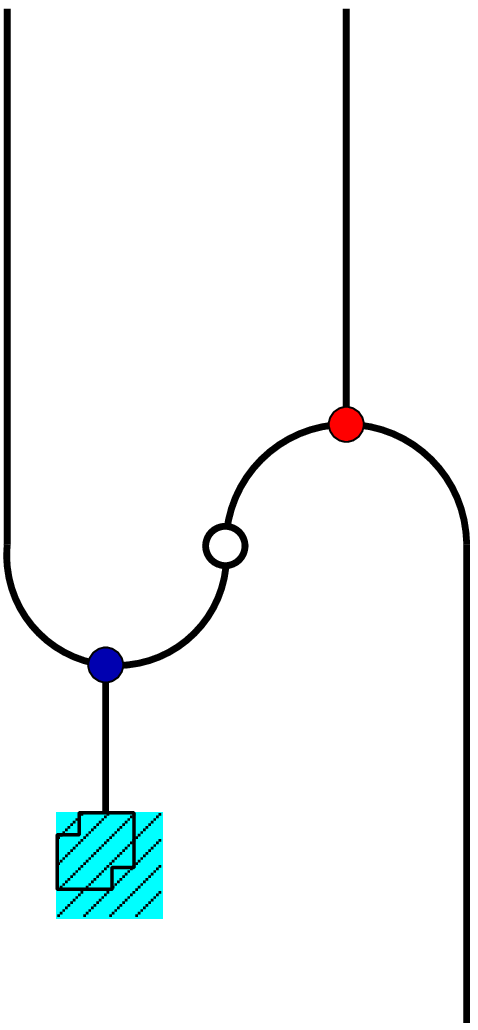}}
  \put(14.2,13) {$\sse\Lambda$}
  \put(17.6,52) {$\sse\apo$}
  \put(75,50)   {$=$}
  \put(110,0) {\Includepichopf{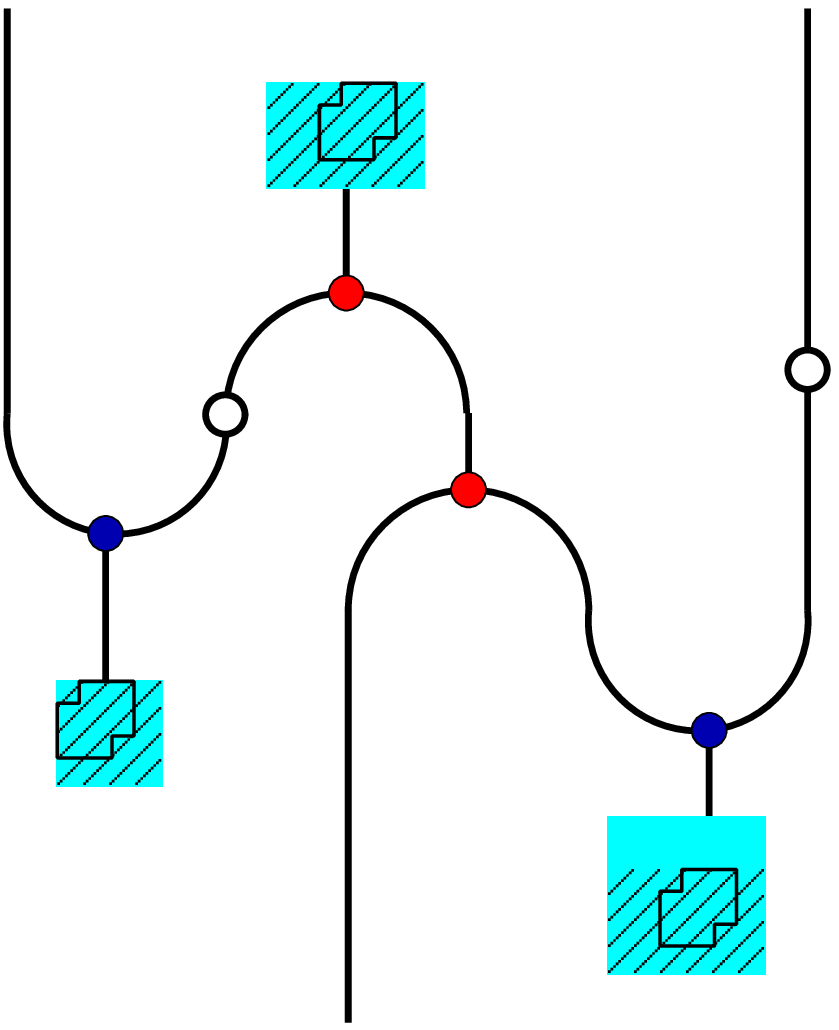}}
  \put(220,50)  {$=$}
  \put(255,0) {\Includepichopf{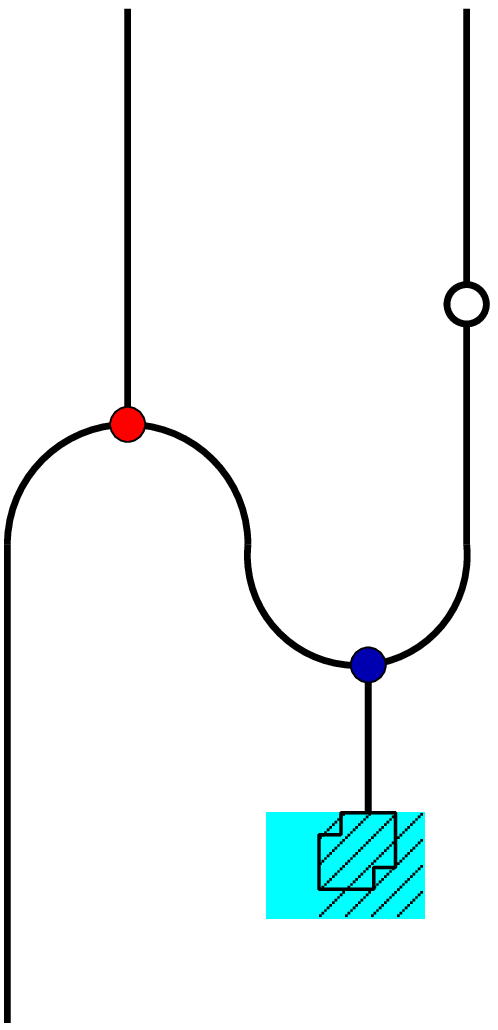}}
  }


 \newpage 

 \newcommand\wb{\,\linebreak[0]} \def\wB {$\,$\wb}
 \newcommand\Bi[2]    {\bibitem[#2]{#1}} 
 \newcommand\inBO[9]  {{\em #9}, in:\ {\em #1}, {#2}\ ({#3}, {#4} {#5}), p.\ {#6--#7} {{\tt [#8]}}}
 \renewcommand\J[7]   {{\em #7}, {#1} {#2} ({#3}) {#4--#5} {{\tt [#6]}}}
 \newcommand\JO[6]    {{\em #6}, {#1} {#2} ({#3}) {#4--#5} }
 \newcommand\JP[7]    {{\em #7}, {#1} ({#3}) {{\tt [#6]}}}
 \newcommand\BOOK[4]  {{\em #1\/} ({#2}, {#3} {#4})}
 \newcommand\prep[2]  {{\em #2}, preprint {\tt #1}}
 \def\adma  {Adv.\wb Math.}
 \def\apcs  {Applied\wB Cate\-go\-rical\wB Struc\-tures}
 \def\aspm  {Adv.\wb Stu\-dies\wB in\wB Pure\wB Math.}
 \def\aste  {Ast\'e\-ris\-que}
 \def\atmp  {Adv.\wb Theor.\wb Math.\wb Phys.}
 \def\bams  {Bull.\wb Amer.\wb Math.\wb Soc.}
 \def\cajm  {Ca\-nad.\wb J.\wb Math.}
 \def\coma  {Con\-temp.\wb Math.}
 \def\comp  {Com\-mun.\wb Math.\wb Phys.}
 \def\cpma  {Com\-pos.\wb Math.}
 \def\cudm  {Current Developments in Mathematics}
 \def\duke  {Duke\wB Math.\wb J.}
 \def\fiic  {Fields\wB Institute\wB Commun.}
 \def\geat  {Geom.\wB and\wB Topol.}
 \def\ijmp  {Int.\wb J.\wb Mod.\wb Phys.\ A}
 \def\imrn  {Int.\wb Math.\wb Res.\wb Notices}
 \def\injm  {Int.\wb J.\wb Math.}
 \def\inma  {Invent.\wb math.}
 \def\joac  {J.\wB Al\-ge\-bra\-ic\wB Com\-bin.}
 \def\joag  {J.\wB Al\-ge\-bra\-ic\wB Geom.}
 \def\joal  {J.\wB Al\-ge\-bra}
 \def\jopa  {J.\wb Phys.\ A}
 \def\jomp  {J.\wb Math.\wb Phys.}
 \def\jktr  {J.\wB Knot\wB Theory\wB and\wB its\wB Ramif.}
 \def\jlms  {J.\wB London\wB Math.\wb Soc.}
 \def\jpaa  {J.\wB Pure\wB Appl.\wb Alg.}
 \def\jram  {J.\wB rei\-ne\wB an\-gew.\wb Math.}
 \def\momj  {Mos\-cow\wB Math.\wb J.} 
 \def\mpcp  {Math.\wb Proc.\wB Cam\-bridge\wB Philos.\wb Soc.}
 \def\nupb  {Nucl.\wb Phys.\ B}
 \def\nyjm  {New\wB York\wB J.\wb Math}
 \def\pajm  {Pa\-cific\wB J.\wb Math.}
 \def\phlb  {Phys.\wb Lett.\ B}
 \def\plms  {Proc.\wB Lon\-don\wB Math.\wb Soc.}
 \def\pnas  {Proc.\wb Natl.\wb Acad.\wb Sci.\wb USA} 
 \def\prja  {Proc.\wB Japan\wB Acad.}
 \def\rvmp  {Rev.\wb Math.\wb Phys.}
 \def\taac  {Theo\-ry\wB and\wB Appl.\wb Cat.}
 \def\taia  {Top\-o\-lo\-gy\wB and\wB its\wB Appl.}
 \def\thmp  {Theor.\wb Math.\wb Phys.}
 \def\trgr  {Trans\-form.\wB Groups}

\end{document}